\documentclass[11pt]{article}
\usepackage{graphicx,color,colortbl}
\usepackage{epstopdf}
\usepackage{subfig}
\usepackage{setspace}
\usepackage{natbib}
\usepackage{enumitem}
\usepackage{pgf,tikz}
\usetikzlibrary{shapes,decorations,arrows,calc,arrows.meta,fit,positioning}
\tikzset{
    -Latex,auto,node distance =1 cm and 1 cm,semithick,
    state/.style ={ellipse, draw, minimum width = 0.7 cm},
    point/.style = {circle, draw, inner sep=0.04cm,fill,node contents={}},
    bidirected/.style={Latex-Latex,dashed},
    el/.style = {inner sep=2pt, align=left, sloped}
}
\usepackage{mathtools}
\usepackage[thin,thinp,thinc]{esdiff}
\usetikzlibrary{positioning}
\usepackage[margin=1in]{geometry}
\usepackage{multibib}
\newcites{SM}{References for Online Appendix}
\usepackage{minitoc} 
\definecolor{mycol}{HTML}{128ba0} 
\definecolor{amber}{rgb}{1.0, 0.49, 0.0}
\definecolor{navyblue}{HTML}{128ba0}
\definecolor{firebrick}{HTML}{b0015a}
\definecolor{mycol}{HTML}{128ba0}
\definecolor{mytheme}{HTML}{b0015a}
\definecolor{mythird}{HTML}{a5900d}
\newcommand{\indep}{\mathop{\perp\!\!\!\!\perp}}
\usepackage{bm}
\definecolor{Gray}{gray}{0.85} 
\newcolumntype{a}{>{\columncolor{Gray}}c} 
\usepackage{etoolbox}
\makeatletter

\patchcmd{\NAT@citex}
  {\@citea\NAT@hyper@{%
     \NAT@nmfmt{\NAT@nm}%
\hyper@natlinkbreak{\NAT@aysep\NAT@spacechar}{\@citeb\@extra@b@citeb}%
     \NAT@date}}
  {\@citea\NAT@nmfmt{\NAT@nm}%
   \NAT@aysep\NAT@spacechar\NAT@hyper@{\NAT@date}}{}{}

\patchcmd{\NAT@citex}
  {\@citea\NAT@hyper@{%
     \NAT@nmfmt{\NAT@nm}%
\hyper@natlinkbreak{\NAT@spacechar\NAT@@open\if*#1*\else#1\NAT@spacechar\fi}%
       {\@citeb\@extra@b@citeb}%
     \NAT@date}}
  {\@citea\NAT@nmfmt{\NAT@nm}%
\NAT@spacechar\NAT@@open\if*#1*\else#1\NAT@spacechar\fi\NAT@hyper@{\NAT@date}}
  {}{}

\makeatother 

\usepackage[colorlinks=true,
            linkcolor=mycol,
            citecolor=mycol,
            urlcolor=mycol
            ]{hyperref}

\usepackage{rotating}
\usepackage{pdfpages}
\usepackage{dcolumn}
\usepackage{amssymb}
\usepackage{amsmath}
\usepackage{multicol}
\usepackage{dsfont}
\newcommand{\E}{\mathbb{E}}
\newcommand{\V}{\mathbb{V}}
\newcommand{\p}{\mathcal{P}}
\newcommand{\A}{\mathcal{A}}

\usepackage{cochineal} 
\usepackage{setspace}
 \begin{titlepage}
 \title{Causal Inference with Ranking Data: Application to Blame Attribution in Police Violence and Ballot Order Effects in Ranked-Choice Voting\thanks{For valuable feedback, I would like to thank Olivier Bergeron Boutin, Benjamin Carter, Charles Crabtree, Amanda Sahar d'Urso, Justin Grimmer, Yusaku Horiuchi, Michael Jankowski, Luke Keele, Jae Yeon Kim, Seo-young Silvia Kim, Yui Nishimura, Brendan Nyhan, Jochen Rehmert, Hikaru Yamagishi, and participants of the 39th Annual Meeting of the Society for Political Methodology, the Junior Americanists Workshop Series, and workshops at Dartmouth College (Department of Government, Program in Quantitative Social Science). I would also like to thank Mattias Agerberg, Simone Dietrich, Anne-Kathrin Kreft, and David Steinberg for sharing their experimental data. This study was approved by the Institutional Review Board (IRB) at Dartmouth College (Study ID: STUDY00032593).}}
 \author{{\Large Yuki Atsusaka\thanks{Yuki Atsusaka is a Postdoctoral Fellow at Dartmouth College. Hanover, NH 03755 (\href{https://atsusaka.org}{\color{mycol} \textsf{https://atsusaka.org}}).}} \\ \vspace{0.3cm}
 \href{mailto:Yuki.Atsusaka@dartmouth.edu}{\Large \color{mycol} \textsf{Yuki.Atsusaka@dartmouth.edu}}}
 \date{\today}
 \end{titlepage}
\begin{document}
\doparttoc 
\faketableofcontents 
\maketitle
\thispagestyle{empty} 
\vspace{-1cm}

\begin{abstract}
\normalsize
\noindent While rankings are at the heart of social science research, little is known about how to analyze ranking data in experimental studies. This paper introduces a potential-outcomes framework to perform causal inference when outcome data are ranking data. It introduces a class of causal estimands tailored to ranked outcomes and develops methods for estimation and inference. Furthermore, it extends the framework to partially ranked data. I show that partial rankings can be considered a selection problem and propose nonparametric sharp bounds for the treatment effects. Using the methods, I reanalyze the recent study on blame attribution in the Stephon Clark shooting, finding that people's attitudes toward officer-involved shootings are robust to contextual information. I also apply the framework to study ballot order effects in three ranked-choice voting (RCV) elections in 2022, proposing a new theory of pattern rankings in RCV. Finally, I present three applications in international relations.\\

\end{abstract} 

\noindent \textsf{Word Count: 11999}

\newpage
\setcounter{page}{1} 
\doublespacing
\section*{Introduction}


Rankings are at the heart of social sciences \citep{luce1959individual,arrow2012social,regenwetter2006behavioral,train2009discrete}. Political scientists, for example, have used the concept of rankings and ranking data (e.g., 15243) to study diverse topics across subfields, including American politics \citep[e.g.,][]{kaufman2021measure}, comparative politics \citep[e.g.,][]{baldwin2010economic}, and international relations \citep[e.g.,][]{kelley2015politics} (see Online Appendix \ref{app:application} for forty applications).  Moreover, a growing body of experimental works has analyzed ranking data to study an array of political phenomena, such as voting and social sanctions \citep{gerber2016people}, political values \citep{reeskens2021stability}, racism and media framing \citep{nelson1997media}, police violence \citep{boudreau2019police}, gender and candidate selection \citep{jankowski2022}, indigenous recognition \citep{mcmurry2022recognition}, blame attribution \citep{malhotra2008attributing}, distributive justice \citep{frohlich1987choices,frohlich1990choosing},  ballot design \citep{horiuchi2019randomized}, economic coercion \citep{gueorguiev2020impact}, foreign aid \citep{dietrich2016donor}, and sexual violence and military intervention \citep{agerberg2022sexual}. In these works, researchers study the treatment effects on ranking data as multidimensional outcome data \citep{marden1995analyzing,alvo2014statistical,yu2019analysis}. However, despite the promising development, it has not been clear what kinds of treatment effects can be studied from ranking outcome data in political science, let alone how to estimate and perform inference on them.

This paper introduces a potential-outcomes framework of causal inference with ranking outcome data. Moreover, it uses the framework to study two topics in American and comparative politics: (1) people's attitudes toward police violence \citep{davis2017policing,boudreau2019police,mcgowen2020racialized} and (2) ballot order effects in ranked-choice voting (RCV) \citep{orr2002ballot,king2009ballot,curtice2014confused}. Furthermore, Online Appendix \ref{app:IR} presents three applications in international relations by examining conflict-related sexual violence \citep{agerberg2022sexual}, foreign aid \citep{dietrich2016donor}, and U.S.-China currency relations \citep{gueorguiev2020impact}.

This work makes several contributions. First, I define a class of causal estimands tailored to ranking outcome data based on the potential-outcomes framework \citep{imbens2015causal}. I also propose nonparametric estimators for point identification and discuss inferential methods for two causal estimands: average rank effects and average pairwise rank effects. Furthermore, I also extend the framework to \textit{partially ranked data}, where rankers (e.g., voters) do not rank all available items (e.g., candidates) \citep{critchlow2012metric} and thus data are partially missing. Building on the framework of principal stratification \citep{frangakis2002principal}, I show that partial rankings can be considered a selection problem \citep{knox2020administrative}. To address the missing data problem in partially ranked data, I propose nonparametric sharp bounds on the treatment effects based solely on the logical constrain of ranking data and expected directions of treatment effects \citep{horowitz2000nonparametric,manski2009identification}. This paper also discusses a range of causal estimands for ranking data that can be useful in political science. The concluding section offers practical guidance for future research about how to choose appropriate estimands and study more complex hypotheses.

Next, I apply the proposed framework to the research on police violence and RCV. In the first study, I reanalyze the experimental data collected by \citet{boudreau2019police}. This study examines the effects of contextual information on people's responses to an officer-involved shooting in Sacramento, California. In their experiment, survey respondents were asked to rank order seven relevant parties, including an unarmed African American victim and two white police officers who shot the victim, according to their perceived levels of responsibility for the victim's death. Unlike the original study using parametric models, I find that people's responses to police violence are remarkably robust to receiving contextual information. Moreover, respondents of different races show widely different opinions even when they receive the same information \citep[echoing][]{jefferson2021seeing}, reaffirming the importance of jury selection in police violence cases \citep{cook2017biased,futrell2018visibly,davis1994rodney}.

In the second study, I develop an experimental design to examine how ballot order (i.e., candidate order) affects candidate ranks under RCV, where partial rankings frequently appear \citep{burnett2015ballot,kilgour2020prevalence}. I perform three survey experiments in the Oakland mayoral election, the U.S. House election in Alaska, and the U.S. Senate election in Alaska in 2022. The experiments reveal a new puzzle --- while ballot order still matters, it affects how people vote under RCV in a way that cannot be fully explained by previous research \citep[e.g.,][]{alvarez2006much}. To explain the puzzle, I propose a new theory of \textit{pattern rankings} and offer evidence for potential voters who vote by following geometric patterns instead of offering their sincere ranked preferences.

Finally, this paper seeks to unify the literature on ranking data and causal inference. On the one hand, it offers opportunities for scholars to use ranking data with the Neyman-Rubin causal model \citep{neyman1923application,imbens2015causal}. While the ranking data literature has studied the analysis of randomized block designs \citep{bradley1952rank,alvo1993rank,alvo2014statistical}, no practical method has been available for political scientists to analyze ranked outcomes more generally. On the other hand, this work contributes to the causal inference literature by addressing another type of complex outcome data. Recent works have studied the treatment effects on multidimensional and/or structured data, such as text \citep{fong2016discovery,egami2018make}, ordinal data \citep{chiba2018bayesian,lu2018treatment,yamauchi2020difference}, and multivariate outcomes \citep{lupparelli2017causal}. This work sheds light on opportunities and challenges in analyzing multidimensional and structured outcomes in the form of rankings.






\section{Motivating Application: People's Attitudes toward Police Violence}

This section introduces the first motivating research on people's responses to the use of deadly force by police officers toward unarmed Black men. 

\subsection{Motivation, Design, Data, and Hypotheses}


How does contextual information affect people's blame attribution in police violence? With increasing attention to officer-involved shootings \citep{davis2017policing}, its subsequent trials (or lack thereof) \citep{fiarfax2017}, and mass protests against them \citep{reny2021opinion}, answers to this question have important implications for multiple areas of research. Such include the role of media in shaping public opinions on race and law enforcement \citep{porter2018public,jefferson2021seeing}, the role of prosecutors in grand and trial jury proceedings in police violence cases \citep{cook2017biased,futrell2018visibly,davis1994rodney}, and racial differences in perceptions of officer-involved shootings \citep{mullinix2021feedback,mcgowen2020racialized,strickler2022racial}.

\citet{boudreau2019police} address this question by performing a survey experiment. More specifically, the authors examine people's responses to the shooting of Stephon Clark on March 18, 2018, in Sacramento, California. In this incident, Stephon Clark, an unarmed 22-year-old African American man, was shot multiple times and killed by two white police officers from the Sacramento Police Department. After about a year, Sacramento County District Attorney (DA) declined to file criminal charges against the officers. Multiple protests and social unrest followed both the shooting and the DA's announcement. To study whether contextual information shapes people's perceptions of the shooting, the authors conducted a survey using samples from four California counties, including Sacramento.

The survey experiment shows respondents different contextual information about the shooting. Next, it shows respondents a list of seven parties related to the incident, including the victim, the two white police officers, the chief of the Police Department, the mayor of Sacramento, the DA, the governor of California, and two California Senators (including Kamala Harris). Finally, it asks respondents to \textit{rank order} the seven parties according to how much each party is responsible for the death of the victim (i.e., 1 means the most and 7 means the least responsible). 

Table \ref{tab:police_data} displays the first six observations from their replication data. For example, the first respondent reports a ranked outcome \textsf{(7,1,2,6,3,5,4)}. This indicates that the individual thinks that the officers are the most responsible and the police chief is the second most responsible, while the victim is the least responsible. In contrast, the sixth respondent reports a ranked outcome \textsf{(1,2,6,5,7,3,4)}. This suggests that the person believes that the victim is the most responsible and the officers are the second most responsible, while the mayor is the least responsible. Thus, each respondent has a \textit{vector} of values as outcome data (as opposed to a scalar) in this design.

\indent

\begin{table}[h!]
    \centering
    \begin{tabular}{cccccccccc}\hline
        ID & County & Group & Victim & Officers & Chief & DA & Mayor & Governor & Senators  \\\hline
        100001 &  Orange     &	control &   7 & 1 & 2 & 6 & 3 & 5 & 4\\
        100002 &  Sacramento &  pattern &	7 &	1 & 2 &	3 &	5 &	4 & 6\\
        100003 &  Sacramento &	reform &	1 &	2 &	3 &	4 &	5 &	6 &	7\\
        100004 &  Los angeles& 	pattern &	2 &	1 &	3 &	4 &	5 &	7 &	6\\
        100005 &  Riverside  &	control &	1 &	2 &	3 &	5 &	4 &	7 &	6\\
        100006 &  Riverside	 &  pattern &	1 &	2 &	6 &	5 &	7 &	3 &	4\\ \hline
    \end{tabular}
    \caption{\textbf{Rank Outcomes from the Police Violence Data in \citet{boudreau2019police}.}\\\textit{Note}: This table displays the first six observations from replication data of \citet{boudreau2019police}. The variable names were modified for simplicity.}
    \label{tab:police_data}
\end{table}

The treatment in this experiment is the type of contextual information. Respondents are randomly assigned to one of three groups, including the control, ``pattern-of-violence'' treatment, and ``reform'' treatment groups. The control units are exposed to episodic information about the shooting. In contrast, respondents in the first treatment group are exposed to both episodic information \textit{and} additional information, which describe that similar incidents involving police brutality have occurred in Sacramento. Similarly, in the second treatment condition, respondents are given both episodic information \textit{and} additional information, which explains that the police department has implemented an extensive reform to prevent officer-involved shootings. Online Appendix \ref{app:vignette} presents the full vignettes.

The experiment is motivated by two complex hypotheses  \citep[1103]{boudreau2019police}. The first hypothesis states:

\begin{quote}
    ``\textbf{H1.} Citizens who receive thematic information describing a pattern of police violence will place \textbf{less blame} on the victim, state, and federal officials and \textbf{more blame} on the police and local officials who oversee them than citizens who receive only episodic information about
police violence'' (emphasis added).
\end{quote}

\noindent The second hypothesis follows:

\begin{quote}
``\textbf{H2.} Citizens who receive thematic information summarizing reforms will place \textbf{less blame} on the victim, the police, and local officials and \textbf{more blame} on state and federal officials than citizens who receive only episodic information about police violence'' (emphasis added).
\end{quote}

\subsection{Limitations in the Original Analysis}

To test these hypotheses, the study uses a \textit{rank-ordered logit} model, which is an extension of multinomial logit for ranking data \citep{beggs1981assessing,train2009discrete}. Online Appendix \ref{app:discuss} details how the authors use their estimated coefficients to compute their (somewhat ad hoc and non-causal) quantities of interest. Substantively, the study finds evidence that supports its hypotheses, concluding that contextual information alters people's blame attribution in the shooting of Stephon Clark. While the study offers new insights on an important issue with a novel experimental design, there are several limitations in the original analysis. 


First, its model specification does not offer direct evidence for its hypotheses. If properly modeled, the study could have recovered the treatment effects on the \textit{probability that each party will be the most responsible} for the victim's death. However, this still does not answer whether each treatment causes people to put \textit{more blame} on some parties and \textit{less blame} on others. Second, the study does not clearly define its causal quantities of interest and clarify how its model-based approach recovers its target effects. While it is essential to define causal estimands in any study before designing and analyzing experiments, it is more so with ranking data due to its more complex structure than conventional outcome data (e.g., continuous and binary outcomes).





\section{A Potential-Outcomes Framework for Ranking Data}
This section proposes a potential-outcomes framework to overcome the above challenges.

\subsection{Notation}

Rankings arise when people put a sequence of numbers on multiple items according to given criteria (e.g., relative preference and importance) \citep{marden1995analyzing,alvo2014statistical,yu2019analysis}. For example, survey respondents may rank order three issue items $\{\text{Abortion}, \text{Gun},  \text{Inflation}\}$ from the most important (1) to the least important (3) in a pre-election poll. Here, the outcome data for each respondent is multidimensional ordered data as it is a \textit{vector} containing an \textit{ordered} sequence of numbers (e.g., 132). 

Formally, let $\A = \{A_{1},...,A_{J}\}$ be a set of $J$ labelled items $A_{j}$ $(j=1,...,J)$. Let $\p_{J}$ be a permutation space of size $J!$. As a function, \textit{ranking} is defined as a mapping $\textbf{R}: \A \mapsto \p_{J}$ from the finite set of labeled items to the permutation space. 

Let $\textbf{R}_{i} \in \p_{J}$ be a specific ranking provided by person or unit $i$ $(i=1,...,N)$. In the above example, unit $i$'s ranking -- $\textbf{R}_{i}$ -- is an element of the permutation space $\p_{3} = \{(123),(132), (213), (231), (312), (321)\}$. In other words, $\p_{J}$ is the sample space for ranking data with $J$ items.\footnote{Importantly, this distinguishes \textit{ranking data} from rank-based statistics commonly used in nonparametric statistical tests \citep{conover1999practical}, where rankings are used to summarize scalar-based outcomes.} 

Finally, I use $R_{ij}$ to denote a \textit{marginal ranking} of item $j$ assigned by person $i$. For example, when $\textbf{R}_{i}=(312)$, unit $i$'s marginal rankings become $R_{i1}=3$, $R_{i2}=1$, and $R_{i3}=2$ (e.g., ranker $i$ thinks that the third item is the second most important). Similarly, I use $I(R_{ij} < R_{ik})$ to denote an indicator function for \textit{pairwise ranking} that takes 1 if person $i$ prefers item $j$ to item $k$ (e.g., person $i$ likes item $j$ more than item $k$) and 0 otherwise.




\subsection{Potential Outcomes and Causal Estimands}

Consider a completely randomized experiment where $N$ subjects are drawn from an infinite \underline{s}uper \underline{p}opulation of size $N_{sp}$ (where $N_{sp}$ is large relative to $N$) via simple random sampling with replacement. Thus, this paper takes the super-population perspective, where potential outcomes are considered to be stochastic \citep[109-112]{ding2017bridging,imbens2015causal}. Consider also a simple treatment regime in which each individual is assigned to the treatment or control group.

Let $\textbf{R}_{i}(1)$ be the potential ranking under the treatment and $\textbf{R}_{i}(0)$ the potential ranking under the control for unit $i$. Let $D_{i}$ be a binary random variable denoting whether unit $i$ is assigned to the treatment ($D_{i}=1)$ or the control group ($D_{i}=0$). Then, unit $i$'s \textit{observed} ranking can be represented by a function of the two potential rankings and the treatment indicator:
\begin{equation}
    \textbf{R}_{i}^{\text{obs}} = \textbf{R}_{i}(1)D_{i} + \textbf{R}_{i}(0)(1 - D_{i})
\end{equation}





\subsubsection{General Average Treatment Effects for Ranked Outcomes}

Let $g()$ be a known function that maps rankings onto $n$-th dimensional real numbers $g: \mathcal{P}_{J} \mapsto \mathcal{R}^{n}$, where $n \leq J!$. Let $d(\alpha,\beta)$ be a known distance function that quantifies the distance between two vectors $\alpha$ and $\beta$. The general form of average treatment effects on ranking outcome data can be defined as follows:
\begin{equation}
    ATE_{g,d} \equiv \E \big[ d\big( g(\textbf{R}_{i}(1)),  g(\textbf{R}_{i}(0)) \big) \big]\label{eq:general}
\end{equation}

Here, $g()$ can take any form on a spectrum of functions that vary in how much data reduction occurs. For example, in the limit case with no data reduction, it becomes an identity function $g(\textbf{R}_{i}) = \textbf{R}_{i}$ that returns an original ranking of size $J!$. In the other benchmark scenario, it can be an indicator function $g(\textbf{R}_{i}) = I(R_{ij}=1)$ that returns 1 if item $j$ is most preferred and 0 otherwise. All other $g$ functions can be located between the two cases on the spectrum.

When $\alpha$ and $\beta$ are scalars, $d(\alpha,\beta)$ can be a simple difference $\alpha - \beta$. Namely,
\begin{equation}
    ATE_{g} \equiv \E \big[ g(\textbf{R}_{i}(1)) -  g(\textbf{R}_{i}(0))  \big]\label{eq:general2}
\end{equation}

Without loss of generality, the rest of this paper assumes that $d()$ is a simple difference. When $\alpha$ and $\beta$ are vectors, it can be a rank distance function (e.g., the Kendall distance) or any other distance function, such as the Mahalanobis distance \citep{marden1995analyzing,alvo2014statistical}. 

The key insight is that researchers can apply the same principle in the potential-outcomes framework to ranking data once they specify $g()$. Importantly, which $g()$ researchers should use \textit{depends upon their hypothesis}. If the hypothesis concerns whether the treatment increases the rank of a specific item $j$ (e.g., DA), using marginal ranks $g(\textbf{R}_{i}) = R_{ij}$ can be useful. Similarly, if the hypothesis focuses on a pair of items  $j$ and $k$ (e.g., the victim vs. the officers), using pairwise rankings $g(\textbf{R}_{i}) = I(R_{ij}<R_{ik})$ may be informative.

Motivated by the running application, this paper focuses on two $g()$ functions that address average ranking and pairwise ranking. Moreover, I consider different causal estimands below. In \textsf{Practical Guidance for Future Research}, I further discuss the choice of $g()$.

\subsubsection{Average Rank Effect (ARE)}
The first effect measure quantifies how much the treatment increases the average rank of an item $j$.\footnote{\citet{gerber2016people} implicitly used this estimand. Research on conjoint analysis has considered similar quantities \citep{abramson2022we,bansak2022using}.} Let $R_{ij}(1)$ and $R_{ij}(0)$ be a potential marginal ranking that unit $i$ would assign to item $j$ had she been exposed to the treatment and control conditions, respectively. I define the super-population \textit{average rank effect} (ARE) for item $j$ ($j=1,2,...,J$) as follows:
\begin{subequations}
\begin{align}
\tau_{j} & \equiv \E\big[R_{ij}(1) - R_{ij}(0)\big]\label{eq:mre1}\\
     & = \E\big[R_{ij}(1)\big] - \E\big[R_{ij}(0)\big],\label{eq:mre2}
\end{align}
\end{subequations}

\noindent where the expectation is taken over both the randomization distribution and the sampling distribution \citep[99]{imbens2015causal}.

The logical bound for this effect is $\tau_{j} \in (1-J, J-1)$. Equation (\ref{eq:mre2}) means that the ARE is defined as the difference between item $j$'s \textit{average rankings} under the treatment and control conditions. Since the ARE can be defined for each item, there are $J$ causal effects in total $\bm{\tau} = (\tau_{1},...,\tau_{J})^{\top}$. The ARE is informative when researchers are interested in the overall effects of the treatment on each item separately. 

\subsubsection{Average Pairwise Rank Effect (APE)}
A second effect measure quantifies how much the treatment increases the probability that item $j$ is preferred to item $k$. Let $I(R_{ij} < R_{ik})(1)$ denote a potential pairwise ranking under the treatment and $I(R_{ij} < R_{ik})(0)$ under the control. For two items $j$ and $k$, I define a \textit{pairwise ranking effect} (APE) as follows:
\begin{subequations}
\begin{align}
\tau^{P}_{jk} & \equiv \E\big[I(R_{ij} < R_{ik})(1) - I(R_{ij} < R_{ik})(0) \big] \\
&= \E\big[I(R_{ij} < R_{ik})(1)\big] -  \E\big[I(R_{ij} < R_{ik})(0)\big]\\
&= \mathbb{P}((R_{ij} < R_{ik})(1)) - \mathbb{P}((R_{ij} < R_{ik})(0)) 
\end{align}
\end{subequations}

The logical bound for this estimand is $\tau_{jk}^{P} \in (-1,1)$. Substantively, the APE means the difference between the probabilities that one item is preferred to another. Since there are ${J}\choose{2}$ possible pairwise comparisons with $J$ items, there are ${J}\choose{2}$ $\times 2$ causal effects: $\bm{\tau^{P}} = (\tau^{P}_{12},...,\tau^{P}_{J(J-1)})$. For example, when $J=3$, there exist ${{3}\choose{2}} \times 2$ $=(3 \times 2) \times 2 = 12$ possible APEs. The APE is informative when there are several pairs of items that scholars seek to study. When $J=2$, both the ARE and APE become a conventional average treatment effect on binary outcome data (Online Appendix \ref{app:degenerate}). 


\subsection{Alternative Quantities of Interest}

Many other estimands can be useful in different applications. For example, an alternative estimand may address the treatment effect on the \textit{probability that each item has a specific rank}, such as $\mathbb{P}(R_{ij}=1)$. This can be, for example, the probability that indigenous
people in the Philippines rank tribal identity as the most important identity \citep{mcmurry2022recognition}. Moreover, scholars may focus on the \textit{probability that each item is ranked higher than a specific rank}, such as $\mathbb{P}(R_{ij} \leq 3)$. An example can be the probability that citizens of Nigeria and Niger rank religion (Islam) as one of the three most important (e.g., top-3) identity attributes \citep{miles1991nationalism}.

Furthermore, researchers may focus on the \textit{conditional} treatment effect on the average rank of each item when some other items are ranked lower than specific ranks, such as $\E[R_{ij}(1)-R_{ij}(0)\big|R_{ik}>3]$. Similarly, they may also study the treatment effect on the probability that \textit{a pair of items will be ranked higher than specific ranks}, such as $\mathbb{P}(R_{ij}\leq 2 \text{ and } R_{ik}\leq 4)$. Finally, analysts may also be interested in whether the treatment changes the \textit{entire distribution of rankings}. In this situation, they could focus on the \textit{rank distance} \citep{alvo1993rank} between the set of rankings in the treatment and control conditions. Depending on the choice of $g()$ and $d()$, Equation (\ref{eq:general}) offers an array of causal estimands that can be useful in political science.

\section{Identifying Assumptions, Estimators, and Inferential Methods}

This section discusses two estimators for the proposed effects and statistical inference for them.\footnote{Online Appendix \ref{app:partial} also discusses partial identification based on nonparametric bounds.}

\subsection{Estimators}

For the following estimators to be unbiased estimators for their corresponding effects, I make two standard assumptions:\\

\noindent \textbf{Assumption 1} (\textsc{Ignorability}). \textit{Both potential rankings are independent from treatment assignment: $ \big(\textbf{R}_{i}(1),  \textbf{R}_{i}(0) \big) \indep D_{i}$ \citep{rubin1978bayesian}}.\\

\noindent \textbf{Assumption 2} (\textsc{SUTVA}). \textit{Each unit's observed ranking only depends on her treatment assignment, and there is only a single version of the treatment \citep{rubin1980randomization}.} 

\indent

While this paper focuses on unit-wise ignorability and SUTVA, future research may relax the two assumptions by focusing on a particular item or a pair of items.

\subsubsection{Difference-in-Mean-Ranks Estimator}
First, I consider the ARE. Let $R_{ij}^{\text{obs}}$ be unit $i$'s observed marginal ranking for item $j$. A natural estimator for the ARE is the difference-in-mean-ranks estimator:
\begin{subequations}
\begin{align}
    \widehat{\tau}_{j} &  = \widehat{\E}[R_{ij}(1)] - \widehat{\E}[R_{ij}(0)]\\
    & = \widehat{\E}[R_{ij}^{\text{obs}}|D_{i}=1] - \widehat{\E}[R_{ij}^{\text{obs}}|D_{i}=0]\\
    & = \frac{\sum_{i=1}^{N}R_{ij}^{\text{obs}}D_{i}}{\sum_{i=1}^{N}D_{i}} - \frac{\sum_{i=1}^{N}R_{ij}^{\text{obs}}(1-D_{i})}{\sum_{i=1}^{N}(1-D_{i})},
\end{align}
\end{subequations}

\noindent where $\frac{\sum_{i=1}^{N}R_{ij}^{\text{obs}}D_{i}}{\sum_{i=1}^{N}D_{i}} \in (1, J)$ is the \textit{mean rank} of item $j$ for treated units and  $\frac{\sum_{i=1}^{N}R_{ij}^{\text{obs}}(1-D_{i})}{\sum_{i=1}^{N}(1-D_{i})} \in (1, J)$ is the mean rank of the same item for control units. These values can be easily calculated by taking the average of observed ranks for item $j$ among the treated and control units. The proof is very standard and omitted. A mean rank is an important summary statistic for ranking data, and it represents the centrality of a given item in the sample \citep{yu2019analysis}. As there are $J$ items, researchers can obtain a vector of estimated effects with the length of $J$: $\widehat{\bm{\tau}} = (\widehat{\tau}_{1},..., \widehat{\tau}_{J})^{\top}$.

\subsubsection{Difference-in-Pairwise-Relative-Frequency Estimator}
Next, a natural estimator for the APE is the difference-in-pairwise-relative-frequency estimator:
\begin{subequations}
\begin{align}
    \widehat{\tau}_{jk}^{P} & = \widehat{\mathbb{P}}((R_{ij} < R_{ik})(1)) - \widehat{\mathbb{P}}((R_{ij} < R_{ik})(0)) \\ 
    & = \widehat{\mathbb{P}}(R_{ij}^{\text{obs}} < R_{ik}^{\text{obs}}|D_{i}=1) - \widehat{\mathbb{P}}(R_{ij}^{\text{obs}} < R_{ik}^{\text{obs}}|D_{i}=0) \\
    & = \frac{\sum_{i=1}^{N}I(R_{ij}^{\text{obs}} < R_{ik}^{\text{obs}})D_{i}}{\sum_{i=1}^{N}D_{i}} - \frac{\sum_{i=1}^{N}I(R_{ij}^{\text{obs}} < R_{ik}^{\text{obs}})(1-D_{i})}{\sum_{i=1}^{N}(1-D_{i})},
\end{align}
\end{subequations}

\noindent where $ \frac{\sum_{i=1}^{N}I(R_{ij}^{\text{obs}} < R_{ik}^{\text{obs}})D_{i}}{\sum_{i=1}^{N}D_{i}}$ is the proportion that item $j$ is ranked higher than item $k$ in the treatment group, whereas $\frac{\sum_{i=1}^{N}I(R_{ij}^{\text{obs}} < R_{ik}^{\text{obs}})(1-D_{i})}{\sum_{i=1}^{N}(1-D_{i})}$ is the same proportion for control units. These values can be easily calculated by creating a binary indicator that item $j$ is preferred to item $k$ for each unit and taking the average of all indicators in the sample. Again, the proof is standard and omitted. Since there are ${J}\choose{2}$ $\times 2$  effects, researchers can obtain a vector of estimated APEs: $\widehat{\bm{\tau}}^{P} = (\widehat{\tau}^{P}_{jk},...,\widehat{\tau}^{P}_{J, J-1})^{\top}$.

\subsection{Constructing Standard Errors}
To construct confidence intervals for each effect, I use a normal approximation of its sampling distribution in large samples. Online Appendix \ref{app:sim_approx} illustrates simulation studies that report that the normal approximation appears reasonable for the proposed estimators.

\subsection{Hypothesis Testing}

Hypothesis testing for AREs and APEs faces a unique challenge for two reasons. First, it usually requires multiple hypothesis testing since researchers can estimate multiple effects at the same time. Second, such multiple tests may not be independent of each other. To account for these challenges, I adopt the Benjamini-Hochberg (BH) procedure to conduct hypothesis testing \citep{benjamini1995controlling} (Online Appendix \ref{app:multiple}), while future research may adopt other methods.


\section{Empirical Illustration: Blame Attribution in Police Violence}

This section reanalyzes the experimental data in \citet{boudreau2019police} using the proposed methods. Contrary to the original study, I find that contextual information does not affect people's attitudes toward which parties are more (and less) responsible for the killing of the Black civilian.\footnote{I analyzed only \textit{non-Sacramento non-African American} subjects because it is for these subjects that the authors find the most consistent evidence for their hypotheses, leading to the same conclusion.}

\subsection{Average Rank Effects}
The first hypothesis in the study is that the ``pattern-of-violence'' treatment causes citizens to put \textit{more} blame on the police officers, the police chief, the mayor, and the DA, while making them put \textit{less} blame on the victim, the governor, and the senators. The second hypothesis states that the ``police reform'' treatment causes citizens to blame the state and federal officials \textit{more},  while making them blame the victim, the police officers, the police chief, and the local officials \textit{less}.

\begin{figure}[h!]
    \centering
    \includegraphics[width=16cm]{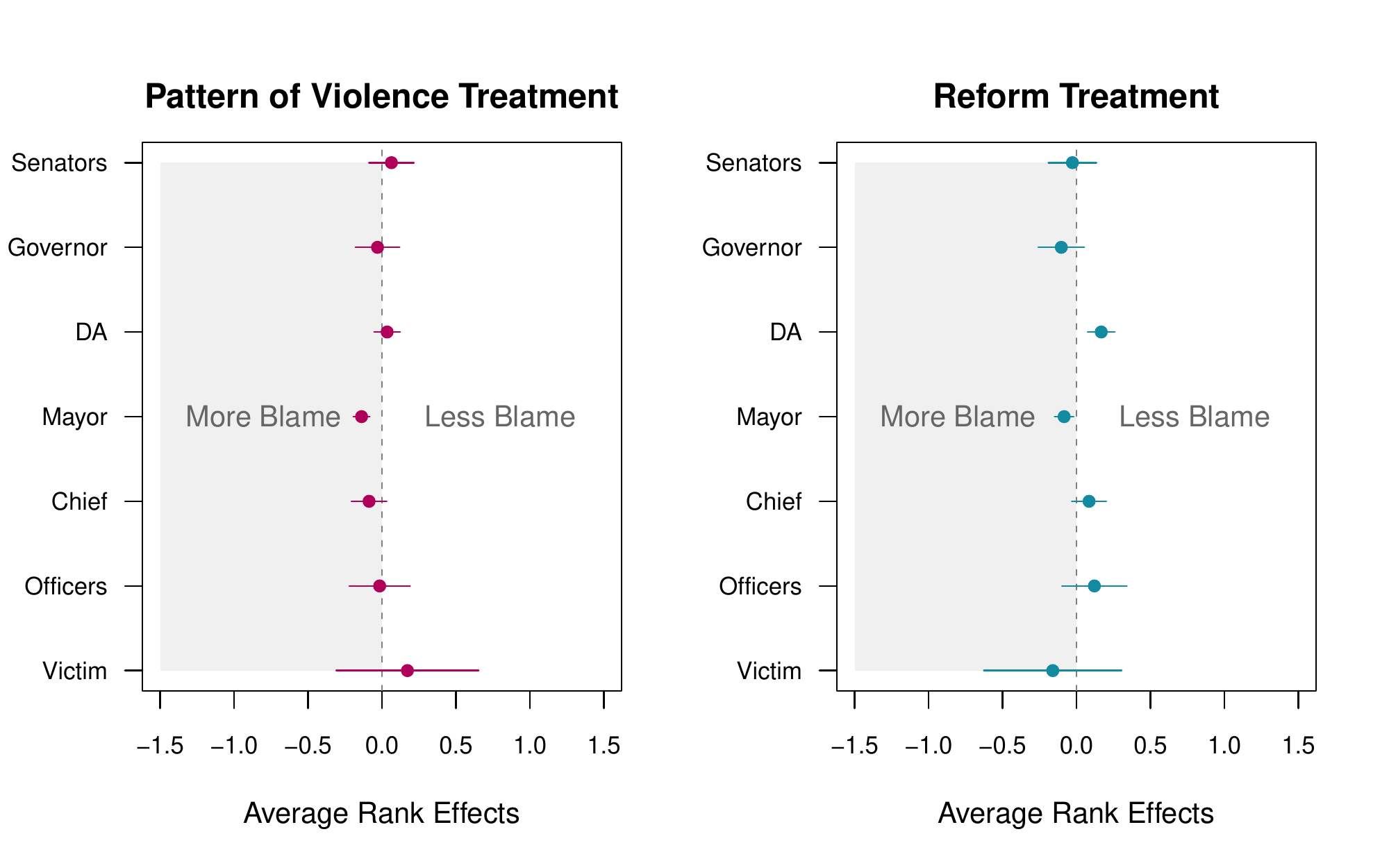}
    \caption{\textbf{Average Rank Effects of the Pattern-of-Violence and Reform Treatments}\\ \textit{Note}: The left (right) panel shows the estimated population average rank effects for seven parties for the pattern-of-violence (reform) treatment.}
    \label{fig:ARE}
\end{figure}

To test these hypotheses, I first examine the AREs of the two treatments. The left panel of Figure \ref{fig:ARE} shows the estimated AREs of the pattern-of-violence treatment for the seven parties. Seeing effects on the left side of the panel (``More Blame'') means that respondents put higher rankings and equivalently more blame on each party. Here, I follow the convention by saying that a rank is \textit{higher} in the treatment group than in the control group when the rank has a \textit{smaller value} in the former than in the latter (e.g., number one has the highest rank). Consequently, \textit{lower} values mean \textit{positive} effects and vice versa. 


I find that the pattern-of-violence treatment did not affect almost all parties. While the direction of most point estimates is consistent with the author's hypothesis, these estimates are not statistically (and substantively) different from zero. One exception is the effect on the mayor, implying that knowing that the Clark shooting is not an isolated incident but belongs to a series of police shootings in the same city makes people blame the mayor slightly more.   

The right panel of Figure \ref{fig:ARE} shows the estimated AREs of the reform treatment. Again, negative values mean positive effects. The result shows that while most estimated effects are in the expected direction (except for the effect on the governor), all effects appear to be statistically insignificant based on the estimated 95\% confidence intervals. One exception is the effect on the DA, suggesting that knowing about the Police Department's effort to reform makes people blame slightly less on the DA. Online Appendix \ref{app:BH2} also performs multiple hypothesis testing, finding substantively similar results.

The above analysis suggests that people's perception of the Clark shooting is robust to the presence and content of additional contextual information. However, it does not account for potential racially heterogeneous effects despite the centrality of race in the discussion of police violence \citep{davis2017policing}. To address this concern, I replicate the above analysis for White, Black, Latino, and Asian respondents separately. Figure \ref{fig:ARE2} shows that across the subpopulations, almost all effects are statistically and substantially insignificant, casting doubt on the presence of heterogeneous effects based on race and ethnicity.

In sum, the reanalysis did not find evidence that supports the hypotheses. Substantively, it implies that how officer-involved shootings are presented (such by media or prosecutors) may \textit{not} alter people's evaluations of police violence.

\subsection{Average Pairwise Rank Effects}

\begin{figure}[h!]
    \centering
    \includegraphics[width=16cm]{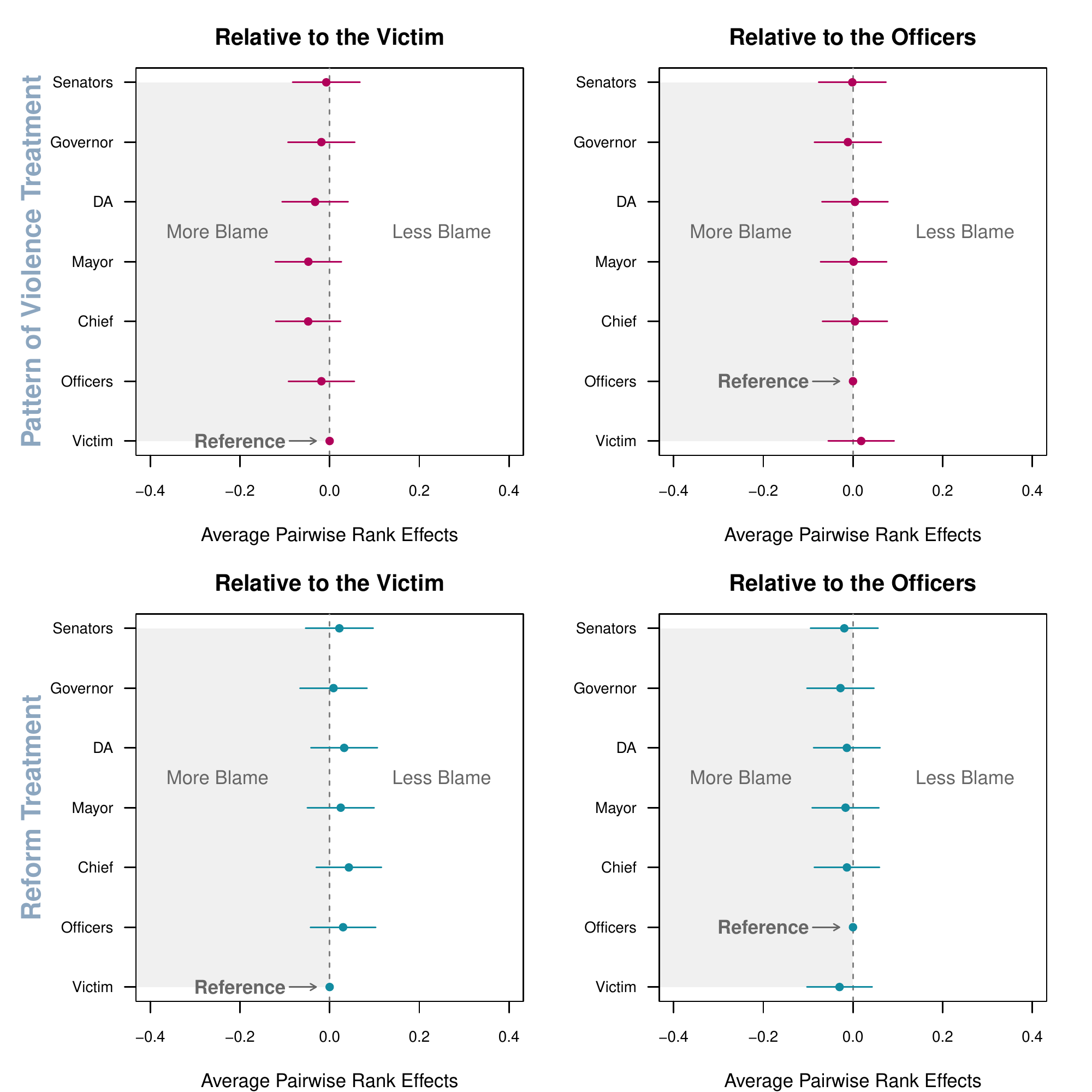}
    \caption{\textbf{Average Pairwise Rank Effects of the Pattern-of-Violence and Police Reform Treatments}\\ \textit{Note}: The upper panels show the estimated APEs of the pattern-of-violence treatment, whereas the lower panels visualize the obtained APEs of the police reform treatment. The left (right) panels present the effects while setting the victim (police officers) as a reference category.}
    \label{fig:PRE}
\end{figure}

The hypotheses focus on each party separately. However, it could also be informative to analyze whether the treatment made respondents blame each party more \textit{relative to} the victim as well as to the police officers --- the central figures in the shooting. With this motivation, I estimate the average pairwise rank effects (APEs) of the two treatments. The upper two panels of Figure \ref{fig:PRE} show the results for the pattern-of-violence treatment, whereas the lower two panels present the findings for the police reform treatment. 

The upper left panel reports the estimated APEs of the pattern-of-violence treatment with the victim as a \textit{reference category}. Lower values mean that the treatment increases the probability that each party is ranked higher than the victim (i.e., blaming each party more than the victim). Similarly, the upper right panel presents the estimated effects with the officers as a reference category. Both panels suggest that the estimated APEs are not statistically significantly distinct from zero. This offers additional evidence against the original study's expectations. Similar patterns are also observed in the lower two panels, suggesting that the reform treatment does not affect people's pairwise rankings in the experiment.

To examine potential heterogeneous APEs, I also replicate the above analysis by focusing on the samples of Asian, Black, Latino, and White respondents, respectively. Figures \ref{fig:PRE_D1}-\ref{fig:PRE_D4} suggest that my substantive conclusion holds within all racial groups. Combined with the earlier findings, the analysis of APEs reveals that none of the active treatments alters people's attitudes toward officer-involved shootings.

What then determines people's blame attribution in police violence? Recent research suggests that Black and White Americans hold distinct perceptions about officer-involved shootings even when they receive the same information \citep[e.g.,][]{jefferson2021seeing}. Motivated by such research and the above findings, Figure \ref{fig:RacialDiff} shows the estimated average rank of each party for Asian, Black, Latino, and White samples. I find that the way people hold the police force accountable significantly differs by race and ethnicity, reaffirming the importance of jury selection in police violence cases \citep{cook2017biased,futrell2018visibly,davis1994rodney}.

\section{Extension to Partially Ranked Data}

So far, I have assumed that people rank all items. In many political science applications, however, researchers observe \textit{partially ranked data}, where people only rank a subset of available items \citep{critchlow2012metric}. In preferential voting, for example, voters often choose not to rank any candidate beyond their first-choice candidates \citep[150-158]{reilly2001democracy}. Similarly, in the study of postmaterialism, survey respondents are asked to rank the two most important items among four items \citep{inglehart1999measuring}. In such scenarios, the above framework cannot be directly applicable because outcome data are partially missing. To overcome this challenge, this section extends the proposed framework to partially ranked data.


\subsection{Motivating Application: Ballot Order Effects in Ranked-Choice Voting}

As a motivating application, I study ballot order effects in ranked-choice voting (RCV)  \citep{orr2002ballot,king2009ballot,curtice2014confused,robson1974importance,ortega2004position,ortega2008gender}.\footnote{More generally, the discussion here applies any preferential voting system, including Alternative Vote and Single Transferable Vote. For more variants of RCV, see \citet{santucci2021variants}.}

Ballot order effects are the effects of the order in which candidates appear on the ballot on voters' stated preferences and candidate vote shares.\footnote{More broadly, ballot order effects appear in survey research with inattentive respondents \citep[e.g.,][]{atsusaka2021bias}.} To date, especially since \textit{Bush v. Gore} (2000), a body of works has estimated these effects, explained their causes, and discussed their implications under first-past-the-post (FPTP) \citep{ho2006randomization,ho2008estimating,meredith2013causes,grant2017ballot,chen2014impact,lutz2010first,barker1980crucial,koppell2004effects,miller1998impact}, proportional representation \citep{geys2003ballot,marcinkiewicz2014electoral,faas2006importance,blom2016ballot,gulzar2022campaigns}, and other systems \citep{song2022rank}. Most importantly, the literature has documented evidence for ``primacy'' (first position) and ``latency'' (last position) effects in elections \citep{alvarez2006much} and survey research more broadly \citep{kim2015moderators}.

RCV is an electoral system in which voters can vote by ranking multiple candidates \citep{santucci2021variants}. The literature has suggested that, while mostly focusing on voters' first-choice preferences, ballot order effects are present and consequential in RCV too \citep{orr2002ballot,king2009ballot,curtice2014confused,robson1974importance,ortega2004position,ortega2008gender,marcinkiewicz2015ballot,soderlund2021coping}.\footnote{One extreme type of ballot order effect in RCV has been documented as the ``donkey vote'' \citep{orr2002ballot} that has been ``observed at federal elections in Australia, in which some electors simply number sequentially from 1 onward down the ballot paper'' \citep[158]{reilly2001democracy}.} Yet, despite the increasing attention to RCV among scholars \citep{drutman2020breaking} and candidates \citep{peters2022},\footnote{One candidate for the Oakland mayoral election mentioned, ``The [ballot] order is very important, extremely important in elections where there’s no incumbent running. This could force as much as a five percent point differential'' \citep{peters2022}.} there has been almost no research that analyzes the ballot order effects on voters' entire ranked ballot, let alone on \textit{partially ranked ballot}.\footnote{One notable exception is \citet{horiuchi2019randomized}.}

Motivated by this challenge, I develop a \textit{generic} strategy to identify and estimate the treatment effects on any partially ranked outcomes (beyond the running application). While this paper focuses on the average rank effects (ARE), the following strategy can be applied to \textit{any quantity} in the class of estimands discussed in Equation (\ref{eq:general}). 

\subsection{Illustration of Partially Ranked Data}
\begin{figure}[h!]
    \centering
    \includegraphics[width=14cm]{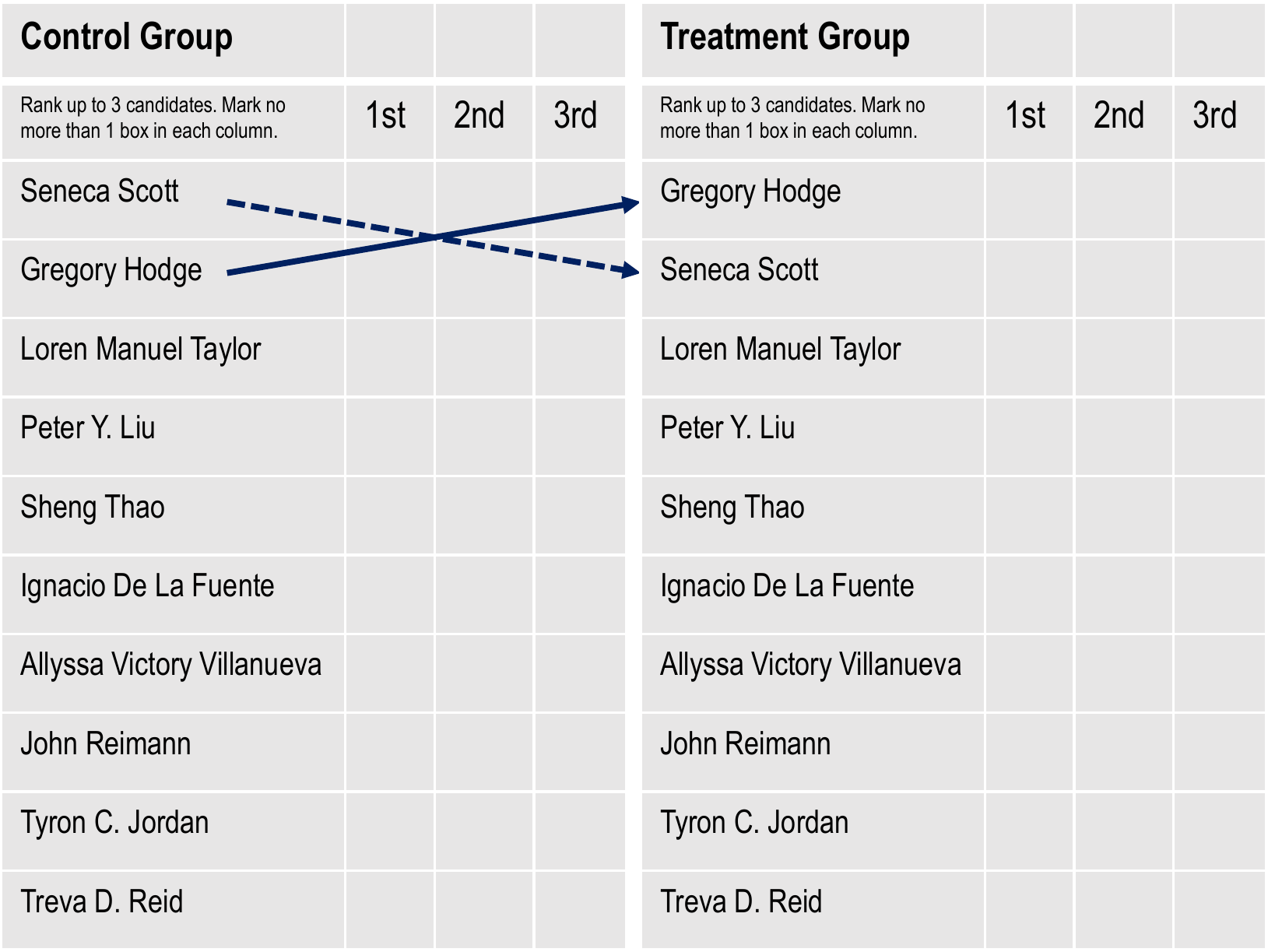}
    \caption{\textbf{Example of Experimental Designs for Studying Ballot Order Effects in RCV}\\ \textit{Note}: The ten candidates are drawn from the 2022 Oakland Mayoral Election.}
    \label{fig:RCV}
\end{figure}

To illustrate partially ranked data, suppose that analysts want to know whether ballot order affects candidate ranks. Suppose that they randomly assign study subjects (hereafter called \textit{voters}) to either the control or treatment group. Figure \ref{fig:RCV} visualizes what voters see under the two conditions in the simplest experimental design. In the control group, voters see a set of candidates in a particular order. In the treatment group, voters see the same list of candidates in a slightly different order. Here, Gregory Hodge now appears in the first position as opposed to the second place, and Seneca Scott appears in the second position as opposed to the first place (i.e., experimental manipulation). When voters rank \textit{all} candidates, researchers can estimate the treatment effect on, for example, Gregory Hodge's average rank using the basic framework discussed above.



\indent

Now, suppose that voters are only allowed to rank \textit{up to three} candidates as in many RCV elections in the U.S. Table \ref{tab:partial} presents an example of partially ranked data that scholars would observe in this condition. The first three units are control, while the last three units are treated. 

The first voter ranked three candidates, including Reinmann, Scott, and Taylor, while not ranking the other candidates (denoted by ``---''). In contrast, the second voter ranked the first three candidates she sees on the ballot, \textit{potentially} engaging in what is known as the ``donkey vote'' (see footnote 11) \citep{orr2002ballot}. Finally, the third voter only ranked two candidates, even though he could have ranked three candidates.

\indent

\begin{table}[h!]
    \centering
    \begin{tabular}{cccacccccccc}\hline
       Unit & Group & Scott &  Hodge & Taylor & Liu & Thao & Fuente & Villanueva & Reinmann & Jordan & Reid \\\hline
        1 & {\color{mythird}Control}  &   2 & --- &    3 &   --- &    --- &   --- & --- & 1 & --- & --- \\
        2 & {\color{mythird}Control}  &	 1  & 2 &    3 &    --- &	--- &	--- & --- & --- & --- & ---\\
        3 & {\color{mythird}Control}  &   1 &   --- &	--- &	--- &	--- &	--- & --- & 2 & ---& ---\\
        4 & {\color{mycol}Treated}  &   --- &   1 &	--- &	--- &	--- &	--- & --- & --- & --- & ---\\
        5 & {\color{mycol}Treated}  &	 2  &   1 &	--- &	--- &	--- &	--- & --- & --- & ---& ---\\
        6 & {\color{mycol}Treated}  &   2  &   1 &	--- &	--- &	--- &	3 & ---   & --- & ---& ---\\ \hline
    \end{tabular}    
    \caption{\textbf{Example of Partially Ranked Outcome Ballot in Randomized Experiments}\\\textit{Note}: The observed stated ranked preferences for ten candidates in a hypothetical experiment. ``---'' means that no rank is observed. Fuente = De La Fuente.}
    \label{tab:partial}
\end{table}

Suppose that researchers wish to estimate the treatment effect on Hodge's average rank. The challenge here is that because the first and third voters did not rank him, Hodge's rank is not observed and missing for these units (denoted by ``---''). Moreover, voters seem more likely to rank Hodge when they are treated than otherwise. How can researchers estimate the treatment effect with partially missing ranking data like this? As discussed below, when the treatment affects the missingness of Hodge's rank, performing listwise deletion on Hodge (i.e., dropping missing observations) leads to a statistical bias with respect to the treatment effect of interest. While the running example focuses on RCV, this is a general problem that arises in causal inference with any partially ranked outcomes.

\subsection{Partial Rankings as a Selection Problem}

To address this problem, I consider partially ranked data as a \textit{selection problem}, in which outcome data are \textit{defined} but \textit{missing} for some units. I also assume that missingness is affected by the treatment assignment. While there are different types of selection problems, the proposed framework is similar to the one discussed for racial bias in policing \citep{knox2020administrative} and different from the ``truncation by death'' problem, where missing outcome data are not defined for truncated units \citep{zhang2003estimation}. 

Figure \ref{fig:DAG} represents the general structure of the problem in a directed-acyclic graph (DAG) \citep{glymour2016causal}. Here, $D_{ij}$ is a treatment indicator (e.g., ballot order) showing whether unit $i$ is treated for item $j$ ($D_{ij}=1$) or not ($D_{ij}=0$). Similarly, $M_{ij}$ is a mediator (or selection) variable denoting whether unit $i$ decides to rank item $j$ ($M_{ij}=1$) or not ($M_{ij}=0$). Finally, $R_{ij} \in \{1,2,...,J\}$ is unit $i$'s marginal ranking for item $j$, whereas $U_{ij}$ is a set of \textit{unobserved} confounders that affect both $M_{ij}$ and $R_{ij}$ (e.g., perceived candidate quality). 

\indent

\begin{figure}[h!]
\centering
\begin{tikzpicture}
    \node (1) at (0,0) {$D_{ij}$ (ballot order)};
    \node (2) [right = of 1] {$M_{ij}$ (decide to rank)};
    \node (3) [right = of 2] {$R_{ij}$ (marginal ranking)};
    \node (4) at (6.2,1.5) {$U_{ij}$ (perceived candidate quality)};
    \path (1) edge  (2);
    \path (2) edge (3);
    \path[dashed] (4) edge (3);    
    \path[dashed] (4) edge (2);        
\end{tikzpicture}
    \caption{\textbf{Directed Acyclic Graph for Partially Ranked Data}\\\textit{Note}: $D_{ij}$, $M_{ij}$, and $R_{ij}$ are observed while $U_{ij}$ is not observed by researchers.}
    \label{fig:DAG}
\end{figure}

Here, a na\"ive approach is to estimate the effect of $D_{ij}$ on $R_{ij}$ after dropping all units with $M_{ij}=0$ (i.e., listwise deletion). However, since $\mathbb{P}(M_{ij}=0)$ itself is affected by $D_{ij}$, only keeping units with $M_{ij}=1$ (selection on the value of $M_{ij}$) means conditioning on a post-treatment variable \citep{montgomery2018conditioning}. A danger for causal inference arises when a set of unmeasured factors $U_{ij}$ affect both $M_{ij}$ and $R_{ij}$. For example, perceived candidate quality (i.e., to what extent voters think a candidate is appealing) may affect \textit{whether} and \textit{how} they rank the candidate.\footnote{Estimating the effect among units whose outcomes are not missing (e.g., $\E[R_{ij}(D_{ij}=1)-R_{ij}(D_{ij}=0)|M_{ij}=1]$) may lead to a non-causal quantity \citep{knox2020administrative}. In Table \ref{tab:partial}, the bias arises because voters who ranked Hodge when they saw him at the top of the ballot (i.e., treated) are \textit{fundamentally different} from voters who ranked him (even) when they do not see him in the first position (i.e., control).}

To study treatment effects with partially ranked data, this paper draws from the framework of principal stratification by  \citet{frangakis2002principal}. For each candidate $j$, I define four subsets of units (called principal strata) based on their pre-treatment characteristics that define how they react to the active treatment \citep{page2015principal}. Among the four strata, I primarily focus on \textit{order-rankers}, which is a group of units that rank item $j$ only when they are in the active treatment condition.

\subsection{Causal Estimands for Partially Ranked Data}

Building on the above argument, I now introduce two causal estimands for partially ranked data. The first estimand is the average rank effect (ARE) for item $j$ ($\tau_j$) as defined in Equation (\ref{eq:mre1}).

The second quantity of interest is what I call the order-ranker ARE. Again, order-rankers refer to people who only rank item $j$ when they are treated. Let $R_{ij}(D_{ij}, M_{ij}(D_{ij}))$ be unit $i$'s potential marginal ranking for item $j$ under the treatment assignment $D_{ij}$ and the potential selection state under the same condition $M_{ij}(D_{ij})$. The order-ranker ARE is defined as follows: 
\begin{subequations}
\begin{align}
\tau_{j,o} & \equiv \E\big[R_{ij}(1,M_{ij}(1)) - R_{ij}(0,M_{ij}(0))|M_{ij}=1\big]\\
     & = \E\big[R_{ij}(1,M_{ij}(1))|M_{ij}=1\big] - \E\big[(0,M_{ij}(0))|M_{ij}=1\big]
\end{align}
\end{subequations}

This estimand, often called a local average treatment effect (LATE) or compiler average causal effect (CASE), is of particular interest in the context of RCV. I focus on this effect because I expect ballot order effects to be present primarily among ``swing'' or ``floating'' voters with no pre-arranged voting schedule. Online Appendix \ref{app:orderrankers} discusses the identification result for the order-ranker ARE.



\subsection{Partial Identification via Nonparametric Bounds}

The fundamental problem in partially ranked data is that $\tau_{j}$ and $\tau_{j,o}$ are never point identified (i.e., cannot be estimated) because the outcome data are partially missing. However, it is still possible to construct nonparametric sharp bounds on (i.e., partially identify) $\tau_{j}$ and $\tau_{j,o}$. Such bounds cover the lowest and highest possible treatment effects consistent with the available information \citep{manski1990nonparametric,horowitz2000nonparametric}.\footnote{More precisely, this strategy imputes potential outcomes for both sampled (but missing) \textit{and} unsampled units under the super-population perspective. Another approach would be constructing confidence sets for bounds \citep{horowitz2000nonparametric,imbens2004confidence}.} To construct the nonparametric bounds, I make two assumptions.

\indent

\noindent \textbf{Assumption 3} \textsc{(Bounded Support)}. \textit{For each ranker, the unobserved marginal ranking of a given item must be lower than any observed marginal ranking and higher than or equal to the lowest possible marginal ranking. Formally, $\max(R_{ij}(D_{ij},1)) < R_{ij}(D_{ij},0) \leq J$ for all $i$, $j$, and $D_{ij}$.} 


\indent

Bounded support is based on the nature of rankings (i.e., logical bound) and thus always holds \citep{manski1990nonparametric}. Table \ref{tab:bounded} shows the logical bound of Hodge's unobserved (and potential) rank for the first voter in Table \ref{tab:partial}. Here, the highest rank for Hodge must be 4 because the highest observed rank is 3. Similarly, the lowest rank for him must be 10 because it is the lowest possible rank when there are ten candidates. Bounded support states that Hodge's potential ranks must be between 4 and 10.

\indent

\begin{table}[h!]
    \centering
    \begin{tabular}{ccaccccccccc}\hline
         & Scott & Hodge & Taylor & Liu & Thao & Fuente & Villanueva & Reinmann & Jordan & Reid \\\hline
         Observed rank  &   2 & {\color{firebrick}---} &    3 &  --- &    --- &   --- & --- & 1& ---& --- \\
        Highest rank   &  2 & {\color{firebrick}4} &    3 &   --- &    --- &   --- & --- & 1 & ---& ---\\
        Lowest rank   &   2 & {\color{firebrick}10} &    3 &   --- &    --- &   --- & --- & 1 & ---& ---\\\hline
    \end{tabular}    
    \caption{\textbf{Illustration of Bounded Support}\\\textit{Note}: The highest rank is 4 because the highest observed rank is 3. The lowest rank is 9 because it is the lowest possible rank given that there are ten candidates. Fuente = De La Fuente.}
    \label{tab:bounded}
\end{table}

\indent

\noindent \textbf{Assumption 4} \textsc{(Positive Average Treatment Effects)}. \textit{Average treatment effects are positive. Formally, $\tau_j > 0$ and $\tau_{j,o} > 0$}.

\indent

With Assumptions 3-4, the proposed strategy first creates two imputed samples: one for the ``best-case'' scenario, where the maximal treatment effect is assumed, and another for the ``worse-case'' scenario, where the minimal treatment effect is assumed \citep[see also][]{coppock2022qualitative}. Let $\overline{R^{*}_{ij}}$ and $\underline{R^{*}_{ij}}$ be the imputed samples based on the maximal and minimal treatment effects. The imputed samples are defined as follows:
\begin{subequations}
\begin{align}
\overline{R^{*}_{ij}} & =\underbrace{M_{ij}R_{ij}}_{\text{observed rank}} + \underbrace{(1-M_{ij})}_{\text{missing rank}}\big\{{\color{firebrick}\underbrace{D_{ij}(\max(R_{ij'})+1)}_{\text{highest rank for treated}}} + {\color{navyblue}\underbrace{(1-D_{ij})J}_{\text{lowest rank for control}}}\big\}\\
\underline{R^{*}_{ij}} & = \underbrace{M_{ij}R_{ij}}_{\text{observed rank}} + \underbrace{(1-M_{ij})}_{\text{missing rank}}\big\{{\color{navyblue}\underbrace{D_{ij}J}_{\text{lowest rank for treated}}} + {\color{firebrick}\underbrace{(1-D_{ij})(\max(R_{ij'})+1)}_{\text{highest rank for control}}}\big\}
\end{align}
\end{subequations}

\indent

Finally, the proposed strategy constructs nonparametric sharp bounds using the two extreme effects.\\

\noindent \textbf{Proposition 1} \textsc{(Nonparametric Bounds on $\tau_{j}$ and $\tau_{jo}$)}. \textit{Let $\underline{\tau_j} = \E[\underline{R_{ij}^{*}}(1,M_{ij}) - \underline{R_{ij}^{*}}(0,M_{ij})]$ be the lowest possible ARE and $\overline{\tau_j} = \E[\overline{R_{ij}^{*}}(1,M_{ij}) - \overline{R_{ij}^{*}}(0,M_{ij})]$ be the highest possible ARE. Then, the ARE can be partially identified with the following bound:}
\begin{subequations}
\begin{align}
\tau_j = \Big[ \underline{\tau_j}, \overline{\tau_j} \Big]
\end{align}

\noindent \textit{Similarly, the order-ranker ARE can be partially identified with the following bound:}\footnote{In practice, $\widehat{\pi}_{o}$ is used instead of $\pi_{o}$. More specifically, both bounds must be within $[-(J-1),J-1]$, although I suppress such constraint for clarity.}
\begin{align}
\tau_{j,o} = \Bigg[ \frac{\underline{\tau_j}}{\pi_{o}}, \frac{\overline{\tau_j}}{\pi_{o}} \Bigg],
\end{align}
\end{subequations}

\noindent where $\pi_o$ is the proportion of order-rankers. To study the properties of the bounds, Online Appendix \ref{app:sim} presents simulation studies. 

\section{Application to Ballot Order Effects in RCV}

I now apply the extended framework to study ballot order effects in RCV.

\subsection{Identification and Estimation of Ballot Order Effects}

I define generalized ballot order effects building on the example in Figure \ref{fig:RCV}. Online Appendix \ref{app:general} discusses the identification and estimation of generalized ballot order effects and three additional assumptions, including ballot order randomization. The key result is that the randomization of candidate positions at the unit level (i.e., random assignment) allows researchers to study the effect of each position on candidate ranks \textit{averaged over} all counterfactual positions \textit{and} the ordering of other candidates.\footnote{The idea is similar to the average marginal component effects in conjoint analysis \citep{hainmueller2014causal}.}






\subsection{Three Survey Experiments}


To study ballot order effects in RCV, I perform survey experiments in (1) the 2022 Oakland mayoral election, (2) the U.S. House of Representatives election in Alaska, and (3) the U.S. Senate election in Alaska. All surveys and analysis plans were pre-registered. These contests vary in the type of office, the level of national attention, voters' partisan leaning, and the novelty of RCV, offering rich contextual variation to study ballot order effects.

\begin{figure}[h!]
    \centering
    \includegraphics[width=12cm]{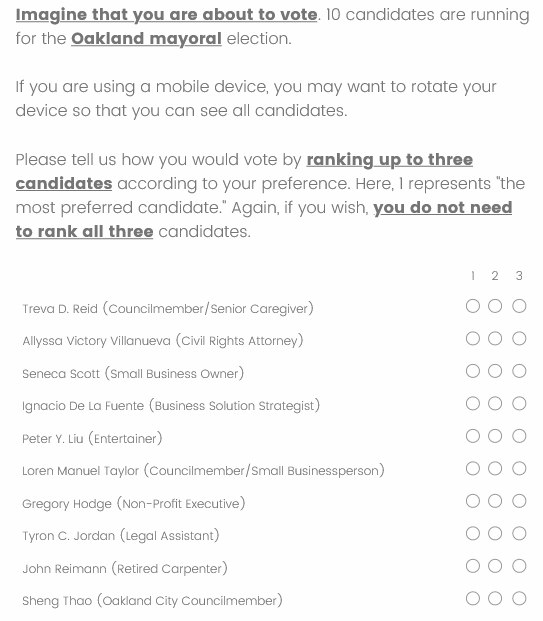}
    \caption{\textbf{An Example of Experimental Survey Question}\\ \textit{Note}: Candidate order (treatment) was randomized at the respondent level.}
    \label{fig:ss_exp}
\end{figure}

Figure \ref{fig:ss_exp} shows an example of the main survey questions. In the experiments, I showed respondents a list of actual candidates (with their occupations in Oakland and registered parties in Alaska). I designed the experimental questions by carefully mocking actual official sample ballots in Oakland and Alaska.\footnote{I used and obtained sample ballot for the last two elections from the \href{https://cao-94612.s3.amazonaws.com/documents/UPDATED-2022-Election-Candidate-List-w-Ballot-Designations.pdf}{City of Oakland website} and the \href{https://www.elections.alaska.gov/election/2022/genr/FEDERAL_ONLY.pdf}{Alaska Division of Elections website}.} Unlike the actual RCV elections, however, the order of candidates was randomized at the respondent level.

Here, the treatment is \textit{whether each candidate appears in a particular position on the ballot ($D_{ij}=1$) or not ($D_{ij}=0$)}. When focusing on the effect of the first position, for example, respondents who see Figure \ref{fig:ss_exp} are treated \textit{with respect to} Treva Reid (but not with respect to the other candidates).

To obtain outcome data, I asked respondents to rank up to three (in Oakland) or four (in Alaska) candidates according to their actual RCV rules by marking radio buttons. Additionally, I also asked them to rank \textit{all} candidates to obtain fully ranked data, which I use as supplemental information below. Online Appendix \ref{app:surveyX} presents a set of survey instructions, attention checks, and experimental questions used in the studies.

\indent

The survey experiments were implemented via the Lucid Marketplace (October 10th -- November 7th, 2022).\footnote{Respondents received \$3 for participating in the surveys.} Unlike many studies using nationally representative samples, I geo-targeted respondents by sampling \textit{within} Oakland and Alaska, respectively. To obtain the largest possible sample while achieving demographic representativeness, I used the 2020 Census to set quotas based on gender, race, and ethnicity in each area. The total number of respondents was 253 (Oakland) and 358 (Alaska), respectively. Only keeping respondents who passed four attention checks I used in the surveys \citep[following][]{aronow2020evidence}, I obtained 169 (Oakland) and 190 (Alaska) respondents. 



\subsection{Study I (2022 Oakland Mayoral Election)}

\begin{figure}[h!]
    \centering
    \includegraphics[width=15cm]{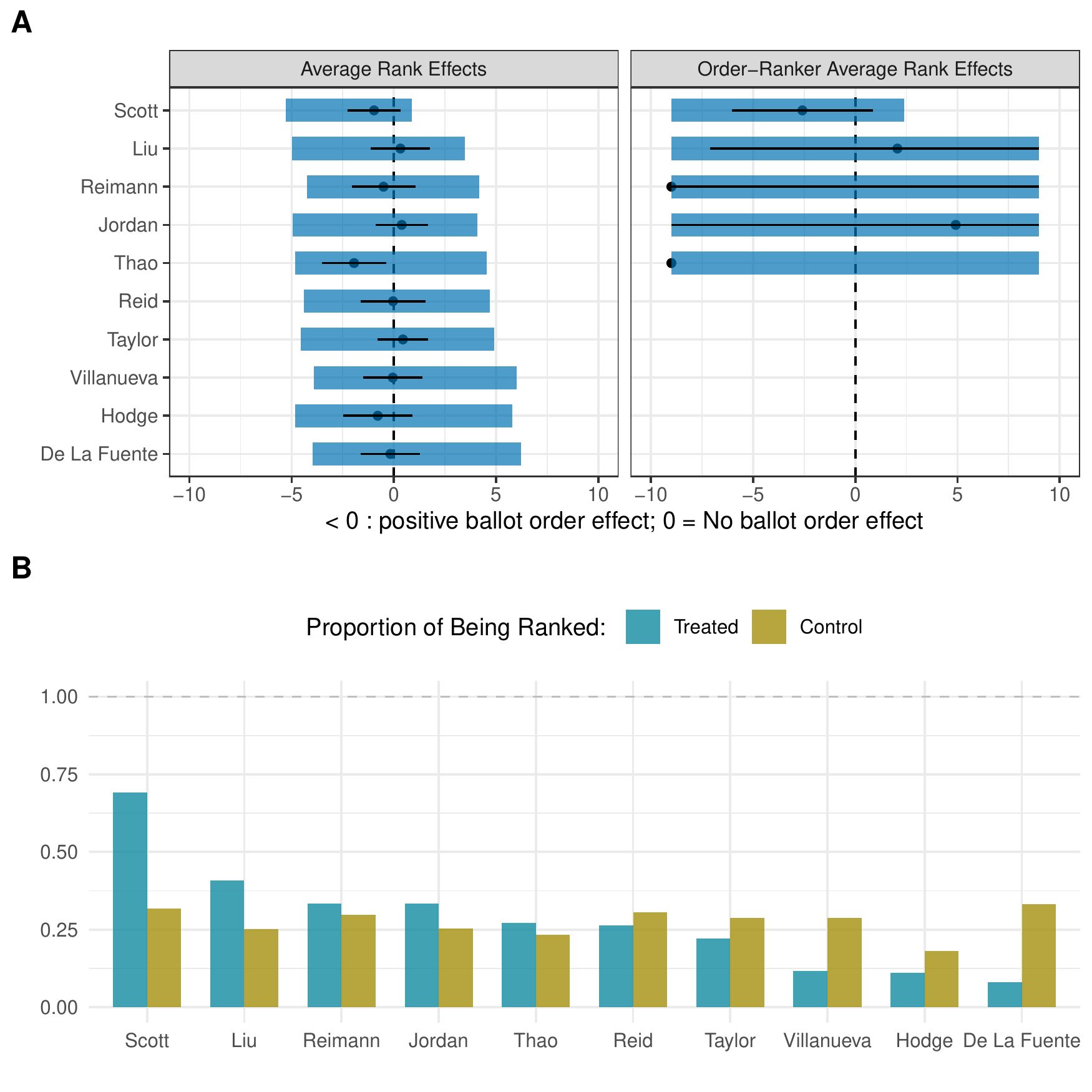}
    \caption{\textbf{First Position Effects in the 2022 Oakland Mayoral Election}\\ \textit{Note}: Panel A: The point estimates and 95\% confidence intervals are based full rankings. Panel B: The proportion of cases with non-missing ranks.}
    \label{fig:bound}
\end{figure}

Figure \ref{fig:bound} shows the effects of the \textit{first position} for ten candidates in the Oakland mayoral election. Panel A visualizes the nonparametric bounds for the ARE (left window) and order-ranker ARE (right window). Taylor, Thao, and De La Fuente are candidates with political experience, while others do not have any office-holding record. There are several notable findings.

\indent

First, all nonparametric bounds on the ARE cover zero. This suggests that none of the ten candidates seem to have received higher average ranks because they appeared in the first position on the ballot. Moreover, I verify this finding using fully ranked data as ground truth data. The point and interval estimates imply that ballot order effects are statistically and substantively insignificant, except for Thao --- one of the three competitive candidates. 

Moreover, five sharp bounds on the order-ranker ARE also cover zero. This means that respondents who only rank these candidates when treated (i.e., seeing them in the first position) did not give higher marginal rankings to the selected candidates. Importantly, the order-ranker ARE was not identified for the lower five candidates due to the violation of Assumption A1 (Ranking Monotonicity). 


Panel B displays the proportion of observed ranks, $\widehat{\mathbb{P}}(M_{ij}=1)$, for all candidates among treated and control units. It shows that respondents are more likely to rank five candidates (Scott, Liu, Reimann, Jordan, and Thao) when they see them at the top of the ballot. By doing so, it documents a novel form of ballot order effects under RCV. Namely, ballot order affects not only \textit{how} people rank each candidate but also \textit{whether} they rank a given candidate. However, respondents are less likely to do so for the other candidates in the same condition (suggesting the presence of ``defiers''). Below, I discuss this puzzling pattern in more detail. 

\begin{figure}[tbh!]
    \centering
    \includegraphics[width=17cm]{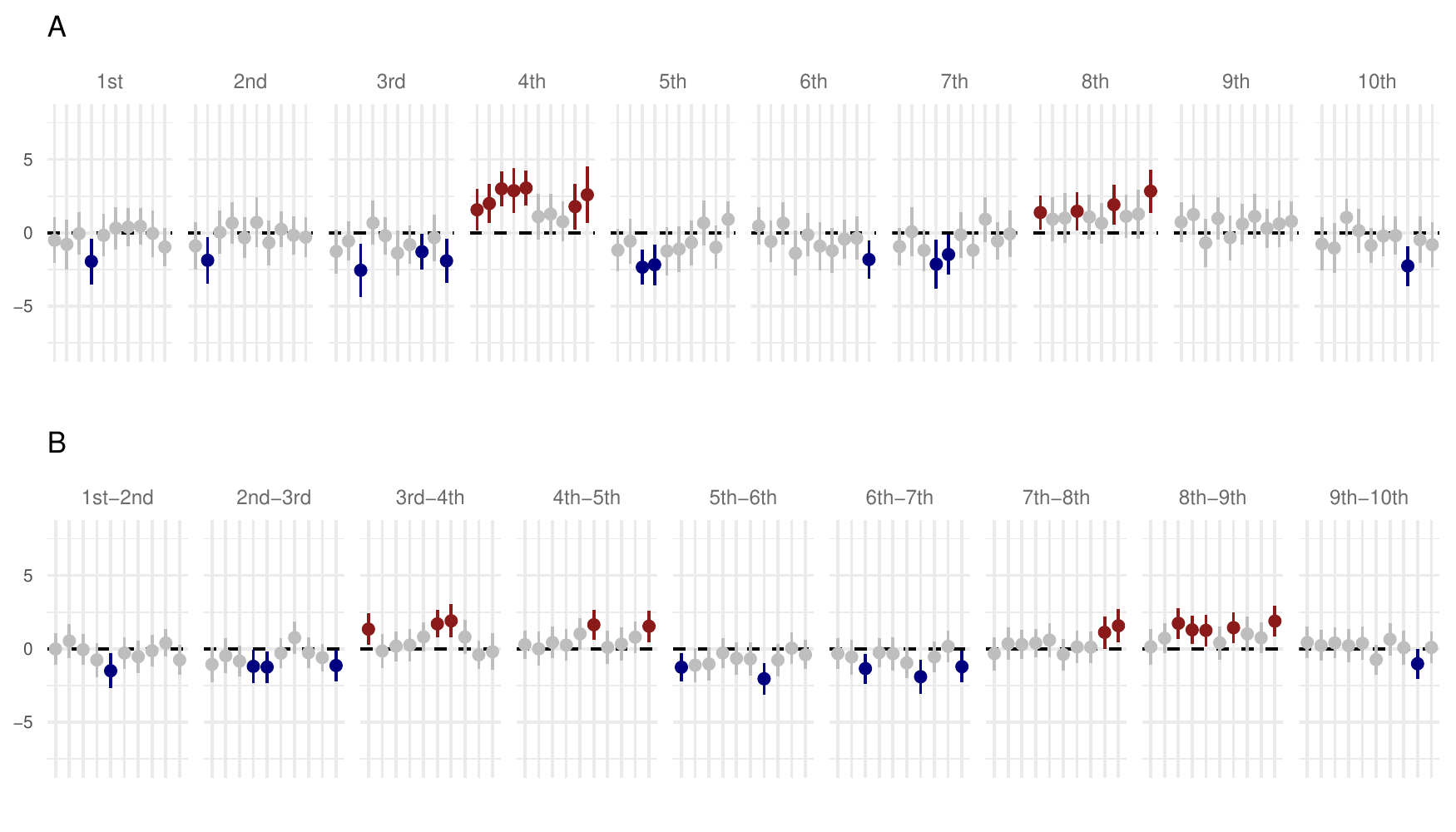}
    \caption{\textbf{Ballot Order Effects Across Positions in Oakland}\\ \textit{Note}: Panel A shows the effect of candidates appearing at each position of the ballot in the Oakland mayoral election data. Panel B shows the effect of the two positions combined. Negative values mean \textit{positive} ballot order effects.}
     \label{fig:general}
\end{figure}

Furthermore, Figure \ref{fig:general} presents the results for all positions (Panel A) and all pairs of adjacent positions (Panel B). It suggests that most of the 190 treatment effects are not statistically and substantively significant, including the effects of the first and the last positions. Indeed, at the $\alpha=0.05$ level, about 10 cases out of the 190 effects could be statistically significant \textit{solely by chance}. This finding seems puzzling since the primacy (first position) and latency (last position) effects are the most common forms of ballot order effects studied in the literature \citep{alvarez2006much,orr2002ballot,curtice2014confused}. Finally, I find evidence that the fourth and eighth positions may yield \textit{lower} ranks for some candidates, which, again, cannot be fully explained by previous research.



\indent

\subsection{Studies II-III (U.S. House and U.S. Senate Elections in Alaska)}

Online Appendix \ref{app:alaska} reports the results for the U.S. House and U.S. Senate elections in Alaska. The overall findings are consistent with the findings from Oakland. Below, I examine the Alaskan data more.

\subsection{New Puzzle}

The experimental results shed light on a new puzzle in the research on ballot order effects in RCV and item order effects in survey research more generally. I find that ballot order still matters, affecting how people might vote by ranking multiple candidates in RCV. However, I do not find any evidence for the primacy (first position) and latency (last position) effects, even though they are the most common forms of ballot order effects discussed in the literature \citep{alvarez2006much,ho2008estimating}. Furthermore, while ballot order seems to affect whether people rank a given candidate, the direction of such effects is inconsistent.

\subsection{Alternative Theory of Ballot Order Effects}

To explain this puzzle, I propose an alternative theory of ballot order effects, which I call \textit{pattern ranking}. The theory states that people vote in RCV and answer rank-order questions by following geometric patterns (instead of revealing their sincere preferences). And this happens when they do not have full ranked preferences and avoid ``looking bad'' (e.g., by engaging in seemingly cheating patterns, such as 1234). One observable implication of the theory is that some geometric patterns are observed with a significantly higher frequency than other patterns.  

\begin{figure}[tbh!]
    \centering
    \includegraphics[width=12cm]{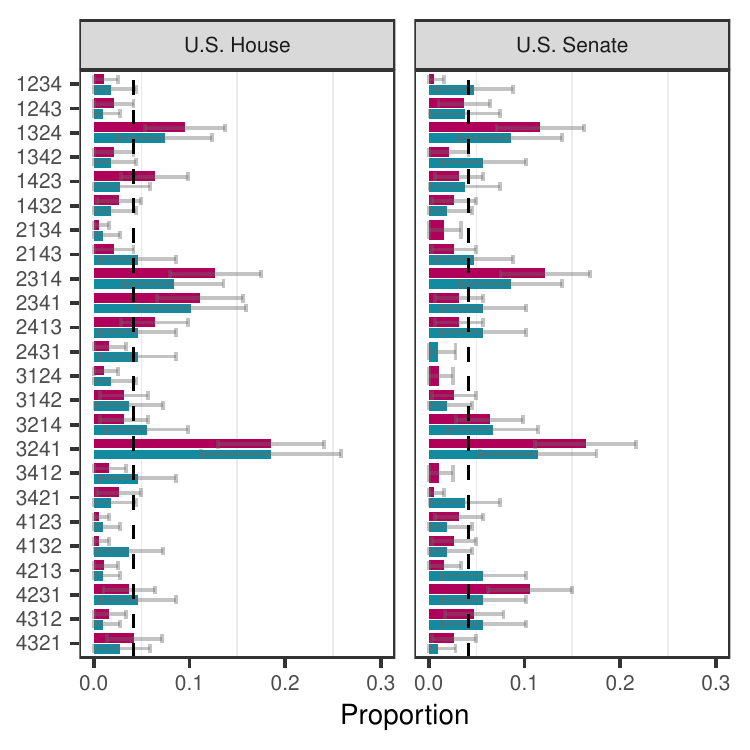}
    \caption{\textbf{The Distributions of Observed Rankings in the U.S. House and Senate elections in Alaska.}\\ \textit{Note}: A dashed line shows the uniform distribution in each panel. Each panel shows the proportion of unique rankings based on partially (red) and fully ranked (blue) data. 95\% confidence intervals are displayed.}
     \label{fig:pattern}
\end{figure}

To test this hypothesis, Figure \ref{fig:pattern} plots the empirical distributions of all observed rankings in the U.S. House (left) and U.S. Senate (right) races in Alaska (with 95\% confidence intervals). It also shows the results from the main experimental question (upper bars) and the supplemental question (lower bars) separately. Recent research shows that when (1) item order is randomized and (2) respondents reveal their sincere ranked preferences, the distribution of observed rankings will be uniform \citep{kimatsusaka2022}. Therefore, non-uniform distributions of observed rankings suggest the presence of ``pattern rankers.''

\indent

The figure confirms the above expectation. Some geometric patterns (e.g., 3241) are selected significantly more often than other patterns (e.g., 1234). Pearson's $\chi^2$ test rejects the null hypothesis that each of the four distributions is uniform. This result is remarkable because it challenges the conventional wisdom that the donkey vote (choosing 1234) is the most common form of ballot order effects in RCV, if any \citep{orr2002ballot,king2009ballot,curtice2014confused,robson1974importance,ortega2004position,ortega2008gender}.\footnote{I also analyze those who have either voted or are likely to vote, finding the same patterns.}

These non-uniform distributions have an important consequence in actual RCV (where candidate order is not randomized). For example, in the Alaskan contests, the most salient ranking profile 3241 corresponds to the ordering Tshibaka (R, Trump-endorsed) $\succ$ Kelley (R, withdrew) $\succ$ Chesbro (D) $\succ$  Murkowski (R, Incumbent)  for the U.S. Senate and Peltora (D, Incumbent) $\succ$ Bye (I) $\succ$ Begich (R) $\succ$ Palin (R, Trump-endorsed) for the U.S. House, respectively ($A\succ B$ means A is preferred to B). 

Furthermore, these pattern rankings were produced by \textit{attentive} respondents who passed the four attention checks in the surveys. When I replicate the analysis for \textit{inattentive} respondents, I find that the resulting distributions are \textit{uniform} (Figure \ref{fig:pattern_inatt}), suggesting that inattentive respondents offer actually random rankings. It is still unclear whether pattern rankings exist only among survey respondents or they also exist among actual voters. Future research must examine the nature and consequences of pattern rankings in RCV and survey research in more depth.

This way, the proposed framework not only promotes new empirical analyses (e.g., people's attitudes toward police violence) but also contributes to the development of novel theories (e.g., ballot order effects in RCV) in political science.

\section{Practical Guidance for Future Research}

This paper introduced a potential-outcomes framework to perform causal inference when outcome data are ranking data. The class of estimands in Equation (\ref{eq:general}) can be applied to various contexts of political science. \textit{How can researchers choose the ``right'' $g()$ (and thus estimand) when analyzing treatment effects on ranking data?}

The choice of $g()$ should be motivated by their hypothesis of interest. For example, if the hypothesis focuses on how the relative importance (i.e., rank) of a \textit{particular item} is affected by the treatment, the ARE can be a helpful estimand. Similarly, the APE can be an appropriate quantity to study if the hypothesis addresses the treatment effect on the relative importance of one or more \textit{pairs of items}. Importantly, researchers do \textit{not} need to hypothesize about \textit{all} available items if their theoretical predictions only concern a subset of available items. Online Appendix \ref{app:IR} illustrates the recommendation by studying conflict-related sexual violence \citep{agerberg2022sexual}, foreign donor \citep{dietrich2016donor}, and economic coercion \citep{gueorguiev2020impact}.

As discussed in \textsf{Alternative Quantities of Interest}, researchers can study many kinds of treatment effects with ranking data. While it is critical that researchers carefully examine their hypotheses at the design stage of experiments, an array of causal estimands can also motivate researchers to find new hypotheses \citep{grimmer2021machine}. One principled way to approach this problem is to (1) start with estimands that only address one or a few items (e.g., average ranks and probabilities of being in top-$k$) and (2) move on to estimands that cover multiple items (e.g., pairwise rankings, rank distances, and conditional effects). This way, causal inference with ranking data sheds light on the possibility that \textit{political scientists can draw more complex hypotheses from their substantive theories} than they do with more traditional outcome data.




\singlespacing
\setlength{\bibsep}{0pt plus 0.3ex}
\bibliography{RankOutcome}
\bibliographystyle{apsr}
\clearpage

\appendix
\setlength{\bibsep}{1pt}

\setcounter{page}{1} 

\part{Online Appendix} 
{{\LARGE For ``Causal Inference with Ranking Data: Application to Blame Attribution in Police Violence and Ballot Order Effects in Ranked-Choice Voting''}}\\

\vspace{0.1cm}

\vspace{0.1cm}

\parttoc\setlength{\bibsep}{0pt plus 0.3ex} 
\renewcommand\thefigure{\thesection.\arabic{figure}}
\renewcommand\thetable{\thesection.\arabic{table}}
\renewcommand\theequation{\thesection.\arabic{equation}}


\newpage
\section{Previous Applications of Rankings in Political Science}\label{app:application}
\setcounter{figure}{0} 
\setcounter{table}{0}  
\setcounter{equation}{0} 
\setcounter{footnote}{0} 

Political scientists have long used rankings and ranking data to study various political phenomena. For example, research in \textbf{American politics} has studied

\begin{itemize}\setlength\itemsep{0em}
    \item political parties \citepSM{aldrich1995parties}
    \item Iowa caucuses \citepSM{brady1990dimensional}
    \item committees \citepSM{stewart1999value,groseclose1998value}
    \item seniority-based nominations \citepSM{cirone2021seniority}
    \item party disciplines \citepSM{mccarty2001hunt}
    \item political texts \citepSM{carlson2017pairwise}
    \item legislative speech \citepSM{quinn2010analyze}
    \item redistricting \citepSM{kaufman2021measure}
    \item representation \citepSM{tate2003black}
    \item political knowledge \citepSM{prior2005news}
    \item political values \citepSM{ciuk2016americans}
    \item political efficacy \citepSM{king2004enhancing}
    \item issue voting \citepSM{rivers1988heterogeneity}
    \item referendum \citepSM{loewen2012testing}
\end{itemize}

Moreover, research in \textbf{comparative politics} has examined

\begin{itemize}\setlength\itemsep{0em}
    \item party-list proportional representation \citepSM{cox2021moral,buisseret2022party}
    \item ranked-choice voting (RCV) \citepSM{tolbert2021editor}
    \item nationalism and ethnic identity \citepSM{miles1991nationalism}
    \item  strategic voting \citepSM{cain1978strategic}
    \item ethnic diversity \citepSM{baldwin2010economic}
    \item postmaterialism \citepSM{duch1993postmaterialism,inglehart1999measuring}
    \item the Chinese Communist Party \citepSM{kung2011tragedy,shih2012getting}
\end{itemize}

Furthermore, studies in \textbf{international relations} have investigated

\begin{itemize}\setlength\itemsep{0em}
    \item human rights \citepSM{mcfarland2005cares}
    \item international governance \citepSM{kelley2015politics,kelley2021governance}
    \item global redulatory behavior \citepSM{doshi2019power}
    \item civil wars and democratic reversals \citepSM{goldstone2010global}
    \item burden-sharing in security governance \citepSM{dorussen2009sharing}
    \item compliance in international agreements \citepSM{dai2005comply}
\end{itemize}

Finally, political scientists have also studied \textbf{other topics} using rankings

\begin{itemize}\setlength\itemsep{0em}
    \item behavioral social choice \citepSM{regenwetter2006behavioral}
    \item policy punctuation \citepSM{jones2003policy}
    \item diffusion in judicial doctrines \citepSM{canon1981patterns}
    \item diffusion in public policy \citepSM{desmarais2015persistent}
    \item local public goods \citepSM{olken2010direct}
    \item academic journals \citepSM{carter2008under}
    \item political science departments \citepSM{hix2004global,frese2022academic}
\end{itemize}

\section{Discussion on \citet{boudreau2019police}}
\setcounter{figure}{0} 
\setcounter{table}{0}  
\setcounter{equation}{0} 
\setcounter{footnote}{0} 

\subsection*{The Experimental Designs}\label{app:vignette}
The full vignette for the control is: ``Stephon Clark, a 22-year-old black man, was shot and killed on the evening of March 18, 2018, by two Sacramento Police Department officers. The officers were looking for a suspect who was breaking windows in the Meadowview neighborhood of Sacramento. They confronted Clark, who they found in the yard of his grandmother’s house, where he lived. The officers ordered Clark to stop and show his hands. Clark then ran from the officers. According to the police, Clark turned and held an object in front of him while he moved toward the officers. The officers believed that Clark was pointing a gun at them and, fearing their lives were in danger, they shot and killed Clark. After the shooting, the police reported that Clark was only carrying a cell phone'' \citepSM[2]{boudreau2019police_OA}.

\indent

The additional vignette for the pattern-of-violence condition is: ``The Stephon Clark shooting is one of several controversial incidents involving police officers and the use of deadly force in Sacramento. In 2017, there were 25 incidents in Sacramento County that involved the use of force resulting in serious injury or death of the suspect, up from 20 in 2016. In July of 2016, two Sacramento Police Department officers shot and killed Joseph Mann, a 51-year old mentally-ill black man who was carrying a pocketknife and acting erratically. Mann refused to comply when the police ordered him to drop his knife and get on the ground. Shortly after, cameras recorded the two officers trying to run over Mann with their police cruiser. When they missed, the officers stopped the cruiser and chased Mann on foot. Moments later, the officers fired 18 times and killed Mann. The Clark and Mann shootings are among a series of incidents where civilians not armed with guns and arguably posing no real imminent threat to officers or others, were shot and killed by the Sacramento police'' \citepSM[3]{boudreau2019police_OA}.

\indent

The additional vignette for the reform group is: ``In response to several controversial incidents involving police officers and the use of deadly force, the Sacramento Police Department has adopted multiple reforms aimed at preventing incidents like the Stephon Clark shooting. Chief among them are the creation of a citizen oversight commission to review police actions and a video-release policy that requires the department to release footage in officer-involved shootings within 30 days. Both reforms go beyond what many other police departments in California have done to improve transparency. The Sacramento Police Department also now requires officers to undergo a 40-hour Crisis Intervention course, which trains police on how to de-escalate encounters with the public. And the Police Department is now equipping all officers with less lethal weapons, including bean bag guns and pepper ball guns. After the Clark shooting, the Police Department also changed its policies for chasing suspects to further emphasize officer and public safety'' \citepSM[5]{boudreau2019police_OA}.

\subsection*{The Original Analysis}\label{app:discuss}

The authors assume the following likelihood function:
\begin{align}
    \mathcal{L}(\text{Parameters}|\text{Rankings of $J$ Items by $N$ Respondents}) \propto \prod_{i=1}^{N} \prod_{j=1}^{J-1}\frac{\exp(\alpha_{ij})}{\sum_{j \in S}\exp(\alpha_{ij})}\label{eq:logit}
\end{align}

\noindent with interactive terms of unit and item-specific covariates for support parameter $\alpha_{ij}$,\footnote{The details of the original specification are available in the Online Appendix (pages 20-24) of \citetSM{boudreau2019police}. Mean rankings by the group are also reported in Table A.1 in their Online Appendix while the authors do not refer to the table in their main text and these tests are not motivated by any causal quantity of interest.} 
\begin{align}
    \alpha_{ij} = \sum_{j=1}^{3}\gamma_{j}Z_{j} + \sum_{j=1}^{3}\sum_{t=1}^{2}\beta_{jt}Z_{j}X_{it} + \sum_{j=1}^{3}\sum_{s=1}^{2}\zeta_{js}S_{is}  + \sum_{j=1}^{3}\sum_{t=1}^{2}\sum_{s=1}^{2}\delta_{jts}Z_{j}X_{it}S_{is}  + \epsilon_{ij},
\end{align}

\noindent where $Z_{j}$ is an item specific binary covariate that takes 1 if item $j$ is either ``Police'' ($j=1$), ``Local officials'' ($j=2$), or ``State or federal officials'' ($j=3$), $X_{it}$ is a binary variable for each of the treatment conditions that takes 1 if unit $i$ is in the pattern-of-violence treatment group ($X_{i1}$) and the reform treatment group ($X_{i2}$). The model also includes a unit specific binary covariate $S_{is}$ that takes 1 if unit $i$ is a resident in Sacramento Country (where the Clark shooting occurred) ($S_{i1}$) or African American ($S_{i2}$). Here, $j \in S$ in Equation (\ref{eq:logit}) represents the set of remaining items following the conventional multinomial logit model.

After estimating the model, the authors compute several quantities of interest by combining the estimated parameters. For example, using $\widehat\gamma_{1}=1.06$ (the coefficient on the police officers dummy), $\widehat\beta_{11}=0.460$ (the coefficient on the interaction of the police officers and the pattern-of-violence treatment dummies), and $\widehat\beta_{12}=-0.269$ (the coefficient on the interaction of the police officers and the reform treatment dummies), the original study reports that the ``likelihood'' that non-Sacramento residents who are not African American \textit{rank the police officers higher than the victim} (i.e., blame more on the police officers than on the victim) is $\widehat\gamma_{1}=1.06$ in the control group, $\widehat\gamma_{1}+\widehat\beta_{11}=1.52$ in the pattern-of-violence group, and $\widehat\gamma_{1}+\widehat\beta_{12}=0.79$ in the reform group, respectively. Finally, the study claims that the causal effects of the two treatments for this particular subpopulation are $1.52-1.06=0.46$ (which is $\widehat\beta_{11}$ by definition) and $0.79-1.06=-0.27$ (which is $\widehat\beta_{12}$ by definition), respectively \citepSM[1106]{boudreau2019police}.

Table \ref{tab:Boudreau} shows the estimated coefficients for non-Sacramento non-African American respondents, for which the authors find supportive evidence for their hypotheses.\footnote{The original study seems to misreport that the results are for ``non-Sacramento respondents'' as opposed to non-Sacramento non-African American respondents. However, I focus on its modeling strategy rather than casting doubt on its conclusion based on this misreporting.} Columns 5-6 show the estimated effects of the two treatments for three chosen categories while setting the victim as a reference category. Based on these results, the original study suggests that, for example, the pattern-of-violence treatment ``increased the likelihood'' that the respondents rank the police officers higher (i.e., blame them more) than the victim by 0.46, which is statistically significant at the 0.05 level. In contrast, the reform treatment ``decreased the likelihood'' that the respondents rank the police officers higher than the victim by 0.27  at the same significance level \citepSM[1106]{boudreau2019police}. Taken together, the original study concludes that it finds supportive evidence for its two hypotheses for this specific subpopulation (but not for other subpopulations).

\indent

\begin{table}[tph!]
    \centering
    \begin{tabular}{l|cccccc}\hline
        Object of Blame & Control & Pattern &  Reform & $\Delta$(Pattern, Control) & $\Delta$(Reform, Control) \\\hline
       Police                  & 1.06  & 1.52  & .79  & .46* & -.27*\\
       Local officials         & .02   & .37   & -.05 & .35* & -.07 \\
       State/federal officials & -.60  & -.33  & -.53 & .27  & .07\\ \hline
    \end{tabular}
    \caption{\textbf{Rank-Ordered Logit Estimates Reported in the officer-involved shootings Study}\\ \textit{Note}: This table shows estimated coefficients for non-Sacramento non-African American respondents reported in \citet[1106]{boudreau2019police}. In Columns 5-6, $*$ indicates that the reported difference is statistically significantly different from zero at the 0.05 level based on the two-tailed test.}
    \label{tab:Boudreau}
\end{table}

\section{Additional Discussion on Causal Estimands}
\setcounter{figure}{0} 
\setcounter{table}{0}  
\setcounter{equation}{0} 
\setcounter{footnote}{0} 

\subsection*{Causal Estimands Degenerate Into Difference-in-Means Estimands for Binary Outcomes}\label{app:degenerate}

\subsubsection*{Average Rank Effects}

To demonstrate, I use $Y_{i} \in \{0,1\}$ to denote a binary outcome and $Y_{i}(1)$ and $Y_{i}(0)$ to represent its potential outcomes under the treatment and control conditions, respectively. A conventional average treatment effect measure is:
\begin{equation}
    \tau \equiv \E[Y_{i}(1) - Y_{i}(0)]
\end{equation}

Now consider that the binary choice is generated by rankings of two items $j$ and $k$ (e.g., NO and YES), where marginal rankings are $R_{ij}=\{1,2\}$. Without changing any substantive meaning, consider a slightly modified version of the marginal rankings $R_{ij}^{*}=\{0,1\}$, where 0 means that item $j$ is preferred to item $k$ (e.g., NO) and 1 means otherwise (e.g., YES). Then, it is straightforward to see $\E[Y_{i}] = \E[R_{ij}]$ (the average number of YESs is equal to the average ranking of NO). Using this relationship, it can be shown that the ARE can be seamlessly related to the conventional effect measure:
\begin{subequations}
\begin{align}
\tau_{1} & = \E[R_{i1}^{*}(1) - R_{i1}^{*}(0)]\\
     & = \E[R_{i1}^{*}(1)] - \E[R_{i1}^{*}(0)]\\
     & = \E[Y_{i}(1)] - \E[Y_{i}(0)]\\
     & = \E[Y_{i}(1) - Y_{i}(0)]\\
     & = \tau
\end{align}
\end{subequations}

\subsubsection*{Average Pairwise Rank Effects}

\begin{subequations}
\begin{align}
\tau_{P} & \equiv \E[1(A_{i,j} < A_{i,k})(1)] - 1(A_{i,j} < A_{i,k})(0) ]\\
         & = \E[Y_{j< k}(1) - Y_{j< k}(0)]\\
         & = \E[R_{ij}(1) - R_{ij}(0)]\\
         & = \tau_{M,j}
\end{align}
\end{subequations}

\section{Partial Identification for Rank Treatment Effects}\label{app:partial}
\setcounter{figure}{0} 
\setcounter{table}{0}  
\setcounter{equation}{0} 
\setcounter{footnote}{0} 

\subsection*{Nonparametric Sharp Bounds}
Sharp bounds on causal effects are determined by the minimum and maximum values in the support of the outcome variable ($\text{Supp}[Y_{i}({d})] \subseteq [Min,Max]$). Below, I consider such nonparametric bounds for AREs and APEs.

\subsubsection*{Sharp Bounds for the ARE}
To present nonparametric bounds for AREs, I consider the minimum and maximum possible values for the marginal ranking $R_{ij}$. Since this work focuses on full rankings (as opposed to partial rankings), the support of the marginal ranking is a vector of $t$ numbers. Hence, $\text{Supp}(R_{ij}) \subseteq \{1,...,t\} \subseteq[1,J]$. 
\begin{subequations}
\begin{align}
\tau_{j}  \in \Big[ & \E[R_{ij}|D_{i}=1]\mathbb{P}(D_{i}=1) + \mathbb{P}(D_{i}=0) - \big(\E[R_{ij}|D_{i}=0]\mathbb{P}(D_{i}=0) + t\mathbb{P}(D_{i}=1) \big),\\
       \big( & \E[R_{ij}|D_{i}=1]\mathbb{P}(D_{i}=1) + t\mathbb{P}(D_{i}=0) - \big(\E[R_{ij}|D_{i}=0]\mathbb{P}(D_{i}=0) + \mathbb{P}(D_{i}=1) \big) \Big]
\end{align}
\end{subequations}

\subsubsection*{Sharp Bounds for the APE}
Next, partial identification for the APE is straightforward as it is based on a set of indicator functions. Since the support of indicator functions is $\text{Supp}[I(R_{ij} < R_{ik})(d)] \subseteq\{0,1\} \subseteq[0,1]$, the sharp bounds for the APE is:
\begin{subequations}
\begin{align}
\tau^{P}_{jk}  \in \Big[ & \mathbb{P}(R_{ij} < R_{ik}|D_{i}=1)\mathbb{P}(D_{i}=1) - \big(\mathbb{P}(R_{ij} < R_{ik}|D_{i}=0)\mathbb{P}(D_{i}=0) + \mathbb{P}(D_{i}=1) \big),\\
       \big( & \mathbb{P}(R_{ij} < R_{ik}|D_{i}=1)\mathbb{P}(D_{i}=1)  + \mathbb{P}(D_{i}=0) \big) - \mathbb{P}(R_{ij} < R_{ik}|D_{i}=0)\mathbb{P}(D_{i}=0) \Big]
\end{align}
\end{subequations}



\section{Results for Simulation Studies}\label{app:simulation}
\setcounter{figure}{0} 
\setcounter{table}{0}  
\setcounter{equation}{0} 
\setcounter{footnote}{0} 

\subsection*{Standard Errors}\label{app:var}

Accordingly, I consider the sampling variance of the estimated ARE for item $j$ as follows:
\begin{align}
    \V \big( \widehat{\tau}_{j} \big) & = \frac{S_{t}^{2}}{\sum_{i=1}^{N}D_{i}} + \frac{S_{c}^{2}}{\sum_{i=1}^{N}(1-D_{i})} - \frac{S_{tc}^{2}}{N},
\end{align}

\noindent where $S_{t}^{2}=\frac{1}{N-1}\sum_{i=1}^{N}(R_{ij}(1) - \overline{R}_{j}(1)),$ $S_{c}^{2}=\frac{1}{N-1}\sum_{i=1}^{N}(R_{ij}(0) -\overline{R}_{j}(0)),$ and $S_{tc}^2 = \frac{1}{N-1}\sum_{i=1}^{N}\big(R_{j}(1) - R_{j}(0) - \tau_{j}\big)^2$. Here, $\overline{R}_{j}(1)$ and $\overline{R}_{j}(0)$ represent the average potential mean ranks for item $j$ under the treatment and control, respectively. To obtain the estimated variance of the estimator for item $j$, I adopt the Neyman estimator of the sampling variance \citepSM[96]{imbens2015causal}:
\begin{align}
    \widehat{\V(\widehat{\tau_{j}})} = \frac{s_{t}^2}{\sum_{i=1}^{N}D_{i}} + \frac{s_{c}^2}{\sum_{i=1}^{N}(1-D_{i})}, 
\end{align}
\noindent where $s_{t}^{2} = \frac{\sum_{i=1}^{N}D_{i}(R_{ij} - \overline R_{tj})^2}{\sum_{i=1}^{N}D_{i} - 1}$ and $s_{c}^{2} = \frac{\sum_{i=1}^{N}(1-D_{i})(R_{ij} - \overline R_{cj})^2}{\sum_{i=1}^{N}(1-D_{i}) - 1}$ are sample variances for sample mean ranks among the treated and control units, respectively, whereas 
$\overline R_{tj}$ and $\overline R_{cj}$ are mean ranks of item $j$ for the treated and control units.

Likewise, the sampling variance of the APE for items $j$ and $k$ is
\begin{align}
    \V \big( \widehat{\tau}^{P}_{jk} \big) & = \frac{S_{t}^{'2}}{\sum_{i=1}^{N}D_{i}} + \frac{S_{c}^{'2}}{\sum_{i=1}^{N}(1-D_{i})} - \frac{S_{tc}^{'2}}{N},
\end{align}

\noindent where $S_{t}^{'2}=\frac{1}{N-1}\sum_{i=1}^{N}(I(R_{ij} < R_{ik})(1) - \overline{R}_{jk}(1)),$ $S_{c}^{'2}=\frac{1}{N-1}\sum_{i=1}^{N}(I(R_{ij} < R_{ik})(0) -\overline{R}_{jk}(0)),$ and $S_{tc}^{'2} = \frac{1}{N-1}\sum_{i=1}^{N}\big(R_{j}(1) - R_{j}(0) - \tau_{j}\big)^2$. Here, $\overline{R}_{jk}(1)$ and $\overline{R}_{jk}(0)$ represent the probability that item $j$ is preferred to item $k$ in the treatment and control groups, respectively. A corresponding estimator of this sampling variance is:
\begin{align}
    \widehat{\V(\widehat{\tau}^{P}_{jk}}) = \frac{s_{t}^{'2}}{\sum_{i=1}^{N}D_{i}} + \frac{s_{c}^{'2}}{\sum_{i=1}^{N}(1-D_{i})}, 
\end{align}
\noindent where $s_{t}^{'2} = \frac{\sum_{i=1}^{N}D_{i}(I(R_{ij} < R_{ik}) - \overline R_{tjk})^2}{\sum_{i=1}^{N}D_{i} - 1}$ and $s_{c}^{'2} = \frac{\sum_{i=1}^{N}(1-D_{i})(I(R_{ij} < R_{ik}) - \overline R_{cjk})^2}{\sum_{i=1}^{N}(1-D_{i}) - 1}$ are sample variances for sample mean ranks among the treated and control units, respectively, whereas 
$\overline R_{tjk}$ and $\overline R_{cjk}$ are the sample proportion that item $j$ is ranked higher than item $k$ in the treatment and control, respectively.

Future research must examine the population and sample variances for the between-population diversity estimator based on the chi-squared approximation \citepSM[68]{alvo2014statistical}.

\subsection*{Plausibility of Normal Approximation of Treatment Effects}\label{app:sim_approx}

\begin{figure}[tph!]
    \centering
    \includegraphics[width=15cm]{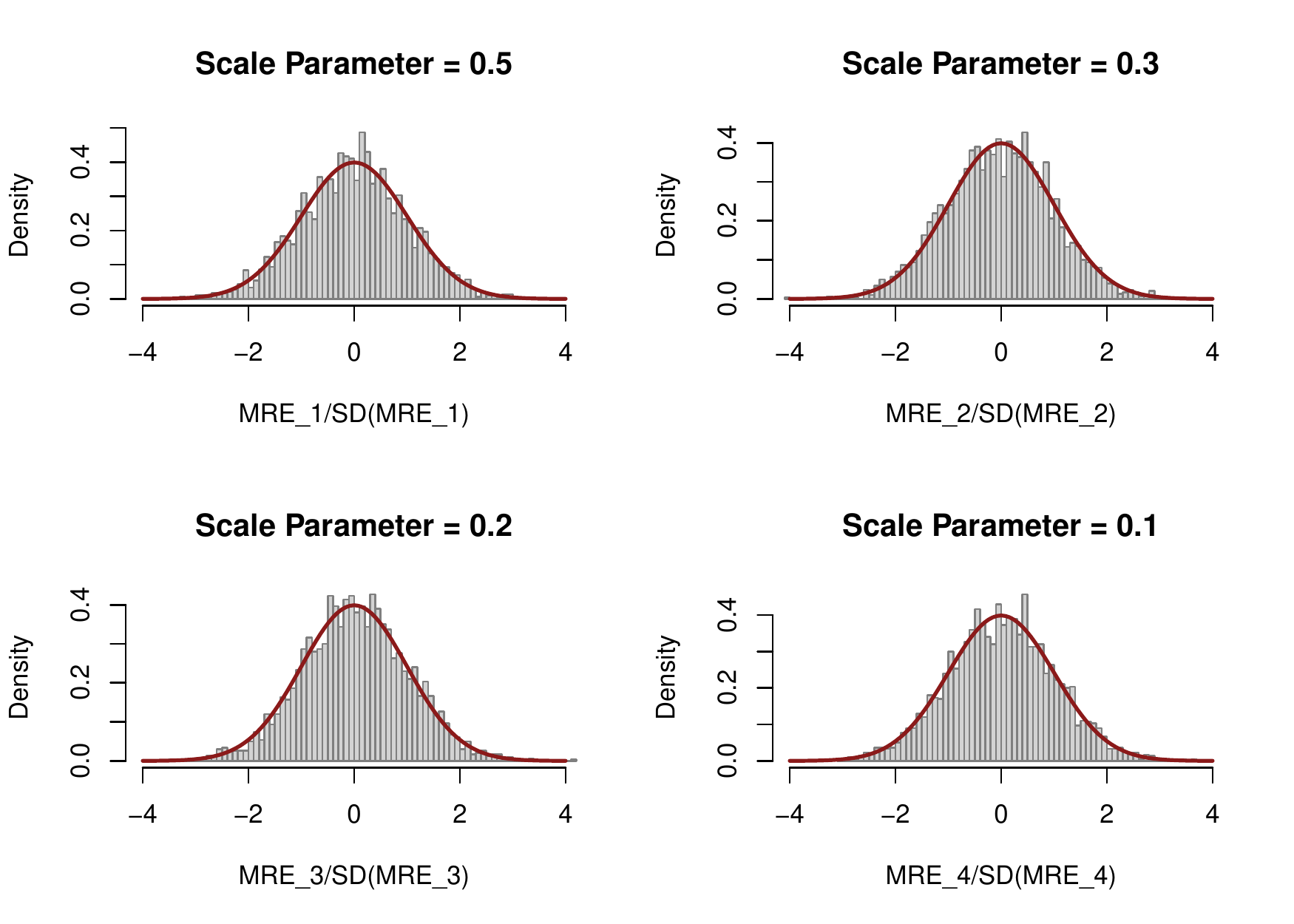}
    \caption{\textbf{Distribution of Null population average rank effects for Four Items}}
    \label{fig:normalARE}
\end{figure}

Here, I present simulation results for the empirical distribution of the proposed treatment effects for ranked outcomes. Figure \ref{fig:normalARE} shows the distribution of null population average rank effects (ARE) for four items with scale parameters (0.5,0.3,0.2,0.1). In this simulation, I draw 2000 samples from a Plackett-Luce model with the specified scale parameters for the quasi-treatment and quasi-control groups. I then compute the ARE using the samples from the two groups and repeat this process 3000 times. Finally, I divide the obtained AREs by their sample standard deviation. Each panel shows that the empirical distribution of a null ARE seems to be well approximated by a standard normal density function, which is represented by a red curve.

\begin{figure}[tph!]
    \centering
    \includegraphics[width=18cm]{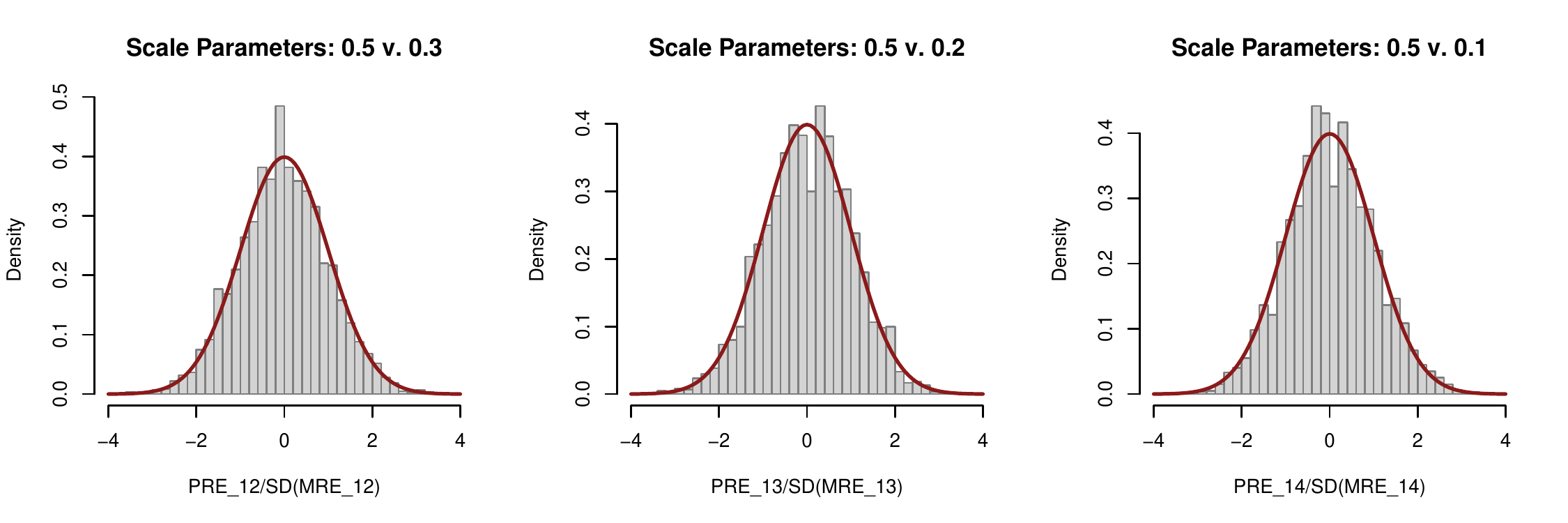}
    \caption{\textbf{Distribution of Null average pairwise rank effects for Four Items}}
    \label{fig:normalPRE}
\end{figure}


\newpage
\section{Additional Discussions on Multiple Hypothesis Testing}\label{app:multiple}
\setcounter{figure}{0} 
\setcounter{table}{0}  
\setcounter{equation}{0} 
\setcounter{footnote}{0} 

The BH method takes the following steps. First, compute $p$-values for the $J$ hypothesis tests and sort the obtained $p$-values by their size: $p_{(1)}<\cdots<p_{(m)}$, where $p_{(j)}$ is the $j$-th smallest $p$-value and $m$ being the highest $p$-value. Next, with a pre-specified level $\alpha$ (e.g., researchers may pre-specify $\alpha=0.05$ to have the false discovery rate less than 5\%), define an \textit{anchor value} $l_{i} = \frac{i\alpha}{\sum_{i=1}^{m}(1/i)m}$ for non-independent $p$-values \citepSM[167]{wasserman2004all}, and set the number of total rejections to the maximum number of tests for which their $p$-values are lower than their corresponding anchors: $r=\max\{i: p_{i}<l_{i}\}$. Finally, define the BH rejection threshold $T=p_{(r)}$ and reject all null hypotheses $H_{0i}$ for which $p_{i} \leq T$ (or simply reject $r$ null hypotheses with the $r$ smallest $p$-values).

Let $m_{0}$ be the number of null hypotheses that are true. Let $U$ be the number of null hypotheses that are true and not rejected, $T$ be the number of null hypotheses that are false and yet not rejected (false discovery), and $V$ be the number of null hypotheses that are true and yet rejected (false rejection). Suppose that there are $m$ hypothesis tests and $r\leq m$ tests in which the null is rejected. I then define the false discovery proportion (FDP) as $\text{FDP} = \frac{V}{r}$. In words, the FDR is a proportion of false rejection among $r$ rejected hypotheses. When $R=0$, the FDR is defined to be 0. This is considered to be an unbiased estimator of the false discovery rate (FDR). The BH method controls the FDR via $\text{FDR} = \E[\text{FDP}] \leq \frac{m_{0}}{m}\alpha \leq \alpha$.

\section{Additional Findings and Discussions on Police Violence}\label{app:morefinding}
\setcounter{figure}{0} 
\setcounter{table}{0}  
\setcounter{equation}{0} 
\setcounter{footnote}{0} 

In this section, I replicate my primary analysis in the main text. 

\subsection*{Multiple Hypothesis Testing for the Reform Treatment}\label{app:BH2}

So far, the reanalysis has focused on the statistical significance of \textit{each} estimated effect based on its (estimated) 95\% confidence interval. Nevertheless, it is critical to perform multiple hypothesis testing to test the authors' hypotheses, considering all seven estimated effects. To do so, I first clarify multiple null hypotheses (that I hope to reject) based on \citet{boudreau2019police}. For example, for the pattern-of-violence treatment, I consider the following composite hypothesis:
\begin{align*}
H_{0v}: \ & \tau_{v} \leq 0 \quad \text{(not less blame for \underline{v}ictim)}\\
H_{0g}: \ & \tau_{g} \leq 0 \quad  \text{(not less blame for \underline{g}overnor)}\\
H_{0s}: \ & \tau_{s} \leq 0 \quad \text{(not less blame for \underline{s}enators)}\\
H_{0o}: \ & \tau_{o} \geq 0 \quad \text{(not more blame for \underline{o}fficers)}\\
H_{0c}: \ & \tau_{c} \geq 0 \quad \text{(not more blame for \underline{c}hief)}\\
H_{0m}: \ & \tau_{m} \geq 0 \quad \text{(not more blame for \underline{m}ayor)}\\
H_{0d}: \ & \tau_{d} \geq 0 \quad \text{(not more blame for \underline{d}istrict attorney)}
\end{align*}

Second, with the above hypothesis, I compute seven $p$-values at the significance level of $\alpha=0.05$ based on the one-tailed test. To illustrate, I first estimate, for example, the effect for the victim as $\widehat{\tau}_{v}=0.172$ with its standard error $\widehat{\sigma}_{v}=0.238$. With these values, I then compute a Wald statistic $w_{v} = \frac{\widehat{\tau}_{v} - \theta_{v0}}{\widehat{\sigma}_{v}} = \frac{0.172}{0.238} = 0.701$, where $\theta_{v0}=0$ is the null target value. Using the normal approximation to the distribution of the Wald statistic, I then compute the $p$-value for the victim as $1-\mathbb{P}(Z>w_{v})=0.242$. Following the same procedure, I compute the $p$-values of all estimated effects (from $H_{0v}$ to $H_{0d}$).

Third, I sort these $p$-values from the smallest to the largest: (0.000, 0.076, 0.204, 0.218, 0.242, 0.347, and 0.441) as shown in Table \ref{tab:BH}. As described earlier, I then compute a vector of anchor values \textbf{l} = (0.007, 0.01, 0.012, 0.014, 0.016, 0.018, and 0.019). This yields the number of total rejections $r = \max\{i: p_{i}<l_{i}\} = 1$, which in turn determines the BH rejection threshold $T=p_{(r)}=0.0004$ (the one smallest $p$-value in the sorted list). Finally, I reject all null hypotheses $H_{0i}$ for which $p_{i} \leq T$. Consequently, I reject the null hypothesis $H_{0m}$ but not the other six null hypotheses. Substantively, this means that I find supporting evidence only for the mayor, which casts doubt on the original finding about the first hypothesis. Table \ref{tab:BH} presents the full list of statistics for the multiple testing. Table \ref{tab:BH_reform} shows the results for the reform treatment.


\indent
\begin{table}[h!]
    \centering
    \begin{tabular}{rrrrrrrr}\hline
        Null     & Party & $\widehat{\tau}$ & $\widehat{\sigma}$ & Wald ($\widehat{\tau}/\widehat{\sigma}$) & $p$-value & Threshold ($T$) & Reject if $p\leq T$   \\\hline
        $H_{0m}$ & 	Mayor     & -0.138   & 0.033 &  -4.869  & 0.000  & 0.000 & Reject   \\
        $H_{0c}$ &  Chief     & -0.087  & 0.060  &  -1.433  & 0.076  & 0.000 & Not reject   \\ 
        $H_{0s}$ & 	Senators  &  0.064  & 0.083  &	0.826   & 0.204  & 0.000 & Not reject 	 \\
        $H_{0d}$ &  DA        &  0.035  & 0.047  &  0.778   & 0.218  & 0.000 & Not reject  \\
        $H_{0v}$ & 	Victim    &  0.172  & 0.238  &	0.701   & 0.242  & 0.000 & Not reject  \\
        $H_{0g}$ &  Governor  & -0.030  & 0.080  &	-0.394  & 0.347  & 0.000 & Not reject \\ 
        $H_{0o}$ & 	Officers  & -0.016  & 0.112  &	-0.149  & 0.441  & 0.000 & Not reject  \\\hline
    \end{tabular}    
    \caption{\textbf{Multiple Hypothesis Testing for the Pattern-of-Violence Treatment}\\\textit{Note}: This table presents a series of statistics required for and the result for the Benjamini-Hochberg method of multiple hypothesis testing. In this reanalysis, $\alpha=0.05$ and $\theta_{0}=0.$}
    \label{tab:BH}
\end{table}

\begin{table}[h!]
    \centering
    \begin{tabular}{rrrrrrrr}\hline
        Null     & Party & $\widehat{\tau}$ & $\widehat{\sigma}$ & Wald ($\widehat{\tau}/\widehat{\sigma}$) & $p$-value & Threshold ($T$) & Reject if $p\leq T$   \\\hline
        $H_{0d}$ &  DA        &  0.168    & 0.047   &  3.559    & 0.000  & 0.006 & Rejected  \\ 
        $H_{0m}$ & 	Mayor     &  -0.084   & 0.033   &  -2.525   & 0.006 & 0.006 & Rejected   \\        
        $H_{0c}$ &  Chief     &  0.085    &  0.060  &   1.426   & 0.077  & 0.006 & Not reject   \\ 
        $H_{0g}$ &  Governor  &  -0.103   &  0.080  &	-1.289  & 0.099 & 0.006 & Not reject \\       
        $H_{0o}$ & 	Officers  &  0.121    & 0.112   &	1.084   & 0.139  & 0.006 & Not reject  \\  
        $H_{0v}$ & 	Victim    & -0.160    & 0.238   &	-0.675  & 0.250  & 0.006 & Not reject  \\
        $H_{0s}$ & 	Senators  &  -0.027   & 0.083   &	 -0.331 & 0.370 & 0.006 & Not reject\\\hline          
    \end{tabular}    
    \caption{\textbf{Multiple Hypothesis Testing for the Reform Treatment}\\\textit{Note}: This table presents a series of statistics required for and the result for the Benjamini-Hochberg method of multiple hypothesis testing. In this reanalysis, $\alpha=0.05$ and $\theta_{0}=0.$}
    \label{tab:BH_reform}
\end{table}

\newpage
\subsection*{Average Rank Effects by Race and Ethnicity}\label{app:ARE_race}

\begin{figure}[h!]
    \centering
    \includegraphics[width=12cm]{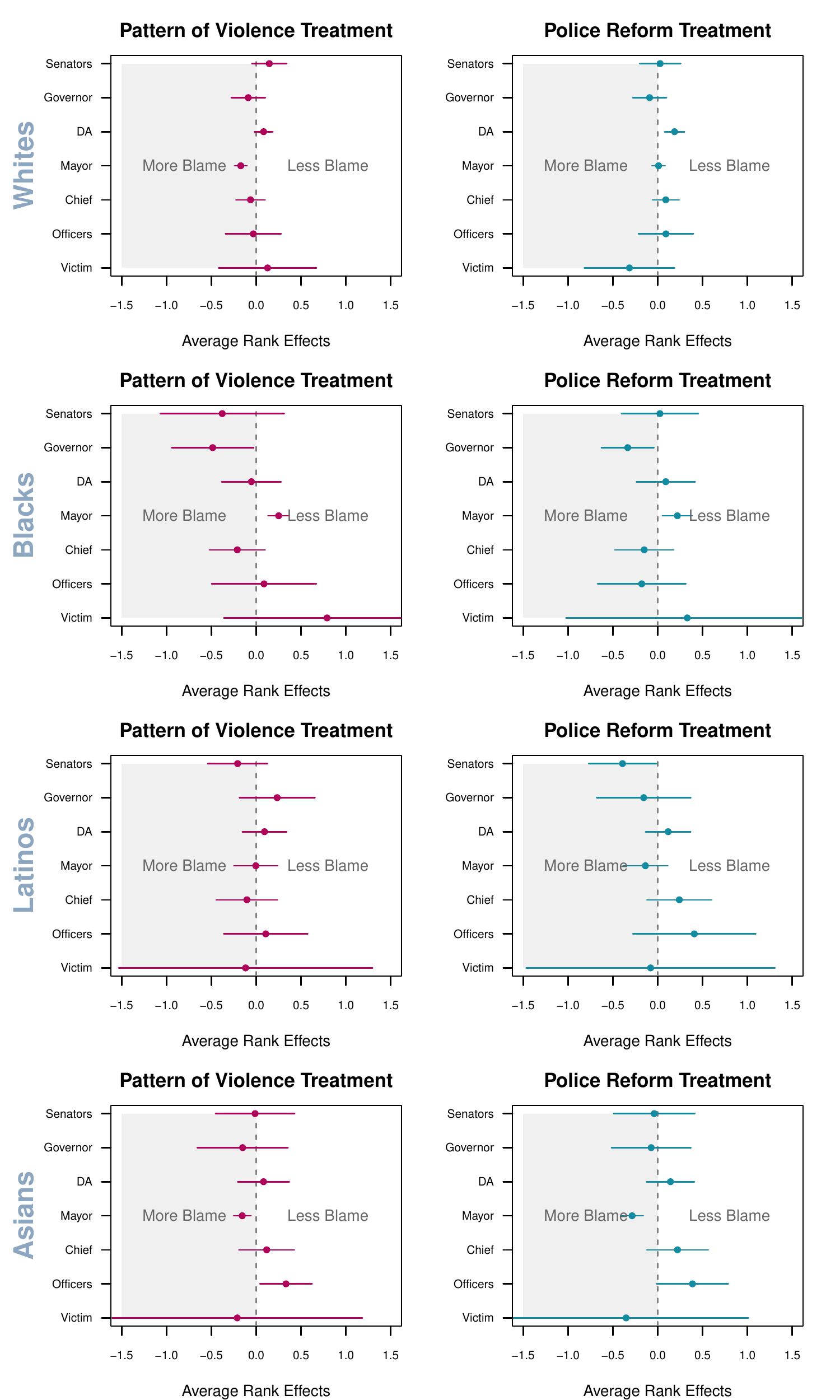}
    \caption{\textbf{Estimated Average Rank Effects by Race and Ethnicity}}
    \label{fig:ARE2}
\end{figure}

\newpage
\subsection*{Average Pairwise Rank Effects by Race and Ethnicity}\label{app:APE_race}

\begin{figure}[h!]
    \centering
    \includegraphics[width=10cm]{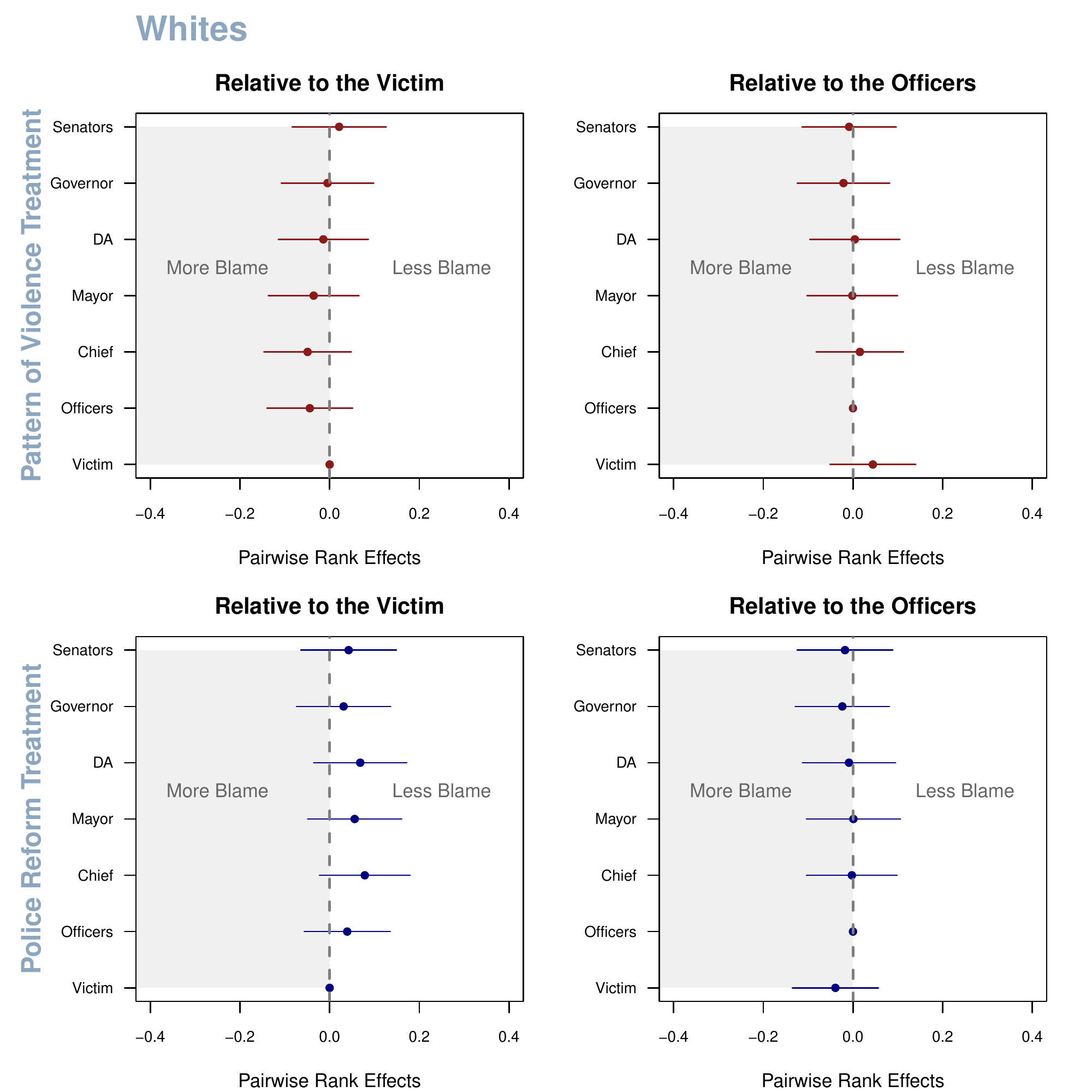}
    \caption{\textbf{Average Rank Effects of the Pattern-of-Violence and Reform Treatments for Whites}}
    \label{fig:PRE_D1}
\end{figure}

\begin{figure}[h!]
    \centering
    \includegraphics[width=10cm]{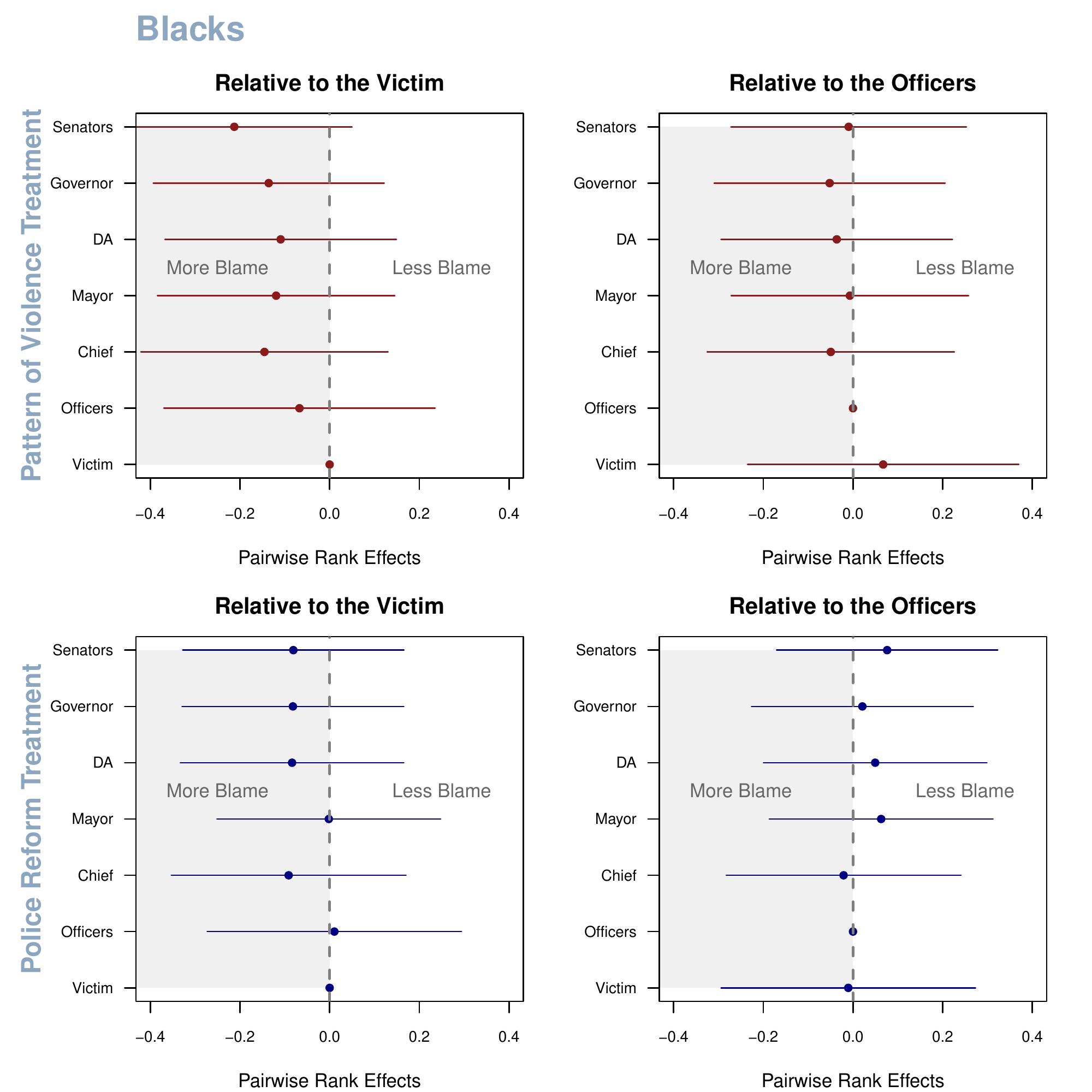}
    \caption{\textbf{Average Rank Effects of the Pattern-of-Violence and Reform Treatments for Blacks}}
    \label{fig:PRE_D2}
\end{figure}

\begin{figure}[h!]
    \centering
    \includegraphics[width=10cm]{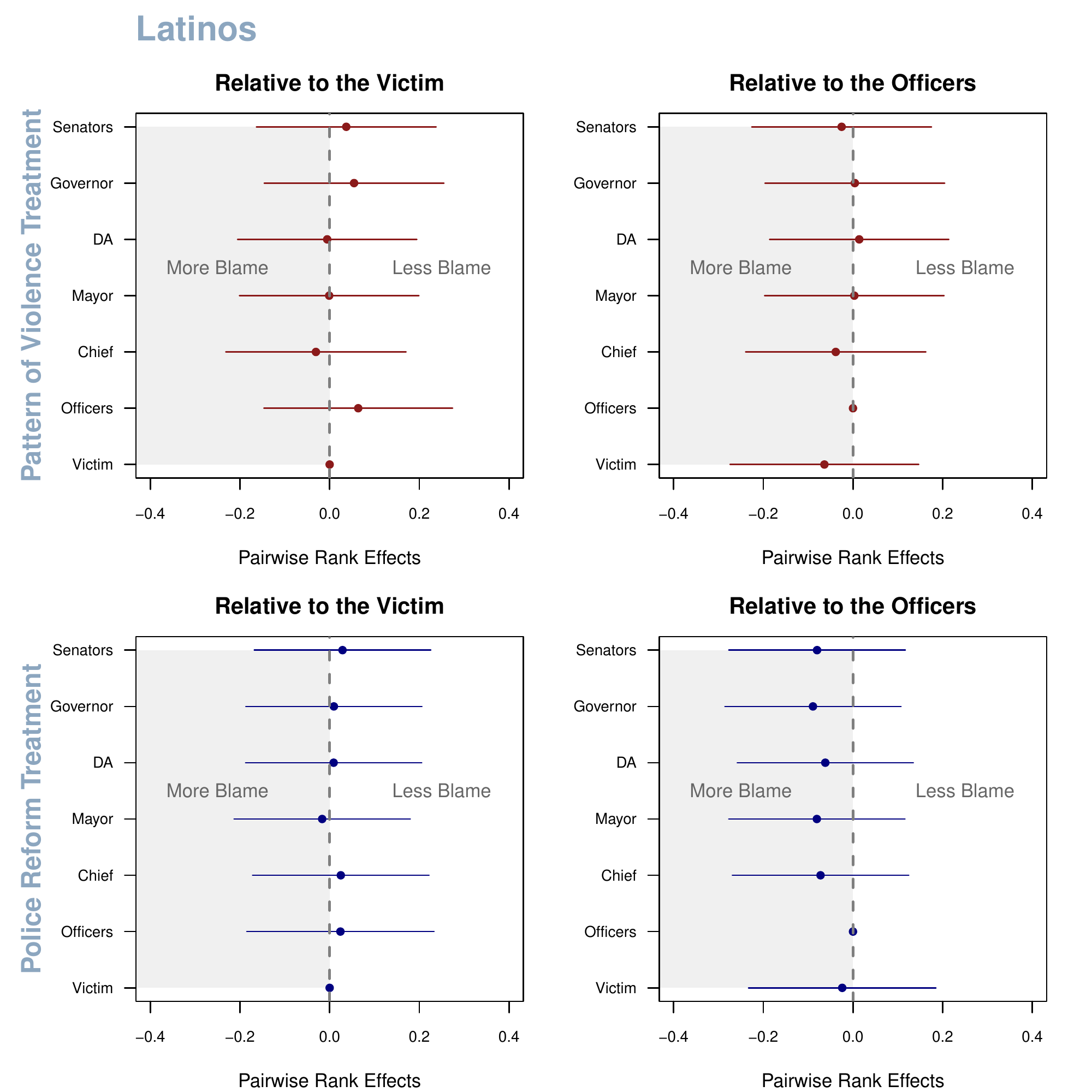}
    \caption{\textbf{Average Rank Effects of the Pattern-of-Violence and Reform Treatments for Latinos}}
    \label{fig:PRE_D3}
\end{figure}

\begin{figure}[h!]
    \centering
    \includegraphics[width=10cm]{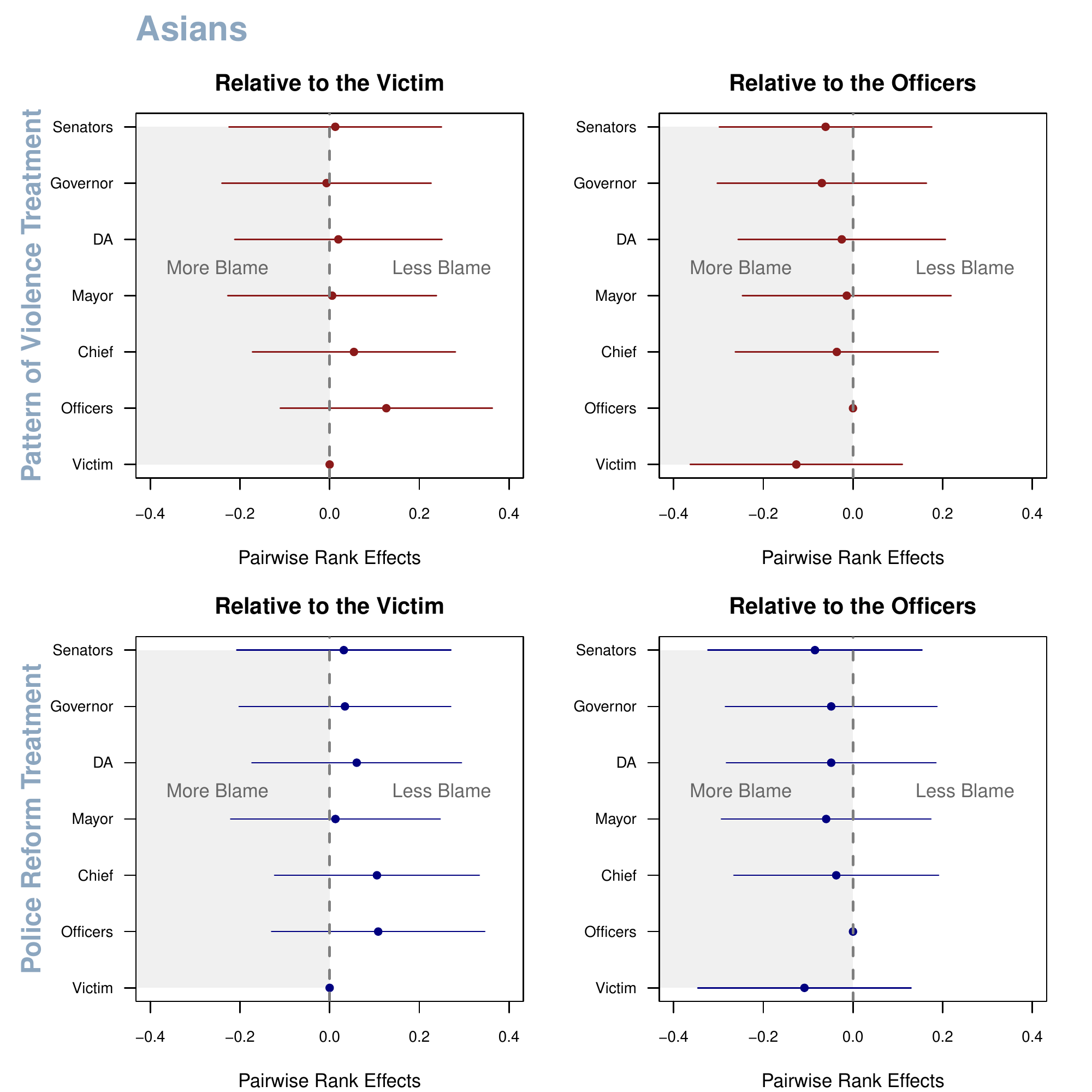}
    \caption{\textbf{Average Rank Effects of the Pattern-of-Violence and Reform Treatments for Asians}}
    \label{fig:PRE_D4}
\end{figure}

Figures \ref{fig:PRE_D1}-\ref{fig:PRE_D4} visualize the estimated PREs for each of the racial and ethnic groups.

\clearpage
\subsection*{Racial Differences in Blame Attribution in Police Violence}\label{app:RacialDiff}
 
\begin{figure}[h!]
    \centering
    \includegraphics[width=12cm]{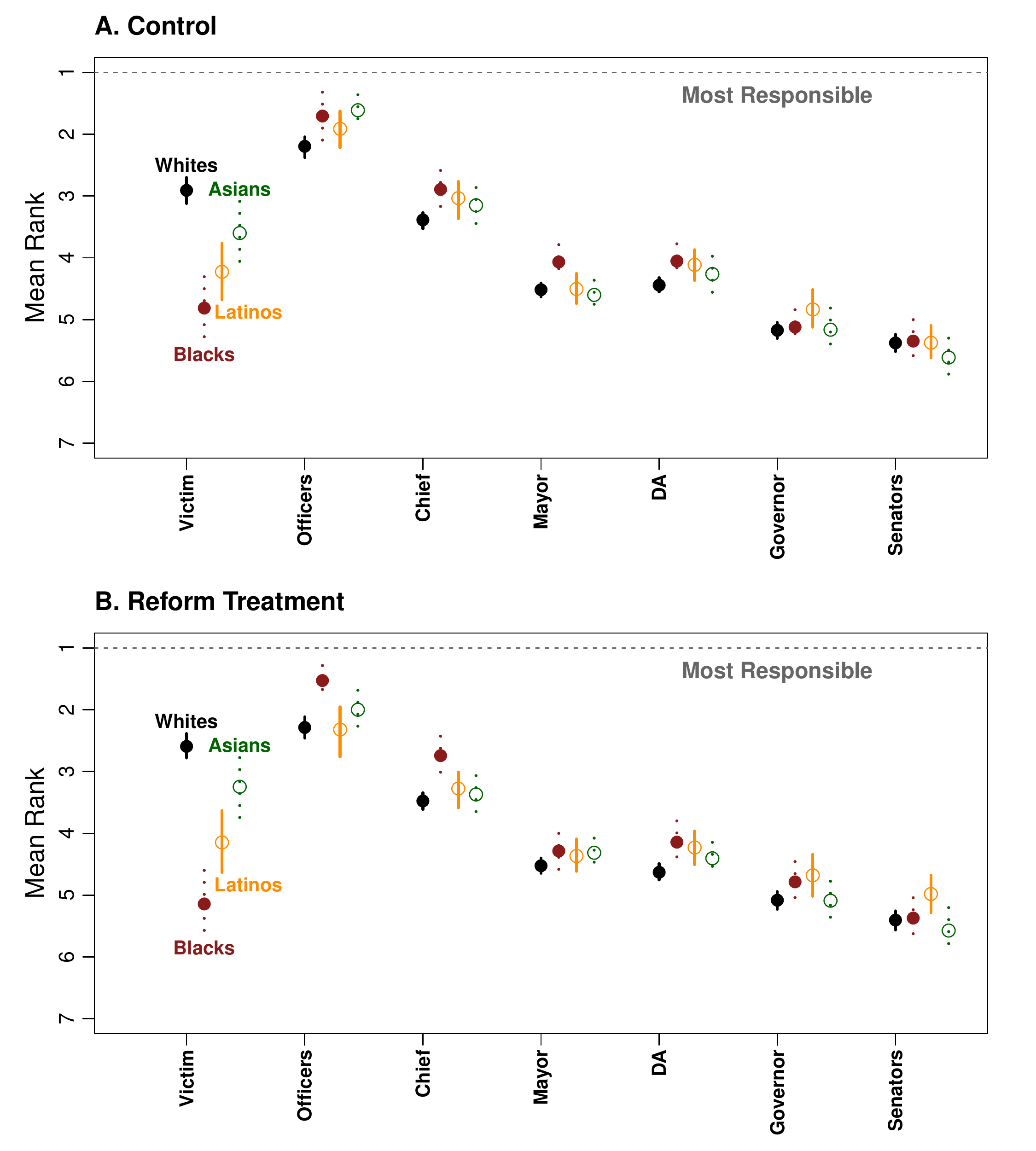}
    \caption{\textbf{Mean Ranks of Seven Parties by Race}\\ \textit{Note}: Each panel shows the estimated average ranks (with 95\% confidence intervals based on bootstrapping) for whites, blacks, Latinos, and Asians for the seven parties of interest, from the victim to the senators, within the same experimental condition. ``1'' in the y-axis means that a party is the most responsible (ranked first) for the shooting. The upper panel displays the result for the subjects in the control condition, whereas the lower panel shows the result for the respondents in the reform treatment group. }
    \label{fig:RacialDiff}
\end{figure}

Figure \ref{fig:RacialDiff} visualizes the estimated average rank (with bootstrapped 95\% confidence interval) of each party for each of the four groups (analyzed in the original study). The figure shows that while all groups think that the two police officers are the most responsible for the victim's death among the seven parties, a stark racial difference appears in their blame attribution for the victim. Within both treatment conditions, whites think that the victim is responsible for his death at the almost same level as the police officers, whereas blacks believe that the victim is one of the most minor responsible parties in the four counties. Both panels show that Latinos are closer to blacks, whereas Asians hold a similar position to whites regarding the victim. This finding suggests that the way people hold the police force accountable, at least in the context of officer-involved shootings, significantly differs by race and ethnicity, which underpins the importance of grand jury selection in police violence cases. 

\clearpage
\section{Identification of Order-Ranker Average Rank Effects}\label{app:orderrankers}
\setcounter{figure}{0} 
\setcounter{table}{0}  
\setcounter{equation}{0} 
\setcounter{footnote}{0} 

First, \textit{always-rankers} is a set of voters who always rank candidate $j$ regardless of the treatment assignment. Second, \textit{order-rankers} is a group of voters who rank candidate $j$ only when they are in the active treatment condition. Third, \textit{never-rankers} mean voters who never rank candidate $j$ regardless of the treatment assignment. Finally, \textit{defiers} is a set of voters who only rank candidate $j$ when they do not see the candidate at the top of the ballot.

\indent

Let $S\in\{\text{order-rankers},\text{always-rankers},\text{never-rankers},\text{defiers}\}$ denote a stratum membership and $s$ denote a specific type of voters in $S$. Let $\pi_{o}$, $\pi_{a}$, $\pi_{n}$, $\pi_{d}$ be the proportion of \underline{o}rder-rankers, \underline{a}lways-rankers, \underline{n}ever-rankers, and \underline{d}efiers, respectively. Here, the candidate index $j$ is suppressed for clarity. 

\indent

To identify the order-ranker ARE, I first rewrite the ARE as a weighted average of strata-specific AREs:
\begin{subequations}
\begin{align}
\tau_{j} & = \E[\tau_{j}|S=s]     \\
         & = \pi_{o}\tau_{j,o} + \pi_{a}\tau_{j,a} +  \pi_{n}\tau_{j,n} +  \pi_{d}\tau_{j,d}
\end{align}
\end{subequations}

Next, to simply the above identity, I introduce two standard assumptions:

\indent

\noindent \textbf{Assumption A1} \textsc{(Ranking Monotonicity)}. \textit{Voter $i$ is at least as likely to rank candidate $j$ when the candidate appears at the top of ballot than otherwise. Formally, $M_{ij}(0) \leq M_{ij}(1)$ for all $i$ and $j$ with probability 1. Equivalently, $\pi_{d}=0$.}

\indent

Ranking monotonicity means that ballot order ($D_{ij}$) has a positive effect on whether voters rank candidate $j$ ($M_{ij}$), and it rules out the presence of defiers \citep[but also see][]{alvarez2006much}. This is a ``refutable'' assumption in that its plausibility can be tested by observed data \citep[46-48]{manski2009identification}.

\indent

\noindent \textbf{Assumption A2} \textsc{(Exclusion Restriction)}. \textit{Ballot order affects candidate ranking only through its effect on voters' decision to rank a given candidate. Consequently, the ARE is zero for always-rankers and never-rankers; $\tau_{a} = \tau_{n}=0$.}

\indent

Exclusion restriction was suggested by Figure \ref{fig:DAG}, where there is no direct causal path from ballot order ($D_{ij}$) to candidate ranking ($R_{ij}$). Because of this, ballot order ($D_j$) can be considered as an instrumental variable with respect to voters' ranking decision ($M_j$) \citep{angrist1996identification}. Exclusion restriction will be violated if always-rankers (never-rankers) put a \textit{higher} (\textit{lower}) rank on a given candidate \textit{because} they saw the candidate at the top of the ballot. I justify this assumption in the running application because always-rankers and never-rankers engage in ``ticket voting,'' in which ``a complete preference ordering could be expressed simply by adopting a party's pre-arranged preference schedule'' \citep[125]{reilly2001democracy}.

When Assumptions A1-A2 hold, it is then possible to simplify the ARE as follows:
\begin{subequations}
\begin{align}
\tau_{j}   \equiv & \ \pi_{o}\tau_{j,o} + \pi_{a}{\color{firebrick}\underbrace{\tau_{j,a}}_{0}} +  \pi_{n}{\color{firebrick}\underbrace{\tau_{j,n}}_{0}} +  {\color{firebrick}\underbrace{\pi_{d}}_{0}}\tau_{j,d}\\
           = & \ \pi_{o}\tau_{j,o}
\end{align}
\end{subequations}
\noindent Rearranging the above, the order-ranker ARE can be expressed as follows:
\begin{align}
\tau_{j,o} = \frac{\tau_j}{\pi_o},\label{eq:tauo}
\end{align}

\noindent where $\pi_{o}$ can be expressed as the difference in the proportions of voters who rank candidate $j$ in the treatment and control conditions:\footnote{This estimand can be thought of the two stage least square estimand in the context of instrumental variables.}
\begin{align}
{\pi}_{o} = \underbrace{{\mathbb{P}}(M_{ij}=1|D_{ij}=1)}_{\text{Prop. of order-rankers and always-rankers}} - \underbrace{{\mathbb{P}}(M_{ij}=1|D_{ij}=0)}_{\text{Prop. of always-rankers}} \label{eq:pio}
\end{align}

Thus, the violation of Assumptions A1-A2 identifies $\tau_{j,o}$ larger than its true value (and the corresponding estimator becomes positively biased).

\section{Discussion on Simulation}\label{app:sim}
\setcounter{figure}{0} 
\setcounter{table}{0}  
\setcounter{equation}{0} 
\setcounter{footnote}{0} 

\subsection*{Set Up}
Based on the Luce-Placket model of ranking data, I assume that each voter has a random utility for each of the six candidate: $V_{A,i} = 2\varepsilon^{'}_{A} + \varepsilon_{A}$, $V_{B,i} = 3\varepsilon^{'}_{B} + \varepsilon_{B}$, $V_{C,i} = \varepsilon^{'}_{C} + \varepsilon_{C}$, $V_{D,i} = 3\varepsilon^{'}_{D} + \varepsilon_{D}$, $V_{E,i} = 3\varepsilon^{'}_{E} + \varepsilon_{E}$, and $V_{F,i} = 4\varepsilon^{'}_{F} + \varepsilon_{F}$, where $\varepsilon^{'}_{j} \sim N(0,1)$ and $\varepsilon_{j} \sim N(0,1)$, respectively. I then define that $V_{A,i} = 2\varepsilon^{'}_{A} + \varepsilon_{A} + \tau$ in the treatment group and $V_{B,i} = 3\varepsilon^{'}_{B} + \varepsilon_{B} + \tau$ in the control group, where $\tau=2$. Building on these utility functions, I then generate $M_{j}$ by simulating a \textit{utility threshold} for each voter. For each voter, I draw a value from $N(0,1)$ and replace each candidate's utility with 0 if the candidate's utility is less than the voter's utility threshold. This represents a behavioral assumption that voters only rank candidates whose utilities are greater than or equal to their internally defined utility benchmark. They first choose which candidates they decide to rank and then rank the candidates. Finally, I generate $R_{j}$ by applying the Placket-Luce model of ranking data, where one candidate is sequentially chosen from a set of remaining candidates. Importantly, the utilities of unranked candidates are set to 0 and thus do not affect the ranking of each ranked candidate.

\subsection*{Simulation Study 1}

\begin{figure}[h!]
    \centering
    \includegraphics[width=14cm]{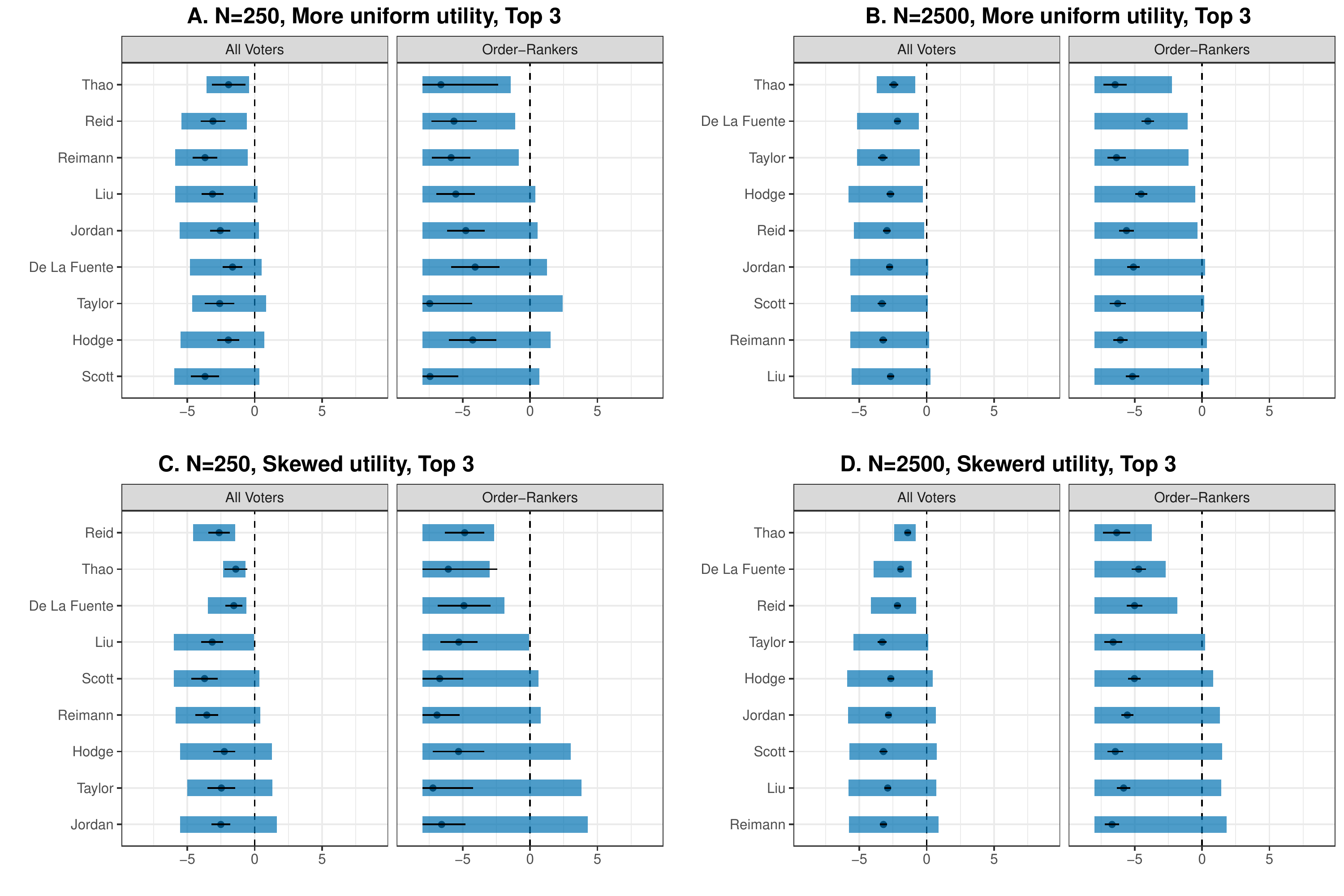}
    \caption{\textbf{Simulation Results by Sample Size and Distribution of Candidate Utilities}}
    \label{fig:Bound_Sim1}
\end{figure}

In Figure \ref{fig:Bound_Sim1}, I vary the number of experimental subjects and the distribution of voters' utilities over available candidates. Here, I fix the number of candidates subjects can rank (i.e., they can express up to their top-3 choice). There are several takeaways. First, the comparison between Panel A ($N=250$) and Panel B ($N=2500$) shows that sample size barely affects the width of the bound. The same pattern appears in the comparison between Panel C ($N=250$) and Panel D ($N=2500$) too. Second, the bound width is highly susceptible to the distribution of candidate utilities. The pattern is confirmed by comparing Panel A and Panel C and comparing Panel B and Panel D. This fits the logic behind the nonparametric bound: the more missing ranks a candidate has, the wider the bound for the candidate becomes (because we have less information we can rely on in the data). Consequently, when voters' support is concentrated around a few candidates, we have more information and narrower bounds for the popular candidates. At the same time, this suggests that we have less information and wider bounds for the remaining candidates.

\subsection*{Simulation Study 2}

\begin{figure}[h!]
    \centering
    \includegraphics[width=14cm]{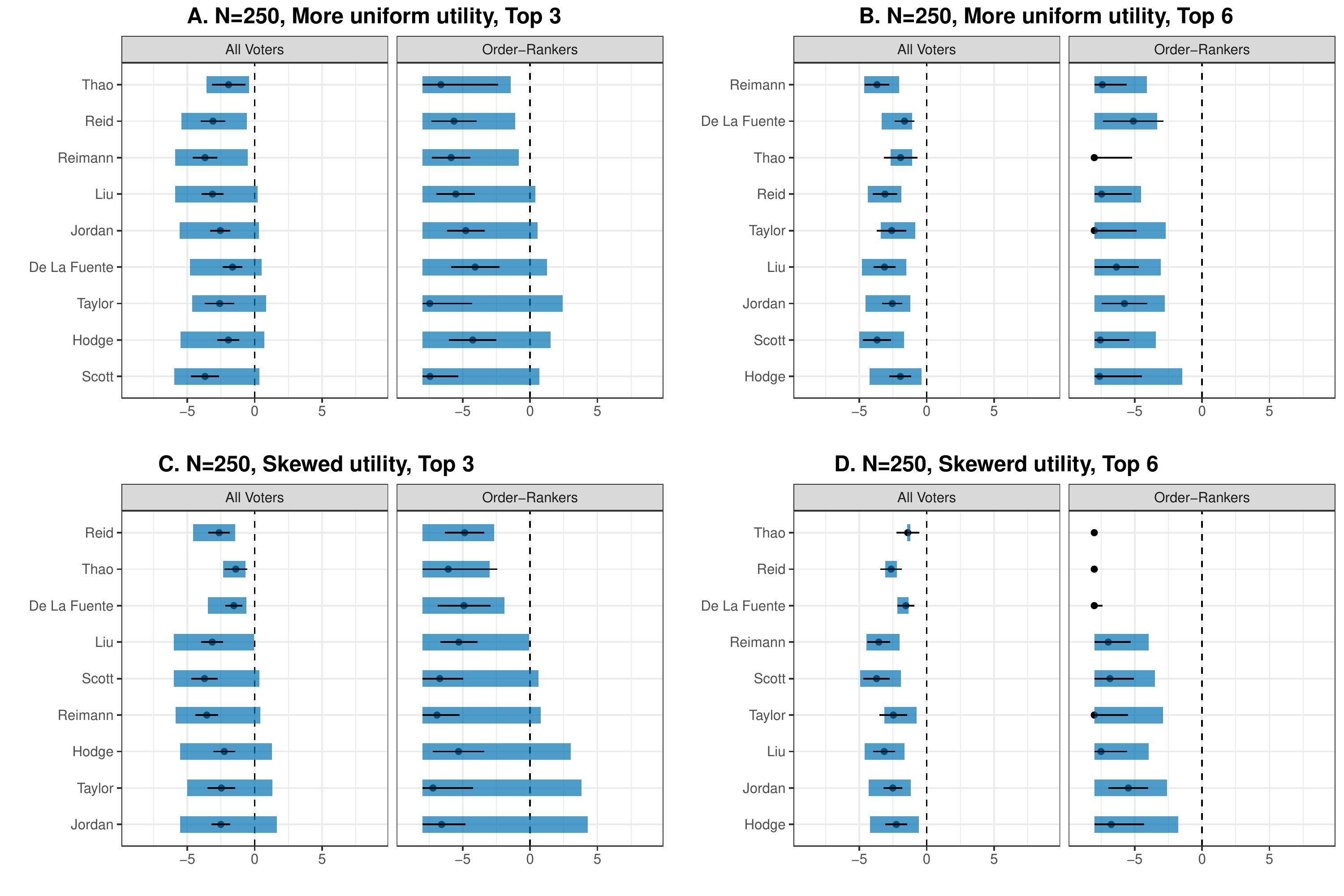}
    \caption{\textbf{Simulation Results by Maximum Number of Ranked Candidates and Distribution of Candidate Utilities}}
    \label{fig:Bound_Sim2}
\end{figure}

In Figure \ref{fig:Bound_Sim2}, I vary the number of candidates voters can rank up to and the distribution of voters' utilities over available candidates. While varying the two parameters, I fix the number of experimental subjects to $N=250$. The comparisons between Panel A and Panel B, as well as Panel C and Panel D, show that the bound width is susceptible to how many candidates voters can rank. This pattern fits the logic behind the nonparametric bound: the more non-missing ranks we have, the narrower the bound becomes.

\clearpage
\section{Identification and Estimation of Generalized Ballot Order Effects}\label{app:general}
\setcounter{figure}{0} 
\setcounter{table}{0}  
\setcounter{equation}{0} 
\setcounter{footnote}{0} 

First, I present a more general experiment design for quantifying ballot order effects. Let $O_{i,j}$ be the position of candidate $j$ that voter $i$ sees in the ordered set of candidates. For example, $O_{i,j}=1$ means that candidate $j$ is the first listed candidate, and $O_{i,j}=J$ suggests that the candidate is in the last position (for voter $i$). Let $\bm{O_{i[-j]}}$ be a random vector denoting the ordering of the other $J-1$ candidates in the set, and $\bm{o^*}$ be a realization of the random vector in the permutation space of $\bm{O}^{J-1}$.

In the above example, I considered the effect of candidate $j$ being listed in the first position as opposed to the second position relative to a particular ordering of other candidates. Formally, this effect can be written as:
\begin{align}
    \tau_{j[2]}^{BO}(\bm{o^{*}}) \equiv \E \left[ R_{ij}(O_{ij}=1) - R_{ij}(O_{ij}=2)|\bm{O_{i[-j]}}=\bm{o^{*}}\right]
\end{align}

More generally, I define the average ballot order effect for candidate $j$ as the average rank effect of being listed first ($O_{ij}=t_{1}$) as opposed to any other position averaged over the joint probability distribution of (1) the other positions (for all $t$) and (2) the ordering of other candidates (for all $\bm{o^*}$):\footnote{More generally, $t_1$ can be defined as any combination of multiple positions (e.g., first and second positions combined).}
\begin{align}
    \tau_{j}^{BO} & \equiv \E \left[ R_{ij}(D_{i,j}=1)  - R_{ij}(D_{i,j}=0)\right]\\
 & = \E \bigg[\E \left[ R_{ij}(O_{ij}=t_{1}, \bm{O_{i[-j]}}=\bm{o^{*}})  - R_{ij}(O_{ij}=t, \bm{O_{i[-j]}}=\bm{o^{*}})\big|\bm{O_{i[-j]}=\bm{o^*}}\right] \bigg]\\    
      & = \sum_{\bm{o^*}\in \bm{O}^{J}} \Bigg\{ \sum_{t=2}^{J}\E \left[ \underbrace{R_{ij}(O_{ij}=t_{1}, \bm{O_{i[-j]}}=\bm{o^{*}})}_{\text{potential rank under the treatment}}  - \underbrace{R_{ij}(O_{ij}=t, \bm{O_{i[-j]}}=\bm{o^{*}})}_{\text{potential rank under the control}}\bigg|\underbrace{\bm{O_{i[-j]}=\bm{o^*}}}_{\text{Ordering of other candidates}}\right] \\
        & \times {\color{firebrick}\underbrace{\mathbb{P}(t_{1}, t|\bm{O_{i[-j]}=\bm{o^*}})}_{\text{conditional probability of candidate $j$ listed in 1st and $t$th positions}}} \Bigg\}\nonumber\\
        & \times \color{navyblue}\underbrace{\mathbb{P}(\bm{O_{i[-j]}=\bm{o^*}})}_{\text{probability of a particular ballot order $o^*$}}  \nonumber
\end{align}

Finally, I make an additional assumption to simplify the identification of the average ballot order effect. Specifically, I assume that it is equally likely that each voter sees a particular comparison between $O_{j}=t$ and $O_{j}=t_{1}$ as well as a particular ordering of other candidates $\bm{O_i} = o^*$. Since there are $J-1$ possible numbers of pairwise comparisons and $(J-1)!$ possible numbers of candidate permutation, I introduce the following:

\indent

\noindent \textbf{Assumption A3} \textsc{(Ballot Order Randomization)}. \textit{Ballot order is randomized at the voter level such that $\mathbb{P}(R_{ij}(O_{j}=1, \bm{O_{i,-j}}=\bm{o^{*}})  - R_{ij}(O_{j}=t, \bm{O_{i,-j}}=\bm{o^{*}})|\bm{O_{i,-j}=\bm{o^*}}) = \frac{1}{J-1}$ and $\mathbb{P}(\bm{O_{i,-j}=\bm{o^*}}) = \frac{1}{(J-1)!}$ Here, it is also assumed that both probabilities are greater than zero and less than one.} 

\indent

With this assumption (and assuming full rankings), I define the following:\\

\noindent \textbf{Proposition A1} \textsc{(Nonparametric Estimator for the General Ballot Order Effect )}. \textit{Given Assumptions 1-7, the average ballot order effect for candidate $j$ is estimated by the following doubly-averaged-difference-in-mean-ranks estimator:}
\begin{align}
    \widehat{\tau}_{j}^{BO} & = \widehat{\E} \left[R_{ij}|D_{i,j}=1, \bm{O_{i}}=\bm{o^{*}}\right]  - \widehat{\E}\left[R_{ij}|D_{i,j}=0, \bm{O_{i}}=\bm{o^{*}}\right]\\
      & = {\color{navyblue}\underbrace{\frac{1}{(J-1)!} \sum_{\bm{o^*}\in \bm{O}^{J-1}}}_{\text{average over all ordering }}} {\color{firebrick}\underbrace{\frac{1}{J-1}\sum_{t=2}^{J}}_{\text{average over all comparison}}} \Bigg\{ \widehat{\E} \left[ R_{ij}|O_{j}=1, \bm{O_{i}}=\bm{o^{*}}\right]  -\widehat{\E}\left[ R_{ij}|O_{j}=t, \bm{O_{i}}=\bm{o^{*}}\right] \Bigg\}
\end{align}

The above quantity can be estimable by the within-strata difference-in-mean ranks as well as by ordinary least square regression. One advantage of this design is that it allows researchers to decompose the general ballot order effect both by (a) the position of comparison and (b) the ordering of other candidates. Meanwhile, one challenge is that the number of possible treatment-control comparisons is quite large. In the running application, for example, there are $(J-1)!(J-1)J = 8!\times 8 \times 9= 2903040$ such comparisons for ten candidates.


One way to regularize the number of possible strata is to assume a constant rank order effect by the remaining ballot order $\bm{O_{i[-j]}}$ as well as candidate $j$'s counterfactual position $t$.

\indent

\noindent \textbf{Assumption A4} \textsc{(Mean Independence on Remaining Ballot Order)}. \textit{$\{\E[R_{ij}(O_{ij}=t_1)],\E[R_{ij}(O_{ij}=t)]\} \indep \bm{O_{i[-j]}}$ for all $i$, $j$, and $t$.}\\

\noindent \textbf{Assumption A5} \textsc{(Mean Independence on Counterfactual Position)}. \textit{$\E[R_{ij}(O_{ij}=t)] =\E[R_{ij}(O_{ij}=t^*)] $ for all $t \neq t_1$ and $t^* \neq t_1$.}\\

These assumptions allow one to pool the information from different treatment-control comparison groups. The validity of each assumption can be empirically tested by estimating (and comparing) the ballot order effect within each stratum. With these assumptions, it is possible to simplify the estimator further.\\

\noindent \textbf{Proposition A2} \textsc{(General Ballot Order Effect via Position and Ordering Mean Independence )}. \textit{Given Proposition 2 and Assumptions 8-9, the average ballot order effect for candidate $j$ is estimated by the following difference-in-mean-ranks estimator:} 
\begin{align}
    \widehat{\tau}_{j}^{BO} & = \frac{R_{ij}I(O_{ij}=t_1)}{N_{j1}} -    \frac{1}{J-1}\sum_{t=2}^{J}\Bigg[\frac{R_{ij}I(O_{ij}=t)}{N_{jt}}\Bigg] \quad \text{(with Assumption A4)}\\
    & = {\color{firebrick}\underbrace{\frac{R_{ij}I(O_{ij}=t_1)}{N_{j1}}}_{\text{mean rank for treated units}}} -    {\color{navyblue}\underbrace{\frac{R_{ij}I(O_{ij}\neq t_1)}{\sum_{t=2}^{J}N_{jt}}}_{\text{mean rank for control units}}} \quad \text{(with Assumptions A4-A5)},
\end{align}

\noindent where $N_{j1} = \sum_{i=1}^{N}I(O_{ij}=t_{1})$ be the number of treated units (in which candidate $j$ appears at the top of the ballot) and $N_{jt} = \sum_{i=1}^{N}I(O_{ij}=t)$ be the number of control units (who see candidate $j$ in the $t$-th position of the ballot) such that $N = N_{j1} + \sum_{t=2}^{J}N_{jt}$.

Finally, in the presence of partial rankings, researchers can construct the nonparametric bounds by applying the estimator to the two imputed samples. 

\clearpage
\section{Survey Experiment Questions}\label{app:surveyX}
\setcounter{figure}{0} 
\setcounter{table}{0}  
\setcounter{equation}{0} 
\setcounter{footnote}{0} 

This section presents the copies of survey instruction and questions for the Oakland 2022 mayoral election. The same set of instructions and questions were used for the US House and US Senate elections in Alaska.

\begin{figure}[h!]
    \centering
    \includegraphics[width=10cm]{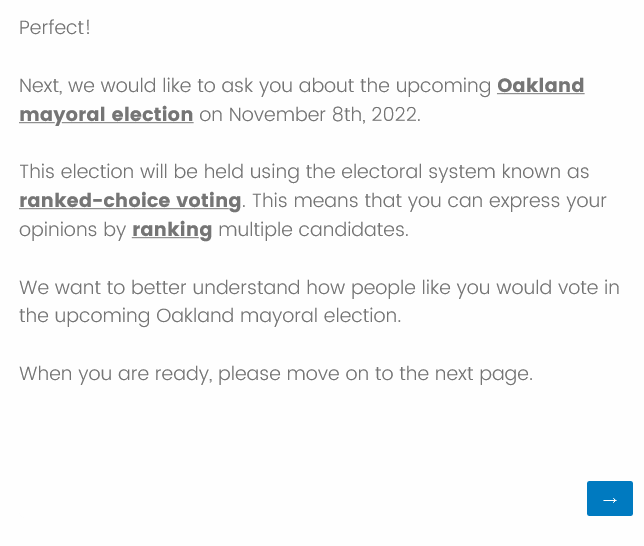}
    \caption{\textbf{Instruction for Experimental Questions}}
    \label{fig:SS_Inst}
\end{figure}

\begin{figure}[h!]
    \centering
    \includegraphics[width=10cm]{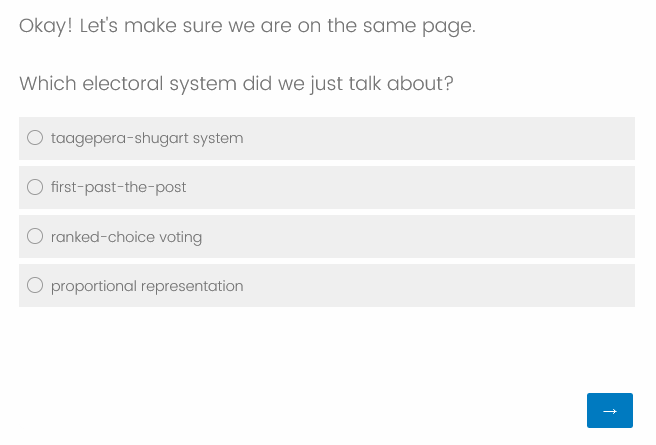}
    \caption{\textbf{Attention Check prior to Experimental Questions}\\ \textit{Note}: The other three attention checks include ``Do you agree to participate?'' (\underline{Yes} or No), ``For our research, careful attention to survey questions is critical! We thank you for your care.'' (\underline{I understand} or I do not understand), and ``People are very busy these days and many do not have time to follow what goes on in the government. We are testing whether people read questions. To show that you've read this much, answer both "extremely interested" and "very interested."'' (\underline{Extremely interested}, \underline{Very interested}, Moderately interested, Slightly interested, or Not interested at all). Attentive respondents are those who select all \underline{underlined items}.}
    \label{fig:SS_Attention}
\end{figure}

\begin{figure}[h!]
    \centering
    \includegraphics[width=8cm]{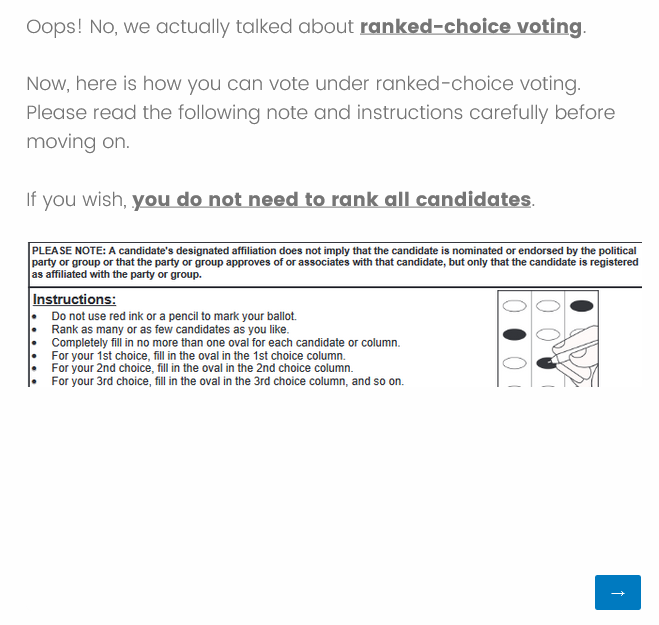}
    \caption{\textbf{Post-Attention Check Instruction}}
    \label{fig:SS_AttentionPost}
\end{figure}

\begin{figure}[h!]
    \centering
    \includegraphics[width=8cm]{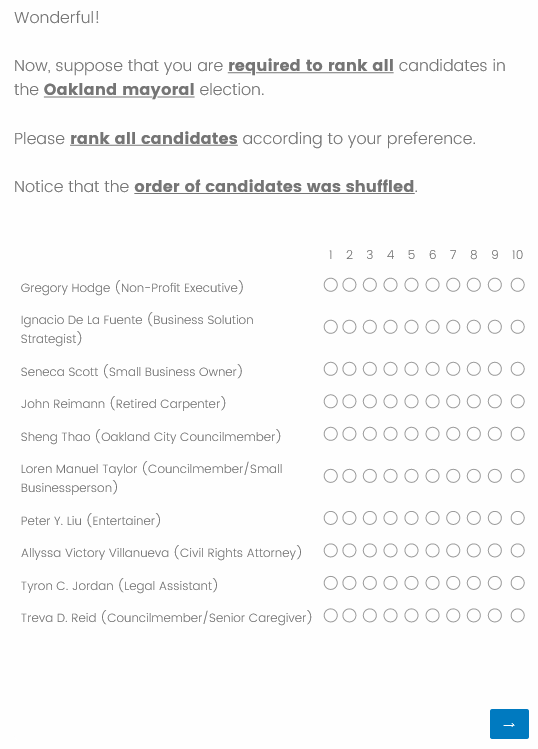}
    \caption{\textbf{Secondary Experimental Question}}
    \label{fig:SS_Sub}
\end{figure}

\clearpage
\section{Additional Results on Ballot Order Effects}\label{app:alaska}
\setcounter{figure}{0} 
\setcounter{table}{0}  
\setcounter{equation}{0} 
\setcounter{footnote}{0} 

This section presents the experimental results for the U.S. House and Senate elections in Alaska.

\begin{figure}[h!]
    \centering
    \includegraphics[width=14cm]{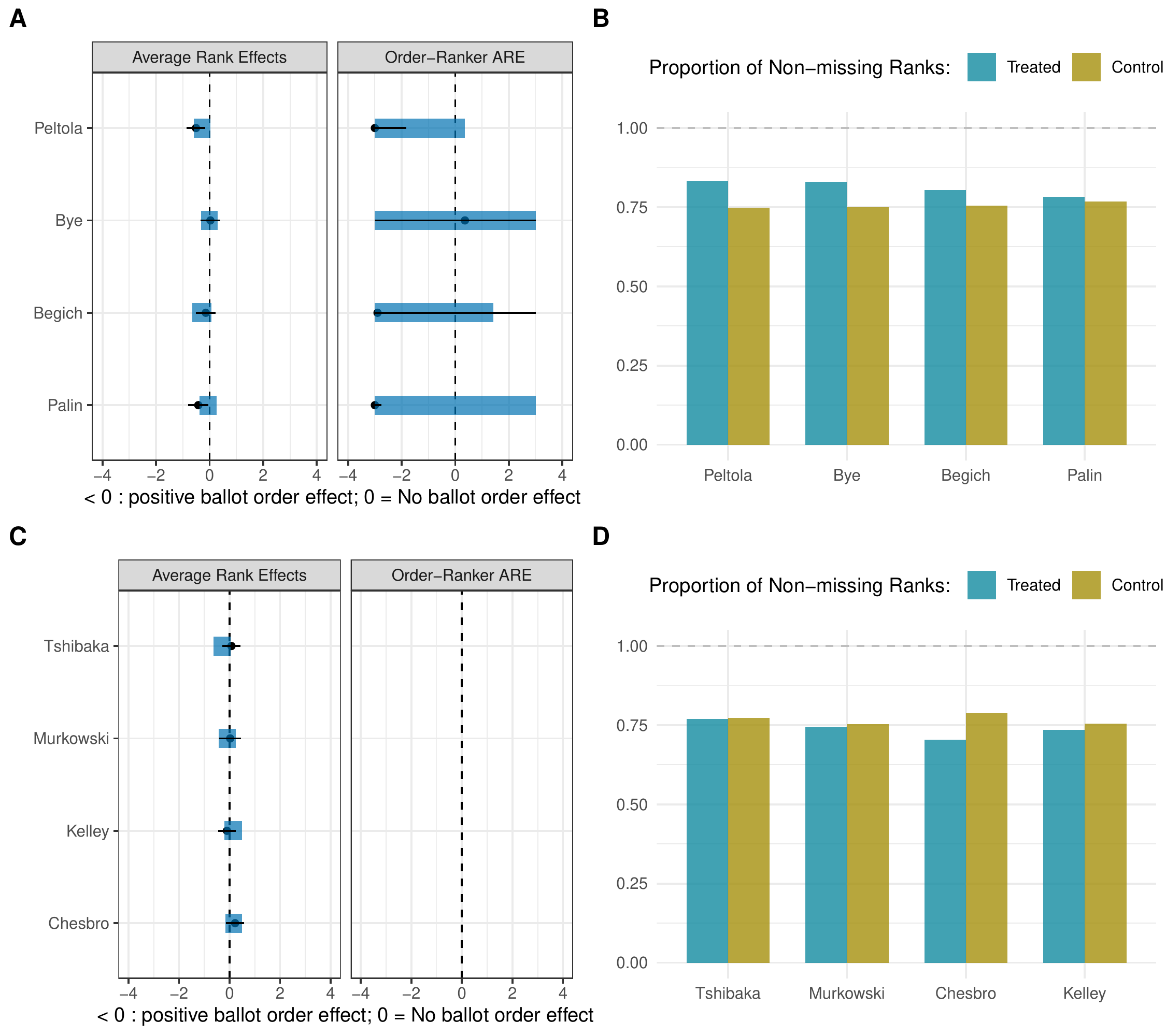}
    \caption{\textbf{Nonparametric Ballot Order Effects in the 2022 U.S. House of Representatives Election in Alaska}\\ \textit{Note}: Panels A-B report results for the U.S. House and Panel C-D present findings for the U.S. Senate elections, respectively.}
    \label{fig:bound2}
\end{figure}

Figure \ref{fig:bound2} shows the results for the U.S. House and Senate elections in Alaska. The house election featured four candidates, including Mary Peltora (Democratic incumbent), Nicholas Begich (Republican), Sarah Palin (Republican), and Chris Bye (Libertarian). The senate race had four candidates, including Senator Lisa Murkowski (Republican Incumbent), Kelly Tshibaka (Republican), Patricia Chesbro (Democrat), and Buzz Kelley (Republican). The design reflects Alaska's design that voters can rank as many or as few candidates as they wish. 


\begin{figure}[h!]
    \centering
    \includegraphics[width=6cm]{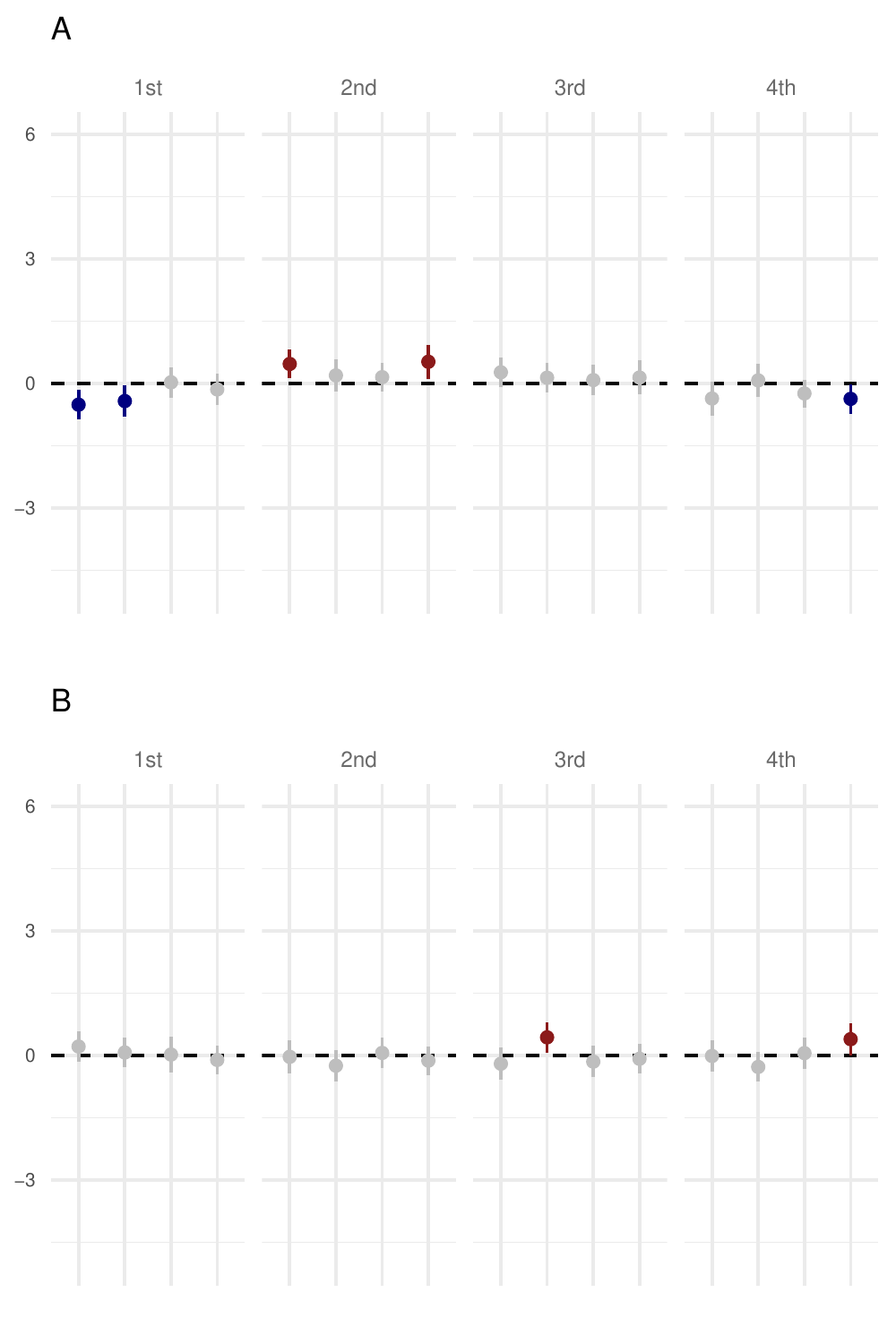}
    \caption{\textbf{Generalized Ballot Order Effects Across Positions in Alaska}\\ \textit{Note}: Panel A is based on the U.S. House of Representatives election in Alaska. Panel B is based don the U.S. Senate election in Alaska.}
     \label{fig:general2}
\end{figure}

In Panels A and C, the nonparametric bounds suggest that the ballot order (i.e., first position) does not give candidates higher ranks in both elections. The fully ranked data confirm this finding while showing that the ballot order effects are statistically significant for Peltola and Pailin in the U.S. House election. Moreover, order-ranker AREs are identified for all candidates in the House race but only for one candidate in the Senate race. Panels B and D show why --- respondents are more likely to rank Begich, Bye, Peltola, Palin, and Tshibaka, but less likely to rank Murkowski, Chesbro, and Kelly when they appear at the top of the ballot. Figure \ref{fig:general2} shows that most ballot order effects were statistically insignificant based on fully ranked data in Alaska.

\indent

\begin{figure}[tbh!]
    \centering
    \includegraphics[width=8cm]{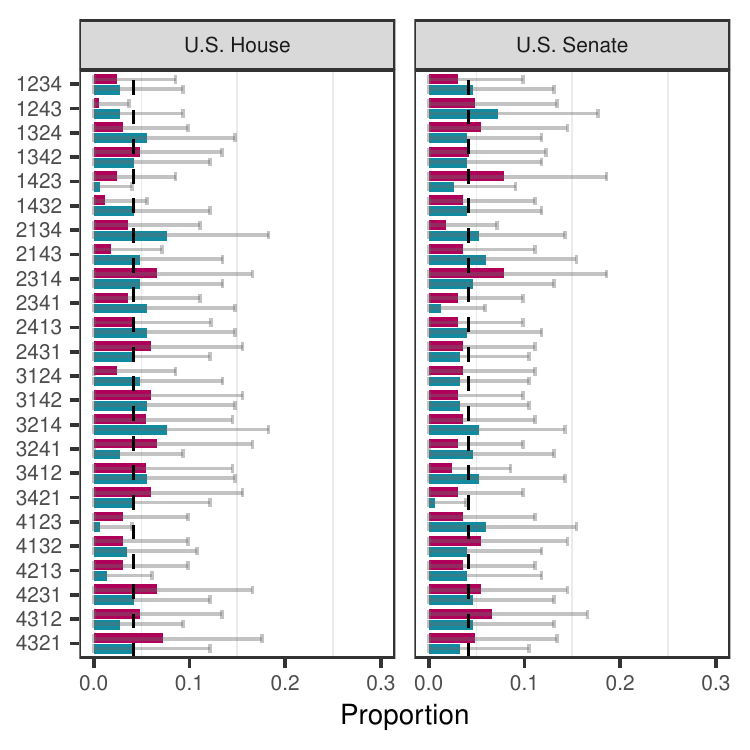}
    \caption{\textbf{The Distributions of Observed Rankings among Inattentive Respondents.}\\ \textit{Note}: A dashed line shows the uniform distribution in each panel. Each panel shows the proportion of unique rankings based on partially (red) and fully ranked (blue) data. 95\% confidence intervals are displayed.}
     \label{fig:pattern_inatt}
\end{figure}

\clearpage
\section{Applications in International Relations}\label{app:IR}
\setcounter{figure}{0} 
\setcounter{table}{0}  
\setcounter{equation}{0} 
\setcounter{footnote}{0} 

\subsection*{Conflict-Related Sexual Violence}

\citetSM{agerberg2022sexual} study whether people are more likely to support humanitarian intervention in armed conflicts when the case of interest involves sexual violence. The authors' \textit{gendered protection norm} theory states that conflict-related sexual violence ``elicits support for intervention because it activates gendered protection norms grounded in benevolent sexist notions of women as innocent, vulnerable victims'' \citepSM[4]{agerberg2022sexual}.

\indent

In their vignette experiment, survey respondents from the U.S., the U.K., and Sweden were exposed to one of three conditions:

\indent

``A poor country has been ravaged by an increasingly violent civil war for the last seven years. After the president refused to step down following his defeat in the presidential election, rebel groups started challenging the government. The conflict soon spiraled out of control, \textbf{[experimental manipulation]}. Several diplomatic attempts at conflict resolution have been unsuccessful. The United Nations have strongly condemned the fighting, and the [widespread violence/ widespread use of sexual violence/
widespread ethnic violence].

\indent

\textbf{Experimental manipulation}
\begin{itemize}
    \item Control: ``resulting in widespread violence''
    \item Main treatment: ``resulting in widespread violence with widespread use of sexual violence by all armed actors''
    \item Auxiliary treatment: ``resulting in widespread ethnic violence''
\end{itemize}

\indent

After the treatment assignment, the authors measure two outcome variables. The secondary outcome variable is based on the following question:

\indent

``In societies emerging from conflict, international organizations, aid agencies and states promote and support a set of initiatives and projects. Yet, especially when facing budget constraints, international actors often disagree about the relative importance of different issues. In the conflict scenario described above, how would you rank the following priorities in order of importance, once the violence has ended? Rank the different priorities using number 1 to 5, with 1 being most important.''

\begin{itemize}
    \item Job generation and employment initiatives
    \item Reconciliation activities between victims and perpetrators of violence
    \item Rebuilding schools and hospitals
    \item Including and empowering women in the peace process
    \item Improving governance and accountability
\end{itemize}

The original analysis (presented in their Online Appendix B) examines the treatment effects of the probability that the inclusion and empowerment of women is the most prioritized or $\mathbb{P}(R_{i,j=4}=1)$. I call this class of effects the \textit{average first-rank effects}. The authors estimate these effects by first estimating logistic regression and computing predicted probabilities that women's inclusion and empowerment is selected as the most important item.

\indent

However, this strategy is sub-optimal \textit{given} their hypotheses of interest,  which concern the \textit{relative} priority of the item over the other four items (emphasis added):

\begin{itemize}
    \item [H2a:] Respondents are \textbf{more likely to prioritize} women’s empowerment and active inclusion in the peace process after a conflict with widespread sexual violence compared to a conflict with overall widespread violence
    \item [H2b:] Respondents are \textbf{more likely to prioritize} women’s empowerment and active inclusion in the peace process after a conflict with widespread sexual violence compared to a conflict with widespread ethnic violence.
\end{itemize}

\begin{figure}[h!]
    \centering
    \includegraphics[width=15cm]{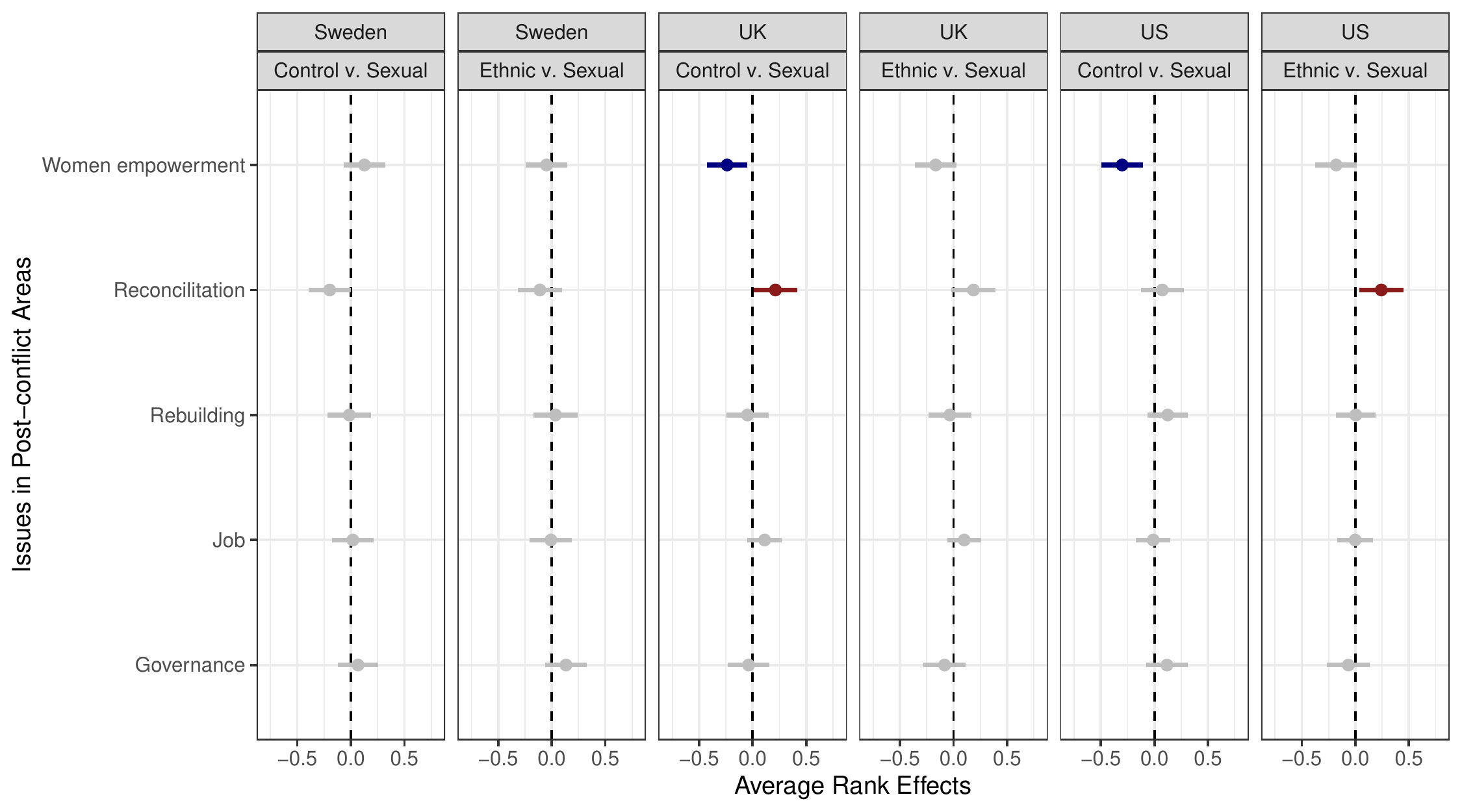}
    \caption{\textbf{Average Rank Effects for Five Issues in Post-Conflict Regions}}
     \label{fig:IR_sexual}
\end{figure}

To better test these hypotheses, I suggest that average rank effects be used for the reanalysis. Figure \ref{fig:IR_sexual} reports the estimated average rank effects for the five issues by the treatment type and across countries. The results support the original conclusion for H2a based on the average first-rank effects: the sexual violence treatment (versus the control) makes people more likely to prioritize women's inclusion and empowerment in post-conflict regions only in the U.S. and the U.K. Moreover, the figure also shows that the sexual violence treatment (as opposed to the control) did not affect the average ranks of the other four items, except for reconciliation in the U.K. While the other four items were not examined in the original study, doing so provides additional support for H2a.

\indent

Regarding H2b, however, the reanalysis did not find statistically significant average rank effects of the sexual violence treatment (versus the ethnic violence treatment). While I find that the ethnic violence treatment (versus the control) does not affect the average rank of women's inclusion and empowerment, the reanalysis implies that there is not enough evidence to confirm or deny H2b.      

\indent

\subsection*{Donor Political Economy and Aid Effectiveness}

\citetSM{dietrich2016donor} examines whether the quality of governance of an aid-receiving country affects the aid decision-makers in an aid-providing country. To test her hypothesis, the author performs a vignette experiment on former or current aid officials from France, Japan, Germany, the U.S., and Sweden. In the experiment, each official was exposed to ``three hypothetical low-income-country scenarios, which differ only in the quality of governance'' (91). These three conditions include:

\begin{itemize}
    \item Control (Country A): a well-performing low-income country with relatively low levels of corruption and strong state institutions.
    \item Treatment I (Country B): weak state institutions—while corruption levels are relatively low
    \item Treatment II (Country C): a
large-scale corruption scandal involving public-sector officials—while state institutions
are relatively strong
\end{itemize}

After exposing to each condition (the order of B and C was randomized), survey respondents are asked to ``rank-order five aid delivery channels'' \citepSM[91]{dietrich2016donor}, including

\begin{itemize}
    \item The recipient government
    \item International organizations
    \item International NGOs
    \item Local NGOs
    \item Private-sector actors
\end{itemize}

Although the study collected ranking data on the five items, the study only analyzed the outcome for the recipient government. The author hypothesizes by noting that ``I expect officials from the United States and Sweden, on average, to change toward greater [support for using the recipient government as a bypass] than their counterparts from France, Japan, and Germany'' (93). Importantly, the study uses the recipient government as the alternative to \textbf{nonstate development
actors}. Thus, ``If respondents rank the recipient government as a 5 it means that they have a clear NO BYPASS preference'' (91).

\indent

To test this hypothesis, the original study performs a ``difference-in-difference''analysis, where (1) it takes the difference between the ranks of the recipient government between the control group and each treatment condition and (2) takes another difference between the political economy type (0 for France, Japan, and Germany; 1 for USA and Sweden) after that. Importantly, the quantity of interest was not clearly defined and thus it is not entirely clear if the estimated ``effects'' support the above hypothesis.

\indent

\begin{figure}[h!]
    \centering
    \includegraphics[width=16cm]{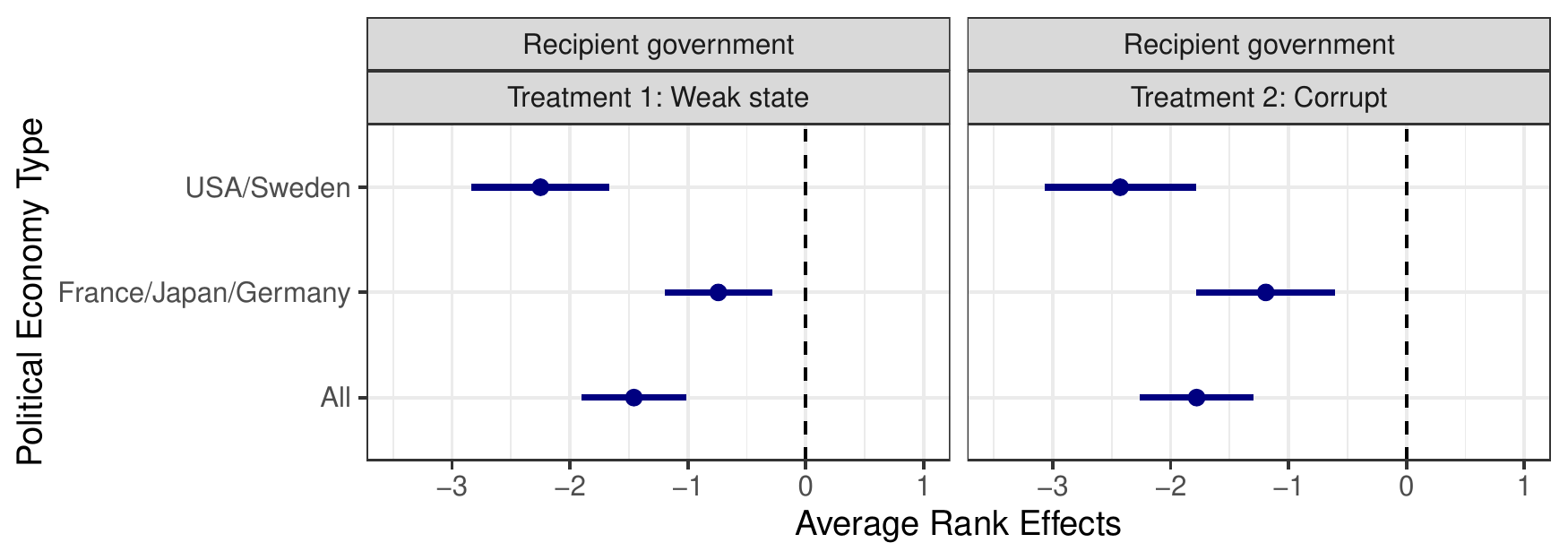}
    \caption{\textbf{Average Rank Effects for Recipient Government as a Preferred Delivery Channel}}
     \label{fig:IR_aid}
\end{figure}

To better test the author's hypothesis, I estimate the \textit{conditional} average rank effect for the recipient government by subsetting the sample based on the political economy type. Figure \ref{fig:IR_aid} presents the estimated effects by the political economy type and the treatment conditions. The figure shows that both treatments lead to higher ranks (negative values) of the recipient government. Moreover, the effect size appears to be larger among aid officials from the US and Sweden than from France, Japan, and Germany, which provides more explicit evidence for the hypothesis.

\indent

However, to further provide evidence for the author's argument, it is essential to examine whether the treatments cause people to prioritize the recipient government over each of the other delivery channels (international organizations, international NGOs, local NGOs, and private-sector actors). Especially, it is informative to study the \textbf{pairwise ranking of the recipient government and private-sector actors} given the argument in the original study. This way, the proposed causal inference framework with ranking data allows researchers to perform more analyses than they conventionally do without the framework.

\indent

\subsection*{Impact of Economic Coercion on Support for Appreciation}
\citetSM{gueorguiev2020impact} study whether economic coercion by the U.S. government affects public opinion --- specifically support for appreciation --- in China. In their survey experiment, respondents were exposed to the following information and one of the three additional explanations.

\indent

''An increase in the value of the RMB relative to other currencies makes imports cheaper whereas a decrease makes China’s exports more competitive in world markets. [Additional information]'' 

\begin{itemize}
    \item Control: (No additional statement)
    \item Encouragement (Treatment I): ``America has encouraged China to increase the value of the RMB relative to the dollar.''
    \item Threat (Treatment II): ``America has threatened to impose taxes on its imports of Chinese-made goods if China does not increase the value of the RMB relative to the dollar.''
\end{itemize}

One of the core analyses is to estimate the average treatment effects on people's support for appreciation \textit{after} subsetting the full sample by how survey respondents perceive US-China currency relations. More specifically, \textit{after} the experimental manipulation, respondents are asked to rank order the following statements:

\begin{itemize}
    \item Adversaries: China and America are adversaries
    \item Competitors: China and America are cooperative competitors
    \item Partners: China and America are international partners
\end{itemize}

The authors then separately estimate the effect of each treatment by focusing on respondents who choose the ``adversaries'' option first and those who do not. However, since the rank-order question followed the experimental manipulation, it is critical to assess whether the treatment does not affect people's rankings on the three statements. If, indeed, the treatment affects their rankings, it is likely that their original conclusion may be susceptible to post-treatment bias. 

\indent

\begin{figure}[h!]
    \centering
    \includegraphics[width=13cm]{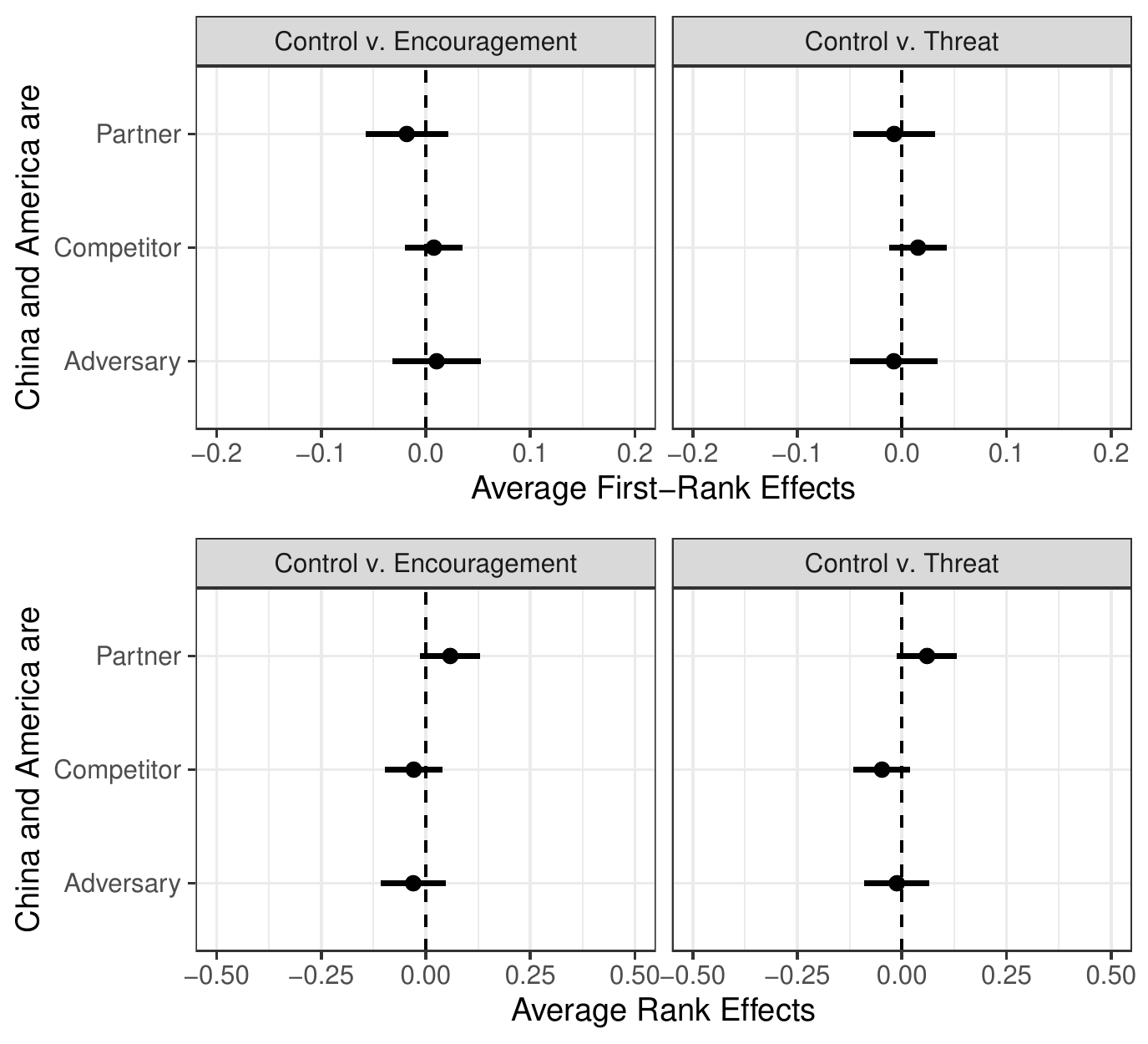}
    \caption{\textbf{Average First-Rank and Average Rank Effects for Three Perceptions of US-China Currency Relations}}
     \label{fig:IR_economic}
\end{figure}

Motivated by this problem, I estimate the average first-rank effect $\mathbb{P}(R_{ij}=1)(1) - \mathbb{P}(R_{ij}=1)(0)$ and average rank effect $\E[R_{ij}(1)] -\E[R_{ij}(0)]$ of each treatment on the three options. The upper panel of Figure \ref{fig:IR_economic} reports the estimated average first-rank effects of all three options. The results show that people's rankings are not affected by the treatments, which suggests that post-treatment bias seems less likely. While the original study subsets the entire sample by using the ``adversary'' option, it is encouraging to confirm that the other two options do not get affected by either treatment. 

\indent

What if each treatment affects their entire rankings while not affecting their first choice? To address this problem, I also estimate the average rank effect of each treatment for all options. The lower panel reports the estimated effects, which turn out to be all statistically and substantively not significant. Taken together, I find that the original analysis is robust to the above additional examination of the treatment effects on people's rankings.

\clearpage
\singlespacing
\bibliographystyleSM{apsr}
\bibliographySM{RankOutcome.bib}

\end{document}